\documentclass[aps,nofootinbib,preprint,superscriptaddress]{revtex4}%
\usepackage{hyperref}
\usepackage{amsmath}
\usepackage{amsfonts}
\usepackage{amssymb}
\usepackage{graphicx,subfig}
\usepackage{color}
\usepackage{graphicx}%
\providecommand{\U}[1]{\protect\rule{.1in}{.1in}}
\begin{document}
\title{Residues of bosonic string scattering amplitudes and the Lauricella functions}
\author{Sheng-Hong Lai}
\email{xgcj944137@gmail.com}
\affiliation{Department of Electrophysics, National Yang Ming Chiao Tung University,
Hsinchu, Taiwan, R.O.C.}
\affiliation{Department of Electrophysics, National Chiao-Tung University, Hsinchu, Taiwan, R.O.C.}
\author{Jen-Chi Lee}
\email{jcclee@cc.nctu.edu.tw}
\affiliation{Department of Electrophysics, National Yang Ming Chiao Tung University,
Hsinchu, Taiwan, R.O.C.}
\affiliation{Department of Electrophysics, National Chiao-Tung University, Hsinchu, Taiwan, R.O.C.}
\author{Yi Yang}
\email{yiyang@mail.nctu.edu.tw}
\affiliation{Department of Electrophysics, National Yang Ming Chiao Tung University,
Hsinchu, Taiwan, R.O.C.}
\affiliation{Department of Electrophysics, National Chiao-Tung University, Hsinchu, Taiwan, R.O.C.}
\date{\today}

\begin{abstract}
We calculate explicitly residues of all $n$-point Koba-Nielsen (KN) amplitudes
by using on-shell recursion relation of string scattering amplitudes (SSA). In
addition, we show that the residues of all SSA including the KN amplitudes can
be expressed in terms of the Lauricella functions. This result demonstrates
the exact $SL(K+3,%
%TCIMACRO{\U{2102} }%
%BeginExpansion
\mathbb{C}
%EndExpansion
)$ symmetry of the tree-level open bosonic string theory. Moreover, we derive
an iteration relation among the residues of a given SSA. This iteration
relation is related to the $SL(K+3,%
%TCIMACRO{\U{2102} }%
%BeginExpansion
\mathbb{C}
%EndExpansion
)$ symmetry and can presumably be used to soften the well-known hard SSA.

\end{abstract}
\maketitle
\tableofcontents

%

%TCIMACRO{\TeXButton{equation number}{\setcounter{equation}{0}
%\renewcommand{\theequation}{\arabic{section}.\arabic{equation}}}}%
%BeginExpansion
\setcounter{equation}{0}
\renewcommand{\theequation}{\arabic{section}.\arabic{equation}}%
%EndExpansion

\section{Introduction}

It is known that a string scattering amplitude (SSA) can be expressed in terms
of\ an infinite sum of simple pole terms with residues which are
well-organized that its hard high energy behavior is very soft and decays as
an exponential fall-off. A well-known simple example is the four tachyon
Veneziano amplitude \cite{GSW}. Presumably, there is a mechanism similar to
the symmetry principle in quantum field theory (QFT) which triggers this soft
behavior for hard SSA. In contrast to QFT, string theory as a quantum theory
of an extended string consists of an infinite number of particles with
arbitrary high spins and masses in the spectrum. It is thus reasonable to
believe that there exists a huge symmetry group \cite{Moore,Moore1,CKT} which
relates these infinite number of couplings and soften the SSA in the hard
scattering limit.

Recently, the present authors calculated a subset of exact $4$-point SSA,
namely, amplitudes of three tachyons and one arbitrary string states, and
expressed them in terms of the $D$-type Lauricella functions \cite{LLY2}. In
addition, it was shown that these Lauricella SSA (LSSA) can be expressed in
terms of the basis functions in the infinite dimensional representation of the
$SL(K+3,C)$ group \cite{Group}. For any fixed positive integer $K$, one has
infinite number of LSSA in the $SL(K+3,C)$ representations \cite{slkc}.
Moreover, it was further shown that there existed $K+2$ recurrence relations
among the $D$-type Lauricella functions. These recurrence relations can be
used to reproduce the Cartan subalgebra and simple root system of the $SL(K+3,%
%TCIMACRO{\U{2102} }%
%BeginExpansion
\mathbb{C}
%EndExpansion
)$ group with rank $K+2$. As a result, the $SL(K+3,%
%TCIMACRO{\U{2102} }%
%BeginExpansion
\mathbb{C}
%EndExpansion
)$ group with its corresponding stringy Ward identities (recurrence relations)
can be used to solve \cite{solve} all the LSSA and express them in terms of
one amplitude. See the recent review \cite{LSSA}.

The LSSA can be used to rederive the solvability of SSA in various scattering
limits \cite{LLY2},\cite{0705,RR1,RR2,RR3,RR4,RR5,RR6,RR7}. One important
application of this solvability was in the hard string scattering limit. In
particular, it was shown that in the hard scattering limit the LSSA can be
used to reproduce \cite{LLY2} infinite linear relations with constant
coefficients among all hard SSA and solve the ratios among them. These
behaviors of high energy string scatterings \cite{GM,GM1,GrossManes} were
first conjectured by Gross \cite{Gross,Gross1} and later corrected
\cite{ChanLee,ChanLee2,CHL} and proved \cite{ChanLee,ChanLee2,CHLTY2,CHLTY1}
by the method of decoupling of zero norm states (ZNS) \cite{Lee,lee-Ov,LeePRL}%
. For details, see the recent review papers \cite{review,over}. More
importantly, since the decoupling of ZNS and thus the infinite linear
relations in the hard scattering limit persist to all string loop orders, it
was believed that the proposed $SL(K+3,%
%TCIMACRO{\U{2102} }%
%BeginExpansion
\mathbb{C}
%EndExpansion
)$ symmetry at string-tree level is also valid for string loop amplitudes. On
the other hand, one believes that by keeping $M$ fixed as a finite constant in
the ZNS calculation, one can obtain more information about the high energy
behavior of string theory in contrast to the tensionless string ($\alpha
^{\prime}\rightarrow\infty$) approach
\cite{less1,less2,less3,less5,less6,less7,less8,less9,less10,less11,less12} in
which all string states are massless.

In this paper, we will apply the string theory extension \cite{stringbcfw},
\cite{bcfw3,bcfw4} of field theory BCFW on-shell recursion relations
\cite{bcfw1,bcfw2} to calculate explicitly residues of all $n$-point
Koba-Nielsen (KN) amplitudes. In addition, we will show that the residues of
all SSA including the KN amplitudes can be expressed in terms of the
Lauricella functions. This result demonstrates the exact $SL(K+3,%
%TCIMACRO{\U{2102} }%
%BeginExpansion
\mathbb{C}
%EndExpansion
)$ symmetry of the tree-level open bosonic string theory. On the other hand,
since the LSSA of three tachyons and one arbitrary string states can be
rederived \cite{LLYT} from the deformed cubic string field theory (SFT)
\cite{Taejin}, we conjecture that the proposed $SL(K+3,%
%TCIMACRO{\U{2102} }%
%BeginExpansion
\mathbb{C}
%EndExpansion
)$ symmetry can be obtained from Witten SFT.

This paper is organized as following. In section II, we give a brief review of
the calculation of the LSSA of three tachyons and one arbitrary string states.
In addition, we review the $SL(K+3,%
%TCIMACRO{\U{2102} }%
%BeginExpansion
\mathbb{C}
%EndExpansion
)$ symmetry group and the corresponding recurrence relations of the LSSA. We
will then show that SSA of four arbitrary string states can also be written in
terms of the Lauricella functions. (details will be given in section V) In
section III, we first explicitly calculate the residues of $5$, $6$ and
$7$-point KN amplitudes. We then extend the calculation of residues to all
$n$-point KN amplitudes. In section IV, we derive an iteration relation among
the residues of a given KN amplitude. This iteration relation is related to
the $SL(K+3,%
%TCIMACRO{\U{2102} }%
%BeginExpansion
\mathbb{C}
%EndExpansion
)$ symmetry and can presumably be used to soften the well-known hard SSA. In
section V, we will first use the shifting method to demonstrate that all
$5$-point SSA with arbitrary five tensor legs can be expressed in terms of the
LSSA. In general, we can use mathematical induction, together with the
on-shell recursion and the shifting principle, to show that all $n$-point SSA
can be expressed in terms of the LSSA. We will see that this multi-tensor
calculation is closely related to the calculation of applying multi-step
string on-shell recursion calculation to KN amplitudes. In section VI, a
conclusion and some open questions are given. Finally, in the end an appendix
is given to prove some mathematical identities used in the text.%

%TCIMACRO{\TeXButton{equation number}{\setcounter{equation}{0}
%\renewcommand{\theequation}{\arabic{section}.\arabic{equation}}}}%
%BeginExpansion
\setcounter{equation}{0}
\renewcommand{\theequation}{\arabic{section}.\arabic{equation}}%
%EndExpansion

\section{The Lauricella string scattering amplitudes}

In this section we give a brief review of the LSSA of three tachyons and one
arbitrary string states in the $26D$ open bosonic string theory and its
associated $SL(K+3,C)$ group. The general states at mass level $M_{2}%
^{2}=2(N_{2}-1)$, $N_{2}=\sum_{n,m,l>0}\left(  nr_{n}^{T}+mr_{m}^{P}%
+lr_{l}^{L}\right)  $ with polarizations on the scattering plane can be
written as \cite{LSSA}
\begin{equation}
\left\vert r_{n}^{T},r_{m}^{P},r_{l}^{L}\right\rangle =\prod_{n>0}\left(
\alpha_{-n}^{T}\right)  ^{r_{n}^{T}}\prod_{m>0}\left(  \alpha_{-m}^{P}\right)
^{r_{m}^{P}}\prod_{l>0}\left(  \alpha_{-l}^{L}\right)  ^{r_{l}^{L}}%
|0,k\rangle\label{state}%
\end{equation}
where $e^{P}=\frac{1}{M_{2}}(E_{2},\mathrm{k}_{2},0)=\frac{k_{2}}{M_{2}}$ the
momentum polarization, $e^{L}=\frac{1}{M_{2}}(\mathrm{k}_{2},E_{2},0)$ the
longitudinal polarization on and $e^{T}=(0,0,1)$ the transverse polarization.
In the CM frame, we define the kinematics {as}%
\begin{align}
k_{1} &  =\left(  \sqrt{M_{1}^{2}+|\vec{k_{1}}|^{2}},-|\vec{k_{1}}|,0\right)
,\\
k_{2} &  =\left(  \sqrt{M_{2}+|\vec{k_{1}}|^{2}},+|\vec{k_{1}}|,0\right)  ,\\
k_{3} &  =\left(  -\sqrt{M_{3}^{2}+|\vec{k_{3}}|^{2}},-|\vec{k_{3}}|\cos
\phi,-|\vec{k_{3}}|\sin\phi\right)  ,\\
k_{4} &  =\left(  -\sqrt{M_{4}^{2}+|\vec{k_{3}}|^{2}},+|\vec{k_{3}}|\cos
\phi,+|\vec{k_{3}}|\sin\phi\right)  \label{CMframe}%
\end{align}
with $M_{1}^{2}=M_{3}^{2}=M_{4}^{2}=-2$ and $\phi$ is the scattering angle.
The Mandelstam variables are defined to be $s=-\left(  k_{1}+k_{2}\right)
^{2}$, $t=-\left(  k_{2}+k_{3}\right)  ^{2}$ and $u=-\left(  k_{1}%
+k_{3}\right)  ^{2}$. We will also use the notation $k_{i}^{X}\equiv
e^{X}\cdot k_{i}$ \ for \ $X=\left(  T,P,L\right)  .$ It is important to note
that SSA of three tachyons and one arbitrary string states with polarizations
orthogonal to the scattering plane vanish. It is interesting to note that the
$\left(  s,t\right)  $ channel of the LSSA can be calculated and expressed in
terms of the Lauricella functions \cite{LSSA}%
\begin{align}
A_{st}^{(r_{n}^{T},r_{m}^{P},r_{l}^{L})} &  =\prod_{n=1}\left[  -(n-1)!k_{3}%
^{T}\right]  ^{r_{n}^{T}}\cdot\prod_{m=1}\left[  -(m-1)!k_{3}^{P}\right]
^{r_{m}^{P}}\prod_{l=1}\left[  -(l-1)!k_{3}^{L}\right]  ^{r_{l}^{L}%
}\nonumber\\
&  \cdot B\left(  -\frac{t}{2}-1,-\frac{s}{2}-1\right)  F_{D}^{(K)}\left(
-\frac{t}{2}-1;R_{n}^{T},R_{m}^{P},R_{l}^{L};\frac{u}{2}+2-N;\tilde{Z}_{n}%
^{T},\tilde{Z}_{m}^{P},\tilde{Z}_{l}^{L}\right)  \label{st1}%
\end{align}
where we have defined%
\begin{equation}
R_{k}^{X}\equiv\left\{  -r_{1}^{X}\right\}  ^{1},\cdots,\left\{  -r_{k}%
^{X}\right\}  ^{k}\text{ \ with \ }\left\{  a\right\}  ^{n}%
=\underset{n}{\underbrace{a,a,\cdots,a}}.
\end{equation}
and%
\begin{equation}
Z_{k}^{X}\equiv\left[  z_{1}^{X}\right]  ,\cdots,\left[  z_{k}^{X}\right]
\text{ \ \ with \ \ }\left[  z_{k}^{X}\right]  =z_{k0}^{X},\cdots,z_{k\left(
k-1\right)  }^{X}.\label{comp}%
\end{equation}
In Eq.(\ref{comp}), we have defined%
\begin{equation}
z_{k}^{X}=\left\vert \left(  -\frac{k_{1}^{X}}{k_{3}^{X}}\right)  ^{\frac
{1}{k}}\right\vert ,\ z_{kk^{\prime}}^{X}=z_{k}^{X}e^{\frac{2\pi ik^{\prime}%
}{k}},\ \tilde{z}_{kk^{\prime}}^{X}\equiv1-z_{kk^{\prime}}^{X}\text{ \ \ for
\ \ }k^{\prime}=0,\cdots,k-1.
\end{equation}

In addition to the mass level $M_{2}$ and $N_{2}$, the integer $K$ in
Eq.(\ref{st1}) is defined to be%
\begin{equation}
\text{ }K=\underset{\{\text{for all }r_{j}^{T}\neq0\}}{\sum j}%
+\underset{\{\text{for all }r_{j}^{P}\neq0\}}{\sum j}+\underset{\{\text{for
all }r_{j}^{L}\neq0\}}{\sum j}.\label{kk}%
\end{equation}
The $D$-type Lauricella function $F_{D}^{(K)}$ in Eq.(\ref{st1}) is one of the
four extensions of the Gauss hypergeometric function to $K$ variables and is
defined to be%
\begin{align}
&  F_{D}^{(K)}\left(  \alpha;\beta_{1},...,\beta_{K};\gamma;x_{1}%
,...,x_{K}\right)  \nonumber\\
&  =\sum_{n_{1},\cdots,n_{K}=0}^{\infty}\frac{\left(  \alpha\right)
_{n_{1}+\cdots+n_{K}}}{\left(  \gamma\right)  _{n_{1}+\cdots+n_{K}}}%
\frac{\left(  \beta_{1}\right)  _{n_{1}}\cdots\left(  \beta_{K}\right)
_{n_{K}}}{n_{1}!\cdots n_{K}!}x_{1}^{n_{1}}\cdots x_{K}^{n_{K}}%
\end{align}
where $(\alpha)_{n}=\alpha\cdot\left(  \alpha+1\right)  \cdots\left(
\alpha+n-1\right)  $ is the Pochhammer symbol. There was an important integral
representation of the Lauricella function \cite{Appell}
\begin{align}
&  F_{D}^{(K)}\left(  \alpha;\beta_{1},...,\beta_{K};\gamma;x_{1}%
,...,x_{K}\right)  \nonumber\\
&  =\frac{\Gamma(\gamma)}{\Gamma(\alpha)\Gamma(\gamma-\alpha)}\int_{0}%
^{1}dt\,t^{\alpha-1}(1-t)^{\gamma-\alpha-1}\cdot(1-x_{1}t)^{-\beta_{1}%
}(1-x_{2}t)^{-\beta_{2}}...(1-x_{K}t)^{-\beta_{K}},
\end{align}
which was used to calculate Eq.(\ref{st1}). One can use the LSSA to calculate
various scattering limits of the SSA \cite{LSSA}.

In the following, we discuss the hard scattering limit of the LSSA. For more
details, see the recent review \cite{LSSA}. It was first observed that for
each fixed mass level $N$ with $M^{2}=2(N-1)$, the following states are of
leading order in energy at the hard scattering limit \cite{CHLTY2,CHLTY1}
\begin{equation}
\left\vert N,2m,q\right\rangle \equiv(\alpha_{-1}^{T})^{N-2m-2q}(\alpha
_{-1}^{L})^{2m}(\alpha_{-2}^{L})^{q}|0,k\rangle. \label{Nmq}%
\end{equation}
Note that in Eq.(\ref{Nmq}), only even powers $2m$ in $\alpha_{-1}^{L}$
\cite{ChanLee,ChanLee2} survive, and the naive energy order of the amplitudes
will drop by an even number of energy powers in general. One important
application of the LSSA presented in Eq.(\ref{st1}) in the hard string
scattering limit is to reproduce \cite{LLY2} the infinite linear relations
with constant coefficients among all hard SSA and solve the ratios among them
\cite{LSSA}
\begin{equation}
\frac{T^{(N,2m,q)}}{T^{(N,0,0)}}=\left(  -\frac{1}{M}\right)  ^{2m+q}\left(
\frac{1}{2}\right)  ^{m+q}(2m-1)!!. \label{lin}%
\end{equation}
The infinite linear relations with constant coefficients and the solvability
of SSA in the hard scattering limit were first conjectured by Gross
\cite{GM,GM1} and later corrected \cite{ChanLee,ChanLee2} and proved
\cite{CHLTY2,CHLTY1} by the method of decoupling of zero norm states (ZNS)
\cite{Lee,lee-Ov,LeePRL}.

\bigskip To reproduce the infinite linear relations in Eq.(\ref{lin}) from
LSSA, we first note that, in the hard scattering limit, the relevant
kinematics reduce to%
\begin{align}
k_{1}^{T}  &  =0\text{, \ \ }k_{3}^{T}\simeq-E\sin\phi,\\
k_{1}^{L}  &  \simeq-\frac{2p^{2}}{M_{2}}\simeq-\frac{2E^{2}}{M_{2}},\\
k_{3}^{L}  &  \simeq\frac{2E^{2}}{M_{2}}\sin^{2}\frac{\phi}{2}%
\end{align}
where $E$ and $\phi$ are the CM frame energy and scattering angle,
respectively. One can calculate%
\begin{equation}
\tilde{z}_{kk^{\prime}}^{T}=1,\ \tilde{z}_{kk^{\prime}}^{L}=1-\left(
-\frac{s}{t}\right)  ^{1/k}e^{\frac{i2\pi k^{\prime}}{k}}\sim O\left(
1\right)  ,
\end{equation}
and the LSSA in Eq.(\ref{st1}) reduces to \cite{LLY2}%
\begin{align}
&  A_{st}^{(r_{n}^{T},r_{l}^{L})}=B\left(  -\frac{t}{2}-1,-\frac{s}%
{2}-1\right) \nonumber\\
&  \cdot\prod_{n=1}\left[  (n-1)!E\sin\phi\right]  ^{r_{n}^{T}}\prod
_{l=1}\left[  -(l-1)!\frac{2E^{2}}{M_{2}}\sin^{2}\frac{\phi}{2}\right]
^{r_{l}^{L}}\nonumber\\
&  \cdot F_{D}^{(K)}\left(  -\frac{t}{2}-1;R_{n}^{T},R_{l}^{L};\frac{u}%
{2}+2-N;\left(  1\right)  _{n},\tilde{Z}_{l}^{L}\right)  .
\end{align}

Since in the hard scattering limit, there was a difference between the naive
energy order and the real energy order corresponding to the $\left(
\alpha_{-1}^{L}\right)  ^{r_{1}^{L}}$ operator in Eq.(\ref{state}), it is
important to pay attention to the corresponding summation and write
\vspace{6pt}
\begin{align}
&  A_{st}^{(r_{n}^{T},r_{l}^{L})}=B\left(  -\frac{t}{2}-1,-\frac{s}%
{2}-1\right) \nonumber\\
&  \cdot\prod_{n=1}\left[  (n-1)!E\sin\phi\right]  ^{r_{n}^{T}}\prod
_{l=1}\left[  -(l-1)!\frac{2E^{2}}{M_{2}}\sin^{2}\frac{\phi}{2}\right]
^{r_{l}^{L}}\nonumber\\
&  \cdot\sum_{k_{r}}\frac{\left(  -\frac{t}{2}-1\right)  _{k_{r}}}{\left(
\frac{u}{2}+2-N\right)  _{k_{r}}}\frac{\left(  -r_{1}^{L}\right)  _{k_{r}}%
}{k_{r}!}\left(  1+\frac{s}{t}\right)  ^{k_{r}}\cdot\left(  \cdots\right)
\end{align}
where $\left(  a\right)  _{n+m}=\left(  a\right)  _{n}\left(  a+n\right)
_{m}$ and $\left(  \cdots\right)  $ are terms which are not relevant to the
following discussion. To proceed, the following formula was proposed
\cite{LLY2}
\begin{align}
&  \sum_{k_{r}=0}^{r_{1}^{L}}\frac{\left(  -\frac{t}{2}-1\right)  _{k_{r}}%
}{\left(  \frac{u}{2}+2-N\right)  _{k_{r}}}\frac{\left(  -r_{1}^{L}\right)
_{k_{r}}}{k_{r}!}\left(  1+\frac{s}{t}\right)  ^{k_{r}}\nonumber\\
=  &  0\cdot\left(  \frac{tu}{s}\right)  ^{0}\!+0\cdot\left(  \frac{tu}%
{s}\right)  ^{-1}\!+\dots+0\cdot\left(  \frac{tu}{s}\right)  ^{-\left[
\frac{r_{1}^{L}+1}{2}\right]  -1}\nonumber\\
&  +C_{r_{1}^{L}}\left(  \frac{tu}{s}\right)  ^{-\left[  \frac{r_{1}^{L}+1}%
{2}\right]  }+\mathit{O}\left\{  \left(  \frac{tu}{s}\right)  ^{-\left[
\frac{r_{1}^{L}+1}{2}\right]  +1}\right\}  . \label{pro}%
\end{align}
where $[$ $]$ stands for the Gauss symbol, $C_{r_{1}^{L}}$ is independent of
energy $E$ and depends on $r_{1}^{L}$ and possibly the scattering angle $\phi
$. When $r_{1}^{L}=2m$\ is an even number, one further proposes that
$C_{r_{1}^{L}}=\frac{\left(  2m\right)  !}{m!}$ and is $\phi$ independent. It
was verified that Eq.(\ref{pro}) is valid for $r_{1}^{L}=0,1,2,\cdots,10$
\cite{LLY2}.

It is interesting to Note that Eq.(\ref{pro}) reduces to the Stirling number
identity by taking the Regge limit ($s\rightarrow\infty$ with $t$ fixed) and
setting $r_{1}^{L}=2m$,
\begin{gather}
\sum_{k_{r}=0}^{2m}\frac{\left(  -\frac{t}{2}-1\right)  _{k_{r}}}{\left(
-\frac{s}{2}\right)  _{k_{r}}}\frac{\left(  -2m\right)  _{k_{r}}}{k_{r}%
!}\left(  \frac{s}{t}\right)  ^{k_{r}}\simeq\sum_{k_{r}=0}^{2m}\left(
-2m\right)  _{k_{r}}\left(  -\frac{t}{2}-1\right)  _{k_{r}}\frac{\left(
-2/t\right)  ^{k_{r}}}{k_{r}!}\nonumber\\
=0\cdot\left(  -t\right)  ^{0}\!+0\cdot\left(  -t\right)  ^{-1}\!+\dots
+0\cdot\left(  -t\right)  ^{-m+1}+\frac{(2m)!}{m!}\left(  -t\right)
^{-m}+\mathit{O}\left\{  \left(  \frac{1}{t}\right)  ^{m+1}\right\}  ,
\label{stirling}%
\end{gather}
which was proposed in \cite{KLY} and proved in \cite{LYAM}. The zero terms in
Eq.(\ref{pro}) correspond to the naive leading energy orders in the hard SSA
calculation. In the hard scattering limit, the true leading order SSA can then
be~identified \vspace{6pt}
\begin{align}
&  A_{st}^{(r_{n}^{T},r_{l}^{L})}\simeq B\left(  -\frac{t}{2}-1,-\frac{s}%
{2}-1\right) \nonumber\\
&  \cdot\prod_{n=1}\left[  (n-1)!E\sin\phi\right]  ^{r_{n}^{T}}\prod
_{l=1}\left[  -(l-1)!\frac{2E^{2}}{M_{2}}\sin^{2}\frac{\phi}{2}\right]
^{r_{l}^{L}}\nonumber\\
&  \cdot C_{r_{1}^{L}}\left(  E\sin\phi\right)  ^{-2\left[  \frac{r_{1}^{L}%
+1}{2}\right]  }\cdot\left(  \cdots\right) \nonumber\\
&  \sim E^{N-\sum_{n\geq2}nr_{n}^{T}-\left(  2\left[  \frac{r_{1}^{L}+1}%
{2}\right]  -r_{1}^{L}\right)  -\sum_{l\geq3}lr_{l}^{L}},
\end{align}
which means that SSA reaches its highest energy when $r_{n\geq2}^{T}%
=r_{l\geq3}^{L}=0$ and $r_{1}^{L}=2m$, an even number. This result is
consistent with the previous result presented in \linebreak Eq.(\ref{Nmq}).

Finally, the leading order SSA in the hard scattering limit, i.e., $r_{1}%
^{T}=N-2m-2q$, $r_{1}^{L}=2m$ and $r_{2}^{L}=q$, can be calculated to be
\cite{LLY2}%
\begin{align}
&  A_{st}^{(N-2m-2q,2m,q)}\nonumber\\
&  \simeq B\left(  -\frac{t}{2}-1,-\frac{s}{2}-1\right)  \left(  E\sin
\phi\right)  ^{N}\frac{\left(  2m\right)  !}{m!}\left(  -\frac{1}{2M_{2}%
}\right)  ^{2m+q}\nonumber\\
&  =(2m-1)!!\left(  -\frac{1}{M_{2}}\right)  ^{2m+q}\left(  \frac{1}%
{2}\right)  ^{m+q}A_{st}^{(N,0,0)}%
\end{align}
which reproduces the ratios in Eq.(\ref{lin}).

To obtain the recurrence relations of the Lauricella functions, we consider
the type-$1$ Appell functions. For the case of $K=2$, the $D$-type Lauricella
functions $F_{D}^{(K)}\left(  \alpha;\beta_{1},...,\beta_{K};\gamma
;x_{1},...,x_{K}\right)  $ reduce to the type-$1$ Appell functions
$F_{1}\left(  \alpha;\beta_{1},\beta_{2};\gamma,x,y\right)  $, and one has $4$
known recurrence relations. It was shown that one can generalize the $4=2+2$
fundamental recurrence relations of the Appell functions $F_{1}$ and prove the
following $K+2$ recurrence relations for the $D$-type Lauricella functions
$(m=1,2,...,K)$ \cite{Group}%
\begin{align}
\left(  \alpha-\underset{i}{\sum}\beta_{i}\right)  F_{D}^{(K)}-\alpha
F_{D}^{(K)}\left(  \alpha+1\right)  +\beta_{1}F_{D}^{(K)}\left(  \beta
_{1}+1\right)  +...+\beta_{K}F_{D}^{(K)}\left(  \beta_{K}+1\right)   &
=0,\label{NNN1}\\
\gamma F_{D}^{(K)}-\left(  \gamma-\alpha\right)  F_{D}^{(K)}\left(
\gamma+1\right)  -\alpha F_{D}^{(K)}\left(  \alpha+1;\gamma+1\right)   &
=0,\label{NNN2}\\
\gamma F_{D}^{(K)}+\gamma(x_{m}-1)F_{D}^{(K)}\left(  \beta_{m}+1\right)
+(\alpha-\gamma)x_{m}F_{D}^{(K)}\left(  \beta_{m}+1,;\gamma+1\right)   &  =0
\label{NNN}%
\end{align}
where for simplicity we have omitted those arguments of $F_{D}^{(K)}$ which
remain the same in the relations. Moreover, these recurrence relations can be
used to reproduce the Cartan subalgebra and simple root system of the $SL(K+3,%
%TCIMACRO{\U{2102} }%
%BeginExpansion
\mathbb{C}
%EndExpansion
)$ group with rank $K+2$\cite{Group}. As a result, the $SL(K+3,%
%TCIMACRO{\U{2102} }%
%BeginExpansion
\mathbb{C}
%EndExpansion
)$ group with its corresponding stringy Ward identities (recurrence relations)
can be used to solve \cite{solve} all the LSSA and express them in terms of
one amplitude.

In particular, it was shown that one can use the above recurrence relations to
decrease the value of $K$ step by step and reduce all the Lauricella functions
$F_{D}^{(K)}$ in the LSSA to the Gauss hypergeometry functions $F_{D}^{(1)}=$
$_{2}F_{1}(\alpha,\beta,\gamma,x)$. One can further reduce the Gauss
hypergeometry functions by deriving a multiplication theorem for them, and
then solve all the LSSA in terms of one single $4$-tachyon amplitude. One of
the reason of this solvability is that all $\beta_{J}$ in the Lauricella
functions of the LSSA take very special values, namely, nonpositive integers.
See the recent review \cite{LSSA}.

\bigskip On the other hand, to obtain the $SL(K+3,%
%TCIMACRO{\U{2102} }%
%BeginExpansion
\mathbb{C}
%EndExpansion
)$ symmetry group of the LSSA, we introduce the variables $a,b_{1}%
,\cdots,b_{K},c$ and define the basis functions \cite{slkc}%
\begin{align}
&  f_{ac}^{b_{1}\cdots b_{K}}\left(  \alpha;\beta_{1},\cdots,\beta_{K}%
;\gamma;x_{1},\cdots,x_{K}\right) \nonumber\\
&  =B\left(  \gamma-\alpha,\alpha\right)  F_{D}^{\left(  K\right)  }\left(
\alpha;\beta_{1},\cdots,\beta_{K};\gamma;x_{1},\cdots,x_{K}\right)  a^{\alpha
}b_{1}^{\beta_{1}}\cdots b_{K}^{\beta_{K}}c^{\gamma}, \label{id2}%
\end{align}
so that the LSSA in Eq.(\ref{st1}) can be written as \cite{Group}
\begin{equation}
A_{st}^{(r_{n}^{T},r_{m}^{P},r_{l}^{L})}=f_{11}^{-(n-1)!k_{3}^{T}%
,-(m-1)!k_{3}^{P},-(l-1)!k_{3}^{L}}\left(  -\frac{t}{2}-1;R_{n}^{T},R_{m}%
^{P},R_{l}^{L};\frac{u}{2}+2-N;\tilde{Z}_{n}^{T},\tilde{Z}_{m}^{P},\tilde
{Z}_{l}^{L}\right)  .
\end{equation}
We then define and introduce the $(K+3)^{2}-1$ generators of $SL(K+3,C)$ group
\cite{slkc,Group}. These are $1$ $E^{\alpha}$, $K$ $E^{\beta_{k}}$, $1$
$E^{\gamma}$,$1$ $E^{\alpha\gamma}$,$K$ $E^{\beta_{k}\gamma}$ and $K$
$E^{\alpha\beta_{k}\gamma}$ which sum up to $3K+3$ raising generators. There
are also $3K+3$ lowering operators. In addition, there are $K\left(
K-1\right)  $ $E_{\beta_{p}}^{\beta_{k}}$ and $K+2$ $\ J$ , $\left\{
J_{\alpha},J_{\beta_{k}},J_{\gamma}\right\}  $, the Cartan subalgebra. In sum,
the total number of generators are $2(3K+3)+K(K-1)+K+2=(K+3)^{2}-1$. The
explicit forms of these generators are
\begin{align}
E^{\alpha}  &  =a\left(  \underset{j}{%
%TCIMACRO{\dsum }%
%BeginExpansion
{\displaystyle\sum}
%EndExpansion
}x_{j}\partial_{j}+a\partial_{a}\right)  ,\nonumber\\
E^{\beta_{k}}  &  =b_{k}\left(  x_{k}\partial_{k}+b_{k}\partial_{b_{k}%
}\right)  ,\nonumber\\
E^{\gamma}  &  =c\left(  \underset{j}{%
%TCIMACRO{\dsum }%
%BeginExpansion
{\displaystyle\sum}
%EndExpansion
}\left(  1-x_{j}\right)  \partial_{x_{j}}+c\partial_{c}-a\partial
_{a}-\underset{j}{%
%TCIMACRO{\dsum }%
%BeginExpansion
{\displaystyle\sum}
%EndExpansion
}b_{j}\partial_{b_{j}}\right)  ,\nonumber\\
E^{\alpha\gamma}  &  =ac\left(  \underset{j}{%
%TCIMACRO{\dsum }%
%BeginExpansion
{\displaystyle\sum}
%EndExpansion
}\left(  1-x_{j}\right)  \partial_{x_{j}}-a\partial_{a}\right)  ,\nonumber\\
E^{\beta_{k}\gamma}  &  =b_{k}c\left[  \left(  x_{k}-1\right)  \partial
_{x_{k}}+b_{k}\partial_{b_{k}}\right]  ,\nonumber\\
E^{\alpha\beta_{k}\gamma}  &  =ab_{k}c\partial_{x_{k}},\nonumber\\
E_{\alpha}  &  =\frac{1}{a}\left[  \underset{j}{%
%TCIMACRO{\dsum }%
%BeginExpansion
{\displaystyle\sum}
%EndExpansion
}x_{j}\left(  1-x_{j}\right)  \partial_{x_{j}}+c\partial_{c}-a\partial
_{a}-\underset{j}{%
%TCIMACRO{\dsum }%
%BeginExpansion
{\displaystyle\sum}
%EndExpansion
}x_{j}b_{j}\partial_{b_{j}}\right]  ,\nonumber\\
E_{\beta_{k}}  &  =\frac{1}{b_{k}}\left[  x_{k}\left(  1-x_{k}\right)
\partial_{x_{k}}+x_{k}\underset{j\neq k}{%
%TCIMACRO{\dsum }%
%BeginExpansion
{\displaystyle\sum}
%EndExpansion
}\left(  1-x_{j}\right)  x_{j}\partial_{x_{j}}+c\partial_{c}-x_{k}%
a\partial_{a}-\underset{j}{%
%TCIMACRO{\dsum }%
%BeginExpansion
{\displaystyle\sum}
%EndExpansion
}b_{j}\partial_{u_{j}}\right]  ,\nonumber\\
E_{\gamma}  &  =-\frac{1}{c}\left(  \underset{j}{%
%TCIMACRO{\dsum }%
%BeginExpansion
{\displaystyle\sum}
%EndExpansion
}x_{j}\partial_{x_{j}}+c\partial_{c}-1\right)  ,\nonumber\\
E_{\alpha\gamma}  &  =\frac{1}{ac}\left[  \underset{j}{%
%TCIMACRO{\dsum }%
%BeginExpansion
{\displaystyle\sum}
%EndExpansion
}x_{j}\left(  1-x_{j}\right)  \partial_{x_{j}}-\underset{j}{%
%TCIMACRO{\dsum }%
%BeginExpansion
{\displaystyle\sum}
%EndExpansion
}x_{j}b_{j}\partial_{b_{j}}+c\partial_{c}-1\right]  ,\nonumber\\
E_{\beta_{k}\gamma}  &  =\frac{1}{b_{k}c}\left[  x_{k}\left(  x_{k}-1\right)
\partial_{x_{k}}+\underset{j\neq k}{%
%TCIMACRO{\dsum }%
%BeginExpansion
{\displaystyle\sum}
%EndExpansion
}\left(  x_{j}-1\right)  x_{j}\partial_{x_{j}}+x_{k}a\partial_{a}%
-c\partial_{c}+1\right]  ,\nonumber\\
E_{\alpha\beta_{k}\gamma}  &  =\frac{1}{ab_{k}c}\left[  \underset{j}{%
%TCIMACRO{\dsum }%
%BeginExpansion
{\displaystyle\sum}
%EndExpansion
}x_{j}\left(  x_{j}-1\right)  \partial_{x_{j}}-c\partial_{c}+x_{k}%
a\partial_{a}+\underset{j}{%
%TCIMACRO{\dsum }%
%BeginExpansion
{\displaystyle\sum}
%EndExpansion
}x_{j}b_{j}\partial_{b_{j}}-x_{k}+1\right]  ,\nonumber\\
E_{\beta_{p}}^{\beta_{k}}  &  =\frac{b_{k}}{b_{p}}\left[  \left(  x_{k}%
-x_{p}\right)  \partial_{z_{k}}+b_{k}\partial_{b_{k}}\right]  ,(k\neq
p),\nonumber\\
J_{\alpha}  &  =a\partial_{a},\nonumber\\
J_{\beta_{k}}  &  =b_{k}\partial_{b_{k}},\nonumber\\
J_{\gamma}  &  =c\partial_{c}. \label{def2}%
\end{align}

It is straightforward to calculate the operation of these generators on the
basis functions ($k=1,2,...K$) $f_{ac}^{b_{1}\cdots b_{K}}\left(  \alpha
;\beta_{1},\cdots,\beta_{K};\gamma;x_{1},\cdots,x_{K}\right)  $%
\begin{align}
E^{\alpha}f_{ac}^{b_{1}\cdots b_{K}}\left(  \alpha\right)   &  =\left(
\gamma-\alpha-1\right)  f_{ac}^{b_{1}\cdots b_{K}}\left(  \alpha+1\right)
,\nonumber\\
E^{\beta_{k}}f_{ac}^{b_{1}\cdots b_{K}}\left(  \beta_{k}\right)   &
=\beta_{k}f_{ac}^{b_{1}\cdots b_{K}}\left(  \beta_{k}+1\right)  ,\nonumber\\
E^{\gamma}f_{ac}^{b_{1}\cdots b_{K}}\left(  \gamma\right)   &  =\left(
\gamma-\underset{j}{%
%TCIMACRO{\dsum }%
%BeginExpansion
{\displaystyle\sum}
%EndExpansion
}\beta_{j}\right)  f_{ac}^{b_{1}\cdots b_{K}}\left(  \gamma+1\right)
,\nonumber\\
E^{\alpha\gamma}f_{ac}^{b_{1}\cdots b_{K}}\left(  \alpha;\gamma\right)   &
=\left(  \underset{j}{%
%TCIMACRO{\dsum }%
%BeginExpansion
{\displaystyle\sum}
%EndExpansion
}\beta_{j}-\gamma\right)  f_{ac}^{b_{1}\cdots b_{K}}\left(  \alpha
+1;\gamma+1\right)  ,\nonumber\\
E^{\beta_{k}\gamma}f_{ac}^{b_{1}\cdots b_{K}}\left(  \beta_{k};\gamma\right)
&  =\beta_{k}f_{ac}^{b_{1}\cdots b_{K}}\left(  \beta_{k}+1;\gamma+1\right)
,\nonumber\\
E^{\alpha\beta_{k}\gamma}f_{ac}^{b_{1}\cdots b_{K}}\left(  \alpha;\beta
_{k};\gamma\right)   &  =\beta_{k}f_{ac}^{b_{1}\cdots b_{K}}\left(
\alpha+1;\beta_{k}+1;\gamma+1\right)  ,\nonumber\\
E_{\alpha}f_{ac}^{b_{1}\cdots b_{K}}\left(  \alpha\right)   &  =\left(
\alpha-1\right)  f_{ac}^{b_{1}\cdots b_{K}}\left(  \alpha-1\right)
,\nonumber\\
E_{\beta_{k}}f_{ac}^{b_{1}\cdots b_{K}}\left(  \beta_{k}\right)   &  =\left(
\gamma-\underset{j}{%
%TCIMACRO{\dsum }%
%BeginExpansion
{\displaystyle\sum}
%EndExpansion
}\beta_{j}\right)  f_{ac}^{b_{1}\cdots b_{K}}\left(  \beta_{k}-1\right)
,\nonumber\\
E_{\gamma}f_{ac}^{b_{1}\cdots b_{K}}\left(  \gamma\right)   &  =\left(
\alpha-\gamma+1\right)  f_{ac}^{b_{1}\cdots b_{K}}\left(  \gamma-1\right)
,\nonumber\\
E_{\alpha\gamma}f_{ac}^{b_{1}\cdots b_{K}}\left(  \alpha;\gamma\right)   &
=\left(  \alpha-1\right)  f_{ac}^{b_{1}\cdots b_{K}}\left(  \alpha
-1;\gamma-1\right)  ,\nonumber\\
E_{\beta_{k}\gamma}f_{ac}^{b_{1}\cdots b_{K}}\left(  \beta_{k};\gamma\right)
&  =\left(  \alpha-\gamma+1\right)  f_{ac}^{b_{1}\cdots b_{K}}\left(
\beta_{k}-1;\gamma-1\right)  ,\nonumber\\
E_{\alpha\beta_{k}\gamma}f_{ac}^{b_{1}\cdots b_{K}}\left(  \alpha;\beta
_{k};\gamma\right)   &  =\left(  1-\alpha\right)  f_{ac}^{b_{1}\cdots b_{K}%
}\left(  \alpha-1;\beta_{k}-1;\gamma-1\right)  ,\nonumber\\
E_{\beta_{p}}^{\beta_{k}}f_{ac}^{b_{1}\cdots b_{K}}\left(  \beta_{k};\beta
_{p}\right)   &  =\beta_{k}f_{ac}^{b_{1}\cdots b_{K}}\left(  \beta_{k}%
+1;\beta_{p}-1\right)  ,\nonumber\\
J_{\alpha}f_{ac}^{b_{1}\cdots b_{K}}\left(  \alpha;\beta_{k};\gamma\right)
&  =\alpha f_{ac}^{b_{1}\cdots b_{K}}\left(  \alpha;\beta_{k};\gamma\right)
,\nonumber\\
J_{\beta_{k}}f_{ac}^{b_{1}\cdots b_{K}}\left(  \alpha;\beta_{k};\gamma\right)
&  =\beta_{k}f_{ac}^{b_{1}\cdots b_{K}}\left(  \alpha;\beta_{k};\gamma\right)
,\nonumber\\
J_{\gamma}f_{ac}^{b_{1}\cdots b_{K}}\left(  \alpha;\beta_{k};\gamma\right)
&  =\gamma f_{ac}^{b_{1}\cdots b_{K}}\left(  \alpha;\beta_{k};\gamma\right)
\label{op2}%
\end{align}
and show explicitly the $SL(K+3,%
%TCIMACRO{\U{2102} }%
%BeginExpansion
\mathbb{C}
%EndExpansion
)$ symmetry of the LSSA \cite{LSSA}. In Eq.(\ref{op2}), we have omitted those
arguments in $f_{ac}^{b_{1}\cdots b_{K}}$ that remain the same after the
operation. Indeed, with the following identifications \cite{slkc}%
\begin{align}
E^{\alpha}  &  =\mathcal{E}_{12},E_{\alpha}=\mathcal{E}_{21},E^{\beta_{k}%
}=\mathcal{E}_{k+3,3},E_{\beta}=\mathcal{E}_{3,k+3},\nonumber\\
E^{\gamma}  &  =\mathcal{E}_{31},E_{\gamma}=\mathcal{E}_{13},E^{\alpha\gamma
}=\mathcal{E}_{32},E_{\alpha\gamma}=\mathcal{E}_{23},\nonumber\\
E^{\beta_{k}\gamma}  &  =-\mathcal{E}_{k+3,1},E_{\beta_{k}\gamma}%
=-\mathcal{E}_{1,k+3},E_{\alpha\beta_{k}\gamma}=-\mathcal{E}_{k+3,2}%
,\nonumber\\
E_{\alpha\beta_{k}\gamma}  &  =-\mathcal{E}_{2,k+3},J_{\alpha}^{\prime}%
=\frac{1}{2}\left(  \mathcal{E}_{11}-\mathcal{E}_{22}\right)  ,J_{\beta_{k}%
}^{\prime}=\frac{1}{2}\left(  \mathcal{E}_{k+3,k+3}-\mathcal{E}_{33}\right)
,J_{\gamma}^{\prime}=\frac{1}{2}\left(  \mathcal{E}_{33}-\mathcal{E}%
_{11}\right)  ,
\end{align}
one can calculate the Lie algebra commutation relations of the $SL(K+3,%
%TCIMACRO{\U{2102} }%
%BeginExpansion
\mathbb{C}
%EndExpansion
)$
\begin{equation}
\left[  \mathcal{E}_{ij},\mathcal{E}_{kl}\right]  =\delta_{jk}\mathcal{E}%
_{il}-\delta_{li}\mathcal{E}_{kj};\text{ \ }1\leqslant i,j\leqslant K+3.
\end{equation}
It is important to note that instead of finite dimensional representation,
here we encounter infinite dimensional representation of the noncompact Lie
group $SL(K+3,C)$. This means that for any fixed positive integer $K$, one has
infinite number of LSSA in the $SL(K+3,C)$ representations.

Finally, it is straightforward to generalize the LSSA of three tachyons and
one arbitrary string states and show that all SSA of four arbitrary string
states can still be expressed in terms of the $D$-type Lauricella functions
$F_{D}^{(K)}.$ See section V for more details. Indeed, for the general
multi-tensor $4$-point functions, there are new terms with finite number of
contractions among $\partial^{n}X...$ and $\partial^{m}X...$, and one obtains
more $D$-type Lauricella functions with different values of $K$. In general,
the LSSA of four arbitrary string states can be expressed as sum over finite
terms of the $D$-type Lauricella functions and can be written as%
\begin{equation}
LSSA=\underset{J}{\sum}c_{J}(k_{i})\int_{0}^{1}dt\,t^{\alpha-1}(1-t)^{\gamma
-\alpha-1}\cdot P_{J}(t)\label{p}%
\end{equation}
where $P_{J}(t)$ is a polynomial with coefficients depending on $k_{i}$ and
polarizations $\zeta$, and $c_{J}(k_{i})$ is a $k_{i}$ and $\zeta$ dependent
coefficient. One simple example \cite{KLT} is the two vectors, two tachyons
bosonic open SSA%
\begin{align}
A\left(  \zeta_{1},k_{1};k_{2};\zeta_{3},k_{3};k_{4}\right)   &  =\left(
\zeta_{1}\cdot k_{2}\right)  \left(  \zeta_{3}\cdot k_{4}\right)  \frac
{\Gamma\left(  \frac{-s}{2}-1\right)  \Gamma\left(  \frac{-t}{2}-1\right)
}{\Gamma\left(  -\frac{s}{2}-\frac{t}{2}-2\right)  }\nonumber\\
&  \times F_{D}^{\left(  2\right)  }\left(  \frac{-s}{2}-1;-1,-1;-\frac{s}%
{2}-\frac{t}{2}-2;-\frac{\zeta_{1}\cdot k_{3}}{\zeta_{1}\cdot k_{2}}%
,-\frac{\zeta_{3}\cdot k_{1}}{\zeta_{3}\cdot k_{4}}\right)  \nonumber\\
&  +\left(  \zeta_{1}\cdot\zeta_{3}\right)  \frac{\Gamma\left(  \frac{-s}%
{2}\right)  \Gamma\left(  \frac{-t}{2}\right)  }{\Gamma\left(  -\frac{s}%
{2}-\frac{t}{2}\right)  }F_{D}^{\left(  0\right)  }\left(  \frac{-s}%
{2};;-\frac{s}{2}-\frac{t}{2};\right)  ,\label{KK}%
\end{align}
which contains two Lauricella functions with $K=0,2$. Note that in contrast to
Eq.(\ref{st1}), all physical polarizations $\zeta_{i}^{\mu}$ except ZNS
contribute for the case of multi-tensor SSA as in Eq.(\ref{KK}). More examples
will be given in section V.%

%TCIMACRO{\TeXButton{equation number}{\setcounter{equation}{0}
%\renewcommand{\theequation}{\arabic{section}.\arabic{equation}}}}%
%BeginExpansion
\setcounter{equation}{0}
\renewcommand{\theequation}{\arabic{section}.\arabic{equation}}%
%EndExpansion

\section{Residues of the Koba-Nielsen (KN) amplitudes}

In this section, we will use the string theory extension \cite{bcfw3,bcfw4} of
field theory BCFW on-shell recursion relations \cite{bcfw1,bcfw2} to calculate
in details the residues of all $n$-point Koba-Nielsen (KN) amplitudes. In
section V, we will show that the residues of all $n$-point KN amplitudes
calculated in this section can be expressed in terms of the Lauricella
functions. Moreover, we will use the shifting principle to show that all SSA
including the KN amplitudes can be expressed in terms of the Lauricella functions.

\subsection{The Veneziano amplitude}

\bigskip We begin with the discussion on four tachyon amplitude which can be
written as%
\begin{equation}
A_{4}=\int_{0}^{1}dzz^{k_{1}\cdot k_{2}}\left(  1-z\right)  ^{k_{2}\cdot
k_{3}}=\int_{0}^{1}dzz^{k_{1}\cdot k_{2}}e^{k_{2}\cdot k_{3}\ln\left(
1-z\right)  }=\int_{0}^{1}dzz^{k_{1}\cdot k_{2}}e^{-k_{2}\cdot k_{3}\left(
z+\frac{z^{2}}{2}+\frac{z^{3}}{3}+\cdots\right)  }, \label{Vene}%
\end{equation}
or%
\begin{align}
&  A_{4}=\int_{0}^{1}dzz^{k_{1}\cdot k_{2}}\exp\left[
\begin{array}
[c]{c}%
1+T\left(  z+\frac{z^{2}}{2}+\frac{z^{3}}{3}+\cdots\right)  +\frac{T^{2}}%
{2}\left(  z+\frac{z^{2}}{2}+\frac{z^{3}}{3}+\cdots\right)  ^{2}\\
+\frac{T^{3}}{3!}\left(  z+\frac{z^{2}}{2}+\frac{z^{3}}{3}+\cdots\right)
^{3}+\cdots
\end{array}
\right] \nonumber\\
&  =\underset{1}{\underbrace{\int_{0}^{1}dzz^{k_{1}\cdot k_{2}}}%
}+\underset{\alpha_{-1}}{\underbrace{T}}\int_{0}^{1}dzz^{k_{1}\cdot k_{2}%
+1}+\left(  \underset{\alpha_{-2}}{\underbrace{\frac{T}{2}}}+\underset{\alpha
_{-1}^{2}}{\underbrace{\frac{T^{2}}{2!}\frac{1}{1^{2}}}}\right)  \int_{0}%
^{1}dzz^{k_{1}\cdot k_{2}+2}\nonumber\\
&  +\left(  \underset{\alpha_{-3}}{\underbrace{\frac{T}{3}}}+\underset{\alpha
_{-1}\alpha_{-2}}{\underbrace{\frac{T^{2}}{1!1!}\frac{1}{1^{1}}\frac{1}{2^{1}%
}}}+\underset{\alpha_{-1}^{3}}{\underbrace{\frac{T^{3}}{3!}}}\right)  \int%
_{0}^{1}dzz^{k_{1}\cdot k_{2}+3}\nonumber\\
&  +\left(  \underset{\alpha_{-4}}{\underbrace{\frac{T}{4}}}+\underset{\alpha
_{-3}\alpha_{-1}}{\underbrace{\frac{T^{2}}{1!1!}\frac{1}{1^{1}}\frac{1}{3^{1}%
}}}+\underset{\alpha_{-2}^{2}}{\underbrace{\frac{T^{2}}{2!}\frac{1}{2^{2}}}%
}+\underset{\alpha_{-1}^{2}\alpha_{-2}}{\underbrace{\frac{T^{3}}{2!1!}\frac
{1}{1^{2}2^{1}}}}+\underset{\alpha_{-1}^{4}}{\underbrace{\frac{T^{4}}{4!}%
\frac{1}{1^{4}}}}\right)  \int_{0}^{1}dzz^{k_{1}\cdot k_{2}+4}+\cdots
\label{sum2}%
\end{align}
where $T=-k_{2}\cdot k_{3}$. With the definition of the Mandelstam variables
$s=-\left(  k_{1}+k_{2}\right)  ^{2}$, $t=-\left(  k_{2}+k_{3}\right)  ^{2}$,
we can express the amplitude in terms of a series of simple pole terms with
residues
\begin{align}
&  A_{4}=\underset{1}{\underbrace{\frac{2}{-s-2}}}+\underset{\alpha
_{-1}}{\underbrace{T\frac{2}{-s+0}}}+(\underset{\alpha_{-2}}{\underbrace{\frac
{T}{2}}}+\underset{\alpha_{-1}^{2}}{\underbrace{\frac{T^{2}}{2!}\frac{1}%
{1^{2}})}}\frac{2}{-s+2}\nonumber\\
&  +\underset{\alpha_{-3}}{\underbrace{(\frac{T}{3}}}+\underset{\alpha
_{-1}\alpha_{-2}}{\underbrace{\frac{T^{2}}{1!1!}\frac{1}{1^{1}}\frac{1}{2^{1}%
}}}+\underset{\alpha_{-1}^{3}}{\underbrace{\frac{T^{3}}{3!})}}\frac{2}%
{-s+4}\nonumber\\
&  \underset{\alpha_{-4}}{\underbrace{(\frac{T}{4}}}+\underset{\alpha
_{-3}\alpha_{-1}}{\underbrace{\frac{T^{2}}{1!1!}\frac{1}{1^{1}}\frac{1}{3^{1}%
}}}+\underset{\alpha_{-2}^{2}}{\underbrace{\frac{T^{2}}{2!}\frac{1}{2^{2}}}%
}+\underset{\alpha_{-1}^{2}\alpha_{-2}}{\underbrace{\frac{T^{3}}{2!1!}\frac
{1}{1^{2}2^{1}}}}+\underset{\alpha_{-1}^{4}}{\underbrace{\frac{T^{4}}{4!}%
\frac{1}{1^{4}})}}\frac{2}{-s+6}+\cdots\label{sum}\\
&  =\overset{\infty}{\underset{n=0}{\sum}}-\frac{(\alpha(t)+1)(\alpha
(t)+2)\cdots(\alpha(t)+n)}{n!}\frac{1}{\alpha(s)-n}%
\end{align}
where $\alpha(t)=\alpha^{\prime}t+\alpha(0)$ with $\alpha^{\prime}=1/2$ and
$\alpha(0)=1$ \cite{GSW}. In Eq.(\ref{sum}) or Eq.(\ref{sum2}), we can see
various residues at each pole of the Veneziano amplitude. We will find
generalization of these residues for higher point KN amplitudes in section IV.

\subsection{The five-point KN amplitude}

For the $5$-point KN amplitude, a simple generalization of Eq.(\ref{sum}) is
to use the string theory extension \cite{stringbcfw} of field theory BCFW
on-shell recursion relations \cite{bcfw1,bcfw2} to calculate the infinite
number of residues. Moreover, we will show that all residues can be expressed
in terms of the Lauricella functions. The $5$-point KN amplitude can be
written as \cite{stringbcfw}%

\begin{align}
A_{5}^{KN}  &  =\int_{0}^{1}dz_{3}\int_{0}^{z_{3}}dz_{2}z_{2}^{k_{2}\cdot
k_{1}}z_{3}^{k_{3}\cdot k_{1}}\left(  1-z_{3}\right)  ^{k_{4}\cdot k_{3}%
}\left(  1-z_{2}\right)  ^{k_{4}\cdot k_{2}}\left(  z_{3}-z_{2}\right)
^{k_{3}\cdot k_{2}}\\
&  =\underset{a,b,c=0}{\overset{\infty}{%
%TCIMACRO{\dsum }%
%BeginExpansion
{\displaystyle\sum}
%EndExpansion
}}\frac{\left(  -k_{2}\cdot k_{3}\right)  _{a}}{a!}\frac{\left(  -k_{2}\cdot
k_{4}\right)  _{b}}{b!}\frac{\left(  -k_{3}\cdot k_{4}\right)  _{c}}{c!}%
\frac{2}{-s_{12}+2\left(  a+b-1\right)  }\frac{2}{-s_{123}+2\left(
b+c-1\right)  } \label{24}%
\end{align}
where $s_{12}=-\left(  k_{1}+k_{2}\right)  ^{2},s_{123}=-\left(  k_{1}%
+k_{2}+k_{3}\right)  ^{2}$. We now consider the BCFW deformation with pair
$(k_{1},k_{5})=(1,5)$, and set%
\begin{equation}
\hat{k}_{1}\left(  z\right)  =k_{1}+zq,\hat{k}_{5}\left(  z\right)
=k_{5}-zq,q^{2}=k_{1}\cdot q=k_{5}\cdot q=0.
\end{equation}
For $s_{12}$ and $s_{123}$, the poles are located at
\begin{align}
z_{2,M}  &  =\frac{\left(  k_{1}+k_{2}\right)  ^{2}+2\left(  M-1\right)
}{-2q\cdot\left(  k_{1}+k_{2}\right)  },M=a+b=0,1,2,...;\\
z_{3,M}  &  =\frac{\left(  k_{1}+k_{2}+k_{3}\right)  ^{2}+2\left(  M-1\right)
}{-2q\cdot\left(  k_{1}+k_{2}+k_{3}\right)  },M=b+c=0,1,2,...
\end{align}
respectively. By using string on-shell recursion relation, we get
\cite{stringbcfw}%
\begin{equation}
A_{5}^{KN}=\underset{M}{\sum}\frac{2}{-s_{12}+2\left(  M-1\right)  }%
R_{3,4}^{M}+\underset{_{M}}{\sum}\frac{2}{-s_{123}+2\left(  M-1\right)
}R_{4,3}^{M} \label{RS}%
\end{equation}
where the residues $R_{3,4}^{M}$ and $R_{4,3}^{M}$ can be calculated to be%
\begin{align}
R_{3,4}^{M}  &  =\underset{a,b,c=0,M=a+b}{\overset{\infty}{%
%TCIMACRO{\dsum }%
%BeginExpansion
{\displaystyle\sum}
%EndExpansion
}}\frac{\left(  -k_{2}\cdot k_{3}\right)  _{a}}{a!}\frac{\left(  -k_{2}\cdot
k_{4}\right)  _{b}}{b!}\frac{\left(  -k_{3}\cdot k_{4}\right)  _{c}}{c!}%
\frac{2}{-\hat{s}_{123}\left(  z_{N}\right)  +2\left(  b+c-1\right)  },\\
R_{4,3}^{M}  &  =\underset{a,b,c=0,M=b+c}{\overset{\infty}{%
%TCIMACRO{\dsum }%
%BeginExpansion
{\displaystyle\sum}
%EndExpansion
}}\frac{\left(  -k_{2}\cdot k_{3}\right)  _{a}}{a!}\frac{\left(  -k_{2}\cdot
k_{4}\right)  _{b}}{b!}\frac{\left(  -k_{3}\cdot k_{4}\right)  _{c}}{c!}%
\frac{2}{-\hat{s}_{12}\left(  w_{M}\right)  +2\left(  a+b-1\right)  }.
\end{align}
The residues $R_{3,4}^{M}$ and $R_{4,3}^{M}$ above can be further calculated
to be%
\begin{align}
R_{3,4}^{M}  &  =\underset{a,b=0,M=a+b}{\overset{\infty}{%
%TCIMACRO{\dsum }%
%BeginExpansion
{\displaystyle\sum}
%EndExpansion
}}\frac{\left(  -k_{2}\cdot k_{3}\right)  _{a}}{a!}\frac{\left(  -k_{2}\cdot
k_{4}\right)  _{b}}{b!}\int_{0}^{1}dz_{2}z_{2}^{\left(  \hat{k}_{1}%
+k_{2}\right)  \cdot k_{3}-a}\left(  1-z_{2}\right)  ^{k_{3}\cdot k_{4}},\\
R_{4,3}^{M}  &  =\underset{b,c=0,M=b+c}{\overset{\infty}{%
%TCIMACRO{\dsum }%
%BeginExpansion
{\displaystyle\sum}
%EndExpansion
}}\frac{\left(  -k_{2}\cdot k_{4}\right)  _{b}}{b!}\frac{\left(  -k_{3}\cdot
k_{4}\right)  _{c}}{c!}\int_{0}^{1}dz_{2}z_{2}^{\hat{k}_{1}\cdot k_{2}%
+b}\left(  1-z_{2}\right)  ^{k_{2}\cdot k_{3}}. \label{444}%
\end{align}
We can now use the identity derived in the appendix%
\begin{equation}
\underset{b+c=M,b,c=0}{\sum}\frac{\left(  T_{1}\right)  _{b}}{b!}\frac{\left(
T_{2}\right)  _{c}}{c!}x_{1}^{b}x_{2}^{c}=\underset{\left\{  N_{r}\right\}
}{\sum}\prod_{r=1}\left(  \frac{1}{N_{r}!r^{N_{r}}}\left[  T_{1}x_{1}%
^{r}+T_{2}x_{2}^{r}\right]  ^{N_{r}}\right)
\end{equation}
to obtain%
\begin{align}
R_{3,4}^{M}  &  =\underset{\left\{  N_{r}\right\}  }{\sum}\left(  -k_{2}\cdot
k_{3}\right)  ^{N}\int_{0}^{1}dz_{2}z_{2}^{\left(  \hat{k}_{1}+k_{2}\right)
\cdot k_{3}-M}\left(  1-z_{2}\right)  ^{k_{3}\cdot k_{4}}\prod_{r=1}\frac
{1}{N_{r}!r^{N_{r}}}\left[  1-\left(  -\frac{k_{2}\cdot k_{4}}{k_{2}\cdot
k_{3}}\right)  z_{2}^{r}\right]  ^{N_{r}},\label{33}\\
R_{4,3}^{M}  &  =\underset{\left\{  N_{r}\right\}  }{\sum}\left(  -k_{3}\cdot
k_{4}\right)  ^{N}\int_{0}^{1}dz_{2}z_{2}^{\hat{k}_{1}\cdot k_{2}}\left(
1-z_{2}\right)  ^{k_{2}\cdot k_{3}}\prod_{r=1}\frac{1}{N_{r}!r^{N_{r}}}\left[
1-\left(  -\frac{k_{2}\cdot k_{4}}{k_{3}\cdot k_{4}}\right)  z_{2}^{r}\right]
^{N_{r}}. \label{44}%
\end{align}
In Eq.(\ref{33}) and Eq.(\ref{44}) the sums are over partitions of $N$ into
$\left\{  N_{r}\right\}  $ with $M=\underset{r=1}{\sum}rN_{r}$ and
$N=\underset{r=1}{\sum}N_{r}$. Thus the string on-shell recursion relation
reduces the $5$-point KN amplitude into two products of subamplitudes
$3\otimes4$ and $4\otimes3$ with $3$-point and $4$-point functions. One can
now easily express the residues in Eq.(\ref{33}) and Eq.(\ref{44}) in terms of
sums of the Lauricella functions%
\begin{align}
R_{3,4}^{M}  &  =\underset{\left\{  N_{r}\right\}  }{\sum}\left(  -k_{2}\cdot
k_{3}\right)  ^{N}B\left(  \left(  \hat{k}_{1}+k_{2}\right)  \cdot
k_{3}-M+1,k_{3}\cdot k_{4}+1\right)  \left(  \prod_{r=1}\frac{1}%
{N_{r}!r^{N_{r}}}\right) \nonumber\\
&  \cdot F_{D}^{(K)}\left(
\begin{array}
[c]{c}%
\left(  \hat{k}_{1}+k_{2}\right)  \cdot k_{3}-M+1;\cdots,\left\{
-N_{r}\right\}  ^{r},\cdots;\\
\left(  \hat{k}_{1}+k_{2}\right)  \cdot k_{3}+k_{3}\cdot k_{4}-M+2;\cdots
,\left[  \left(  -\frac{k_{2}\cdot k_{4}}{k_{2}\cdot k_{3}}\right)
_{r}\right]  ,\cdots
\end{array}
\right)  ,\\
R_{4,3}^{M}  &  =\underset{\left\{  N_{r}\right\}  }{\sum}\left(  -k_{3}\cdot
k_{4}\right)  ^{N}B\left(  \hat{k}_{1}\cdot k_{2}+1,k_{2}\cdot k_{3}+1\right)
\left(  \prod_{r=1}\frac{1}{N_{r}!r^{N_{r}}}\right) \nonumber\\
&  \cdot F_{D}^{(K)}\left(
\begin{array}
[c]{c}%
\hat{k}_{1}\cdot k_{2}+1;\cdots,\left\{  -N_{r}\right\}  ^{r},\cdots;\\
\hat{k}_{1}\cdot k_{2}+k_{2}\cdot k_{3}+2;\cdots,\left[  \left(  -\frac
{k_{2}\cdot k_{4}}{k_{3}\cdot k_{4}}\right)  _{r}\right]  ,\cdots
\end{array}
\right)  .
\end{align}
where $K=\underset{\{\text{for all }N_{r}\neq0\}}{\sum r}$.

\subsection{The six-point KN amplitude}

For the higher $n$-point KN amplitudes, we define the residues $R_{m+1,n-m+1}%
^{M}$ ($2\leq m\leq n-2$) of $n$-point KN amplitude and express the KN
amplitude as
\begin{equation}
A_{n}^{KN}=\underset{m=2}{\overset{n-2}{\sum}}\underset{M=0}{\overset{\infty
}{\sum}}\frac{2R_{m+1,n-m+1}^{M}}{\left(  k_{1}+k_{2}+...+k_{m}\right)
^{2}+2(M-1)} \label{AJ}%
\end{equation}
where the pole locations $z_{m,M}$ are given by solutions of
\begin{equation}
\left(  \hat{k}_{1}+k_{2}+....+k_{m}\right)  ^{2}+2(M-1)=0,M=0,1,2...\text{.}%
\end{equation}%
%TCIMACRO{\FRAME{ftbpFU}{7.1788in}{1.0153in}{0pt}{\Qcb{Residues of KN
%amplitudes}}{\Qlb{recursion}}{recursion.eps}%
%width 7.1788in;  height 1.0153in;  depth 0pt;  original-width 7.1131in;
%original-height 0.9824in;  cropleft "0";  croptop "1";  cropright "1";
%cropbottom "0";  filename 'recursion.eps';file-properties "XNPEU";}} }%
%BeginExpansion
\begin{figure}[ptb]%
\centering
\includegraphics[
height=1.0153in,
width=7.1788in
]%
{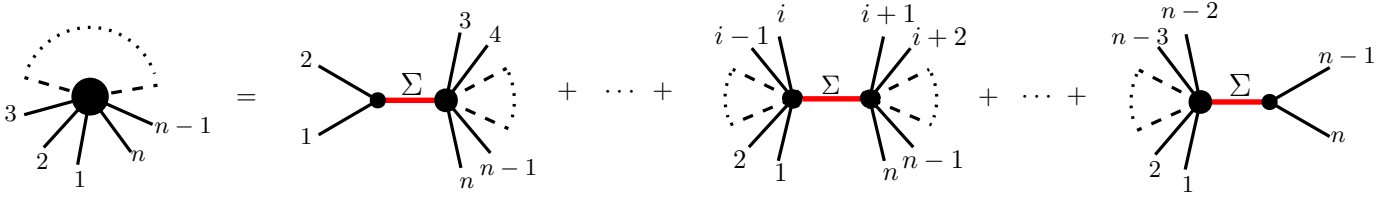}%
\caption{Residues of KN amplitudes}%
\label{recursion}%
\end{figure}
%EndExpansion

For the case of $6$-point KN amplitude. There are three types of residues
$R_{3,5}^{M}$, $R_{4,4}^{M}$ and $R_{5,3}^{M}$. For the residue $R_{3,5}^{M}$,
we \textit{propose} the left and right subamplitudes as following%
\begin{equation}
A\left(  \hat{1},2,-P\right)  =R_{3,5}^{P,L}=\prod_{r=1}\frac{\underset{\sigma
=1}{\overset{N_{r}}{%
%TCIMACRO{\dprod }%
%BeginExpansion
{\displaystyle\prod}
%EndExpansion
}}\epsilon_{r}^{(\sigma)}\cdot k_{2}}{\sqrt{N_{r}!r^{N_{r}}}}, \label{LL}%
\end{equation}%
\begin{align}
&  A\left(  P,3,4,5,\hat{6}\right)  =R_{3,5}^{P,R}\nonumber\\
&  =\int_{0}^{1}dw_{4}\int_{0}^{w_{4}}dw_{3}w_{3}^{k_{P}\cdot k_{3}}%
w_{4}^{k_{P}\cdot k_{4}}\left(  1-w_{3}\right)  ^{k_{3}\cdot k_{5}}\left(
1-w_{4}\right)  ^{k_{4}\cdot k_{5}}\left(  w_{4}-w_{3}\right)  ^{k_{3}\cdot
k_{4}}\nonumber\\
&  \times\prod_{r=1}\frac{1}{\sqrt{N_{r}!r^{N_{r}}}}\underset{\sigma
=1}{\overset{N_{r}}{%
%TCIMACRO{\dprod }%
%BeginExpansion
{\displaystyle\prod}
%EndExpansion
}}\left(  -\epsilon_{r}^{(\sigma)}\right)  \cdot\left[  k_{3}\left(  \frac
{1}{w_{3}}\right)  ^{r}+k_{4}\left(  \frac{1}{w_{4}}\right)  ^{r}+k_{5}\left(
\frac{1}{1}\right)  ^{r}\right]  . \label{RR}%
\end{align}
where $k_{P}=\hat{k}_{1}+k_{2}$ and $P$ is a normalized arbitrary string fock
state%
\begin{equation}
\left\vert P\right\rangle =\prod_{r=1}\frac{\left(  \underset{\sigma
=1}{\overset{N_{r}}{%
%TCIMACRO{\dprod }%
%BeginExpansion
{\displaystyle\prod}
%EndExpansion
}}\epsilon_{r}^{(\sigma)}\cdot\alpha_{-r}\right)  }{\sqrt{N_{r}!r^{Nr}}%
}\left\vert 0,k_{P}\right\rangle . \label{pp}%
\end{equation}
In Eq.(\ref{pp}), it is understood, for example, that the state $\epsilon
_{1}^{(1)}\cdot\alpha_{-1}\epsilon_{1}^{(2)}\cdot\alpha_{-1}\epsilon_{1}%
^{(3)}\cdot\alpha_{-1}$ means $\epsilon_{\mu\nu\lambda}\alpha_{-1}^{\mu}%
\alpha_{-1}^{\nu}\alpha_{-1}^{\lambda}$. In Eq.(\ref{LL}), we have chosen
$z_{1}=0$, $z_{2}=1$ and $z_{P}=\infty$ to fix the $SL(2,R)$ gauge. Similarly,
in Eq.(\ref{RR}), we set $w_{P}=0$, $w_{5}=1$ and $w_{6}=\infty$. After
summing over all $P$, we obtain the residue%

\begin{align}
&  R_{3,5}^{M}=\underset{\left\{  N_{r}\right\}  }{\sum}\int_{0}^{1}dw_{4}%
\int_{0}^{w_{4}}dw_{3}w_{3}^{\left(  \hat{k}_{1}+k_{2}\right)  \cdot k_{3}%
}w_{4}^{\left(  \hat{k}_{1}+k_{2}\right)  \cdot k_{4}}\left(  1-w_{3}\right)
^{k_{3}\cdot k_{5}}\left(  1-w_{4}\right)  ^{k_{4}\cdot k_{5}}\left(
w_{4}-w_{3}\right)  ^{k_{3}\cdot k_{4}}\nonumber\\
&  \times\prod_{r=1}\frac{\left[  \left(  -k_{2}\cdot k_{3}\right)  \left(
\frac{1}{w_{3}}\right)  ^{r}+\left(  -k_{2}\cdot k_{4}\right)  \left(
\frac{1}{w_{4}}\right)  ^{r}+\left(  -k_{2}\cdot k_{5}\right)  \left(
\frac{1}{1}\right)  ^{r}\right]  }{N_{r}!r^{N_{r}}}^{N_{r}}. \label{R35}%
\end{align}

For the residue $R_{4,4}^{M}$, we propose the left and right subamplitudes as
following%
\begin{align}
A\left(  \hat{1},2,3,-P\right)   &  =R_{4,4}^{P,L}=\prod_{r=1}\int_{0}%
^{1}dz_{2}z_{2}^{\hat{k}_{1}\cdot k_{2}}\left(  1-z_{2}\right)  ^{k_{2}\cdot
k_{3}}\frac{\underset{\sigma=1}{\overset{N_{r}}{%
%TCIMACRO{\dprod }%
%BeginExpansion
{\displaystyle\prod}
%EndExpansion
}}\epsilon_{r}^{(\sigma)}\cdot\left[  k_{2}\left(  z_{2}\right)  ^{r}%
+k_{3}\left(  1\right)  ^{r}\right]  }{\sqrt{N_{r}!r^{N_{r}}}},\label{LL2}\\
A\left(  P,4,5,\hat{6}\right)   &  =R_{4,4}^{P,R}=\prod_{r=1}\int_{0}%
^{1}dw_{4}w_{4}^{k_{P}\cdot k_{4}}\left(  1-w_{4}\right)  ^{k_{4}\cdot k_{5}%
}\frac{\underset{\sigma=1}{\overset{N_{r}}{%
%TCIMACRO{\dprod }%
%BeginExpansion
{\displaystyle\prod}
%EndExpansion
}}\left(  -\epsilon_{r}^{(\sigma)}\right)  \cdot\left[  k_{4}\left(  \frac
{1}{w_{4}}\right)  ^{r}+k_{5}\left(  \frac{1}{1}\right)  ^{r}\right]  }%
{\sqrt{N_{r}!r^{N_{r}}}} \label{RR2}%
\end{align}
where $k_{P}=\hat{k}_{1}+k_{2}+k_{3}$. In Eq.(\ref{LL2}), we have chosen
$z_{1}=0$, $z_{3}=1$ and $z_{P}=\infty$ to fix the $SL(2,R)$ gauge. Similarly,
in Eq.(\ref{RR2}), we set $w_{P}=0$, $w_{5}=1$ and $w_{6}=\infty$. After
summing over all $P$, we obtain the residue%
\begin{align}
R_{4,4}^{M}  &  =\underset{\left\{  N_{r}\right\}  }{\sum}\int_{0}^{1}%
dw_{4}\int_{0}^{1}dz_{2}z_{2}^{\hat{k}_{1}\cdot k_{2}}\left(  1-z_{2}\right)
^{k_{2}\cdot k_{3}}w_{4}^{\left(  \hat{k}_{1}+k_{2}+k_{3}\right)  \cdot k_{4}%
}\left(  1-w_{4}\right)  ^{k_{4}\cdot k_{5}}\nonumber\\
&  \times\prod_{r=1}\frac{\left[  \left(  -k_{2}\cdot k_{4}\right)  \left(
\frac{z_{2}}{w_{4}}\right)  ^{r}+\left(  -k_{2}\cdot k_{5}\right)  \left(
z_{2}\right)  ^{r}+\left(  -k_{3}\cdot k_{4}\right)  \left(  \frac{1}{w_{4}%
}\right)  ^{r}+\left(  -k_{3}\cdot k_{5}\right)  1^{r}\right]  ^{N_{r}}}%
{N_{r}!r^{N_{r}}}, \label{R44}%
\end{align}
which is a sum of products of the Lauricella functions after term by term
integrations for a given $M$. On the other hand, to prove that the residue
$R_{3,5}^{M}$ in Eq.(\ref{R35}) (and the residue $R_{5,3}^{M}$ to be discussed
below) can be expressed in terms of LSSA, one needs to do a second on-shell
recursion process. This will be done in section V.

For the residue $R_{5,3}^{M}$, we propose the left and right subamplitudes as
following%
\begin{align}
&  A\left(  -P,5,\hat{6}\right)  =R_{5,3}^{P,R}=\prod_{r=1}\frac{\left(
-\underset{\sigma=1}{\overset{N_{r}}{%
%TCIMACRO{\dprod }%
%BeginExpansion
{\displaystyle\prod}
%EndExpansion
}}\epsilon_{r}^{(\sigma)}\cdot k_{5}\right)  }{\sqrt{N_{r}!r^{N_{r}}}%
}\label{RR3}\\
&  A\left(  \hat{1},2,3,4,P\right)  =R_{5,3}^{P,L}\nonumber\\
&  =\int_{0}^{1}dz_{3}\int_{0}^{z_{3}}dz_{2}z_{2}^{\hat{k}_{1}\cdot k_{2}%
}z_{3}^{\hat{k}_{1}\cdot k_{3}}\left(  z_{3}-z_{2}\right)  ^{k_{2}\cdot k_{3}%
}\left(  1-z_{2}\right)  ^{k_{2}\cdot k_{4}}\left(  1-z_{3}\right)
^{k_{3}\cdot k_{4}}\nonumber\\
&  \times\prod_{r=1}\frac{\underset{\sigma=1}{\overset{N_{r}}{%
%TCIMACRO{\dprod }%
%BeginExpansion
{\displaystyle\prod}
%EndExpansion
}}\epsilon_{r}^{(\sigma)}\cdot\left[  k_{2}\left(  z_{2}\right)  ^{r}%
+k_{3}\left(  z_{3}\right)  ^{r}+k_{4}\left(  1\right)  ^{r}\right]  }%
{\sqrt{N_{r}!r^{N_{r}}}} \label{LL3}%
\end{align}
where we have chosen $z_{1}=0$, $z_{4}=1$ and $z_{P}=\infty$ to fix the
$SL(2,R)$ gauge. Similarly, in Eq.(\ref{RR3}), we set $w_{P}=0$, $w_{5}=1$ and
$w_{6}=\infty$. After summing over all $P$, we can get%

\begin{align}
R_{5,3}^{M}  &  =\underset{\left\{  N_{r}\right\}  }{\sum}\int_{0}^{1}%
dz_{3}\int_{0}^{z_{3}}dz_{2}z_{2}^{\hat{k}_{1}\cdot k_{2}}z_{3}^{\hat{k}%
_{1}\cdot k_{3}}\left(  z_{3}-z_{2}\right)  ^{k_{2}\cdot k_{3}}\left(
1-z_{2}\right)  ^{k_{2}\cdot k_{4}}\left(  1-z_{3}\right)  ^{k_{3}\cdot k_{4}%
}\nonumber\\
&  \times\prod_{r=1}\frac{\left[  \left(  -k_{2}\cdot k_{5}\right)  \left(
z_{2}\right)  ^{r}+\left(  -k_{3}\cdot k_{5}\right)  \left(  z_{3}\right)
^{r}+\left(  -k_{4}\cdot k_{5}\right)  \left(  1\right)  ^{r}\right]  ^{N_{r}%
}}{N_{r}!r^{N_{r}}}. \label{R53}%
\end{align}

To justify the results in Eq.(\ref{R35}), Eq.(\ref{R44}) and Eq.(\ref{R53}),
we follow the standard string on-shell recursion calculation. The $6$-point KN
amplitude can be written as%

\begin{align}
&  A_{6}^{KN}=\int_{0}^{1}dz_{4}\int_{0}^{z_{4}}dz_{3}\int_{0}^{z_{3}}%
dz_{2}z_{2}^{k_{1}\cdot k_{2}}z_{3}^{k_{1}\cdot k_{3}}z_{4}^{k_{1}\cdot k_{4}%
}\nonumber\\
&  \cdot\left(  z_{3}-z_{2}\right)  ^{k_{2}\cdot k_{3}}\left(  z_{4}%
-z_{2}\right)  ^{k_{2}\cdot k_{4}}\left(  1-z_{2}\right)  ^{k_{2}\cdot k_{5}%
}\left(  z_{4}-z_{3}\right)  ^{k_{3}\cdot k_{4}}\left(  1-z_{3}\right)
^{k_{3}\cdot k_{5}}\nonumber\\
&  =\underset{a,b,c,d,e,f=0}{\sum}\frac{\left(  -k_{2}\cdot k_{3}\right)
_{a}}{a!}\frac{\left(  -k_{2}\cdot k_{4}\right)  _{b}}{b!}\frac{\left(
-k_{2}\cdot k_{5}\right)  _{c}}{c!}\frac{\left(  -k_{3}\cdot k_{4}\right)
_{d}}{d!}\frac{\left(  -k_{3}\cdot k_{5}\right)  _{e}}{e!}\frac{\left(
-k_{4}\cdot k_{5}\right)  _{f}}{f!}\nonumber\\
&  \times\frac{2}{\left(  -s_{12}\right)  +2\left(  a+b+c-1\right)  }\frac
{2}{\left(  -s_{123}\right)  +2\left(  b+c+d+e-1\right)  }\frac{2}{\left(
-s_{1234}\right)  +2\left(  c+e+f-1\right)  }%
\end{align}
where%
\begin{equation}
s_{12}=-\left(  k_{1}+k_{2}\right)  ^{2},s_{123}=-\left(  k_{1}+k_{2}%
+k_{3}\right)  ^{2},s_{1234}=-\left(  k_{1}+k_{2}+k_{3}+k_{4}\right)  ^{2}.
\end{equation}
The next step is to do the on-shell recursion deformation%

\begin{equation}
\hat{k}_{1}\left(  z\right)  =k_{1}+zq,q^{2}=k_{1}\cdot q=k_{6}\cdot
q=0,\hat{k}_{6}\left(  z\right)  =k_{6}-zq
\end{equation}
to obtain the poles%
\begin{align}
z_{2,M}  &  =\frac{\left(  k_{1}+k_{2}\right)  ^{2}+2\left(  M-1\right)
}{-2q\cdot\left(  k_{1}+k_{2}\right)  },M=a+b+c=0,1,2,...;\\
z_{3,M}  &  =\frac{\left(  k_{1}+k_{2}+k_{3}\right)  ^{2}+2\left(  M-1\right)
}{-2q\cdot\left(  k_{1}+k_{2}+k_{3}\right)  },M=b+c+d+e=0,1,2,...;\\
z_{4,M}  &  =\frac{\left(  k_{1}+k_{2}+k_{3}+k_{4}\right)  ^{2}+2\left(
M-1\right)  }{-2q\cdot\left(  k_{1}+k_{2}+k_{3}+k_{4}\right)  }%
,M=c+e+f=0,1,2,....
\end{align}
The amplitude can then be written as%
\begin{equation}
A_{6}^{KN}=\underset{M}{\sum}\frac{2}{\left(  -s_{12}\right)  +2\left(
M-1\right)  }R_{3,5}^{M}+\underset{M}{\sum}\frac{2}{\left(  -s_{123}\right)
+2\left(  M-1\right)  }R_{4,4}^{M}+\underset{M}{\sum}\frac{2}{\left(
-s_{1234}\right)  +2\left(  M-1\right)  }R_{5,3}^{M}%
\end{equation}
where the residues can be calculated to be%
\begin{align}
&  R_{3,5}^{M}=\underset{%
\begin{array}
[c]{c}%
a,b,c,d,e,f=0\\
M=a+b+c
\end{array}
}{\sum}\frac{\left(  -k_{2}\cdot k_{3}\right)  _{a}}{a!}\frac{\left(
-k_{2}\cdot k_{4}\right)  _{b}}{b!}\frac{\left(  -k_{2}\cdot k_{5}\right)
_{c}}{c!}\frac{\left(  -k_{3}\cdot k_{4}\right)  _{d}}{d!}\frac{\left(
-k_{3}\cdot k_{5}\right)  _{e}}{e!}\frac{\left(  -k_{4}\cdot k_{5}\right)
_{f}}{f!}\nonumber\\
&  \times\frac{2}{\left(  -\hat{s}_{123}\left(  z_{2,M}\right)  \right)
+2\left(  b+c+d+e-1\right)  }\frac{2}{\left(  -\hat{s}_{1234}\left(
z_{2,M}\right)  \right)  +2\left(  c+e+f-1\right)  },\label{2R35}\\
&  R_{4,4}^{M}=\underset{%
\begin{array}
[c]{c}%
a,b,c,d,e,f=0\\
M=b+c+d+e
\end{array}
}{\sum}\frac{\left(  -k_{2}\cdot k_{3}\right)  _{a}}{a!}\frac{\left(
-k_{2}\cdot k_{4}\right)  _{b}}{b!}\frac{\left(  -k_{2}\cdot k_{5}\right)
_{c}}{c!}\frac{\left(  -k_{3}\cdot k_{4}\right)  _{d}}{d!}\frac{\left(
-k_{3}\cdot k_{5}\right)  _{e}}{e!}\frac{\left(  -k_{4}\cdot k_{5}\right)
_{f}}{f!}\nonumber\\
&  \times\frac{2}{\left(  -\hat{s}_{12}\left(  z_{3,M}\right)  \right)
+2\left(  a+b+c-1\right)  }\frac{2}{\left(  -\hat{s}_{1234}\left(
z_{3,M}\right)  \right)  +2\left(  c+e+f-1\right)  },\label{2R44}\\
&  R_{5,3}^{M}=\underset{%
\begin{array}
[c]{c}%
a,b,c,d,e,f=0\\
M=c+e+f
\end{array}
}{\sum}\frac{\left(  -k_{2}\cdot k_{3}\right)  _{a}}{a!}\frac{\left(
-k_{2}\cdot k_{4}\right)  _{b}}{b!}\frac{\left(  -k_{2}\cdot k_{5}\right)
_{c}}{c!}\frac{\left(  -k_{3}\cdot k_{4}\right)  _{d}}{d!}\frac{\left(
-k_{3}\cdot k_{5}\right)  _{e}}{e!}\frac{\left(  -k_{4}\cdot k_{5}\right)
_{f}}{f!}\nonumber\\
&  \times\frac{2}{\left(  -\hat{s}_{12}\left(  z_{4,M}\right)  \right)
+2\left(  a+b+c-1\right)  }\frac{2}{\left(  -\hat{s}_{123}\left(
z_{4,M}\right)  \right)  +2\left(  b+c+d+e-1\right)  }. \label{2R53}%
\end{align}

Finally, we need to identify Eq.(\ref{R35}) with Eq.(\ref{2R35}),
Eq.(\ref{R44}) with Eq.(\ref{2R44}) and Eq.(\ref{R53}) with Eq.(\ref{2R53})
respectively. To do this on the residue $R_{3,5}^{M}$, we apply Eq.(\ref{id1})
for the case of $n=3$ and rewrite Eq.(\ref{R35}) as%
\begin{align}
&  R_{3,5}^{M}=\int_{0}^{1}dw_{4}\int_{0}^{w_{4}}dw_{3}w_{3}^{\left(  \hat
{k}_{1}+k_{2}\right)  \cdot k_{3}}w_{4}^{\left(  \hat{k}_{1}+k_{2}\right)
\cdot k_{4}}\left(  1-w_{3}\right)  ^{k_{3}\cdot k_{5}}\left(  1-w_{4}\right)
^{k_{4}\cdot k_{5}}\left(  w_{4}-w_{3}\right)  ^{k_{3}\cdot k_{4}}\nonumber\\
&  \times\underset{%
\begin{array}
[c]{c}%
M=a+b+c\\
a,b,c=0
\end{array}
}{\sum}\frac{\left(  -k_{2}\cdot k_{3}\right)  _{a}}{a!}\frac{\left(
-k_{2}\cdot k_{4}\right)  _{b}}{b!}\frac{\left(  -k_{2}\cdot k_{5}\right)
_{c}}{c!}w_{3}^{-a}w_{4}^{-b}1^{-c},
\end{align}
which after binormal expansions and integrations can be further written as

\bigskip%
\begin{align}
R_{3,5}^{M}  &  =\underset{%
\begin{array}
[c]{c}%
M=a+b+c\\
a,b,c,d,e,f=0
\end{array}
}{\sum}\frac{\left(  -k_{2}\cdot k_{3}\right)  _{a}}{a!}\frac{\left(
-k_{2}\cdot k_{4}\right)  _{b}}{b!}\frac{\left(  -k_{2}\cdot k_{5}\right)
_{c}}{c!}\frac{\left(  -k_{3}\cdot k_{4}\right)  _{d}}{d!}\frac{\left(
-k_{3}\cdot k_{5}\right)  _{e}}{e!}\frac{\left(  -k_{4}\cdot k_{5}\right)
_{f}}{f!}\nonumber\\
&  \times\frac{1}{\left(  \hat{k}_{1}+k_{2}\right)  \cdot k_{3}+(d+e-a+1)}%
\frac{1}{\left(  \hat{k}_{1}+k_{2}\right)  \cdot\left(  k_{3}+k_{4}\right)
+k_{3}\cdot k_{4}+\left(  e+f-a+2\right)  }. \label{same}%
\end{align}
We see that Eq.(\ref{same}) above is the same as Eq.(\ref{2R35}) since%
\begin{equation}
2\left(  \hat{k}_{1}+k_{2}\right)  \cdot k_{3}+2\left(  d+e-a+1\right)
=\left(  -\hat{s}_{123}\left(  z_{2,M}\right)  \right)  +2\left(
b+c+d+e-1\right)
\end{equation}
and%
\begin{equation}
2\left(  \hat{k}_{1}+k_{2}\right)  \cdot\left(  k_{3}+k_{4}\right)
+2k_{3}\cdot k_{4}+2\left(  e+f-a+2\right)  =\left(  -\hat{s}_{1234}\left(
z_{2,M}\right)  \right)  +2\left(  c+e+f-1\right)  .
\end{equation}
This completes the proof that the proposals of the subamplitudes in
Eq.(\ref{LL}), Eq.(\ref{RR}) and thus the result of the residue calculated in
Eq.(\ref{R35}) are indeed the correct ones.

For the residue $R_{4,4}^{M}$, we apply Eq.(\ref{id1}) for the case of $n=4$
and rewrite Eq.(\ref{R44}) as%
\begin{align}
R_{4,4}^{M} &  =\int_{0}^{1}dw_{4}\int_{0}^{1}dz_{2}z_{2}^{\hat{k}_{1}\cdot
k_{2}}\left(  1-z_{2}\right)  ^{k_{2}\cdot k_{3}}w_{4}^{\left(  \hat{k}%
_{1}+k_{2}+k_{3}\right)  \cdot k_{4}}\left(  1-w_{4}\right)  ^{k_{4}\cdot
k_{5}}\nonumber\\
&  \underset{%
\begin{array}
[c]{c}%
b,c,d,e=0\\
M=b+c+d+e
\end{array}
}{\sum}\frac{\left(  -k_{2}\cdot k_{4}\right)  _{b}}{b!}\frac{\left(
-k_{2}\cdot k_{5}\right)  _{c}}{c!}\frac{\left(  -k_{3}\cdot k_{4}\right)
_{d}}{d!}\frac{\left(  -k_{3}\cdot k_{5}\right)  _{e}}{e!}\left(  \frac{z_{2}%
}{w_{4}}\right)  ^{b}\left(  z_{2}\right)  ^{c}\left(  \frac{1}{w_{4}}\right)
^{d}%
\end{align}
which after binomial expansions and integrations can be further written as%
\begin{align}
R_{4,4}^{M} &  =\underset{%
\begin{array}
[c]{c}%
M=b+c+d+e\\
a,b,c,d,e,f=0
\end{array}
}{\sum}\frac{\left(  -k_{2}\cdot k_{3}\right)  _{a}}{a!}\frac{\left(
-k_{2}\cdot k_{4}\right)  _{b}}{b!}\frac{\left(  -k_{2}\cdot k_{5}\right)
_{c}}{c!}\frac{\left(  -k_{3}\cdot k_{4}\right)  _{d}}{d!}\frac{\left(
-k_{3}\cdot k_{5}\right)  _{e}}{e!}\frac{\left(  -k_{4}\cdot k_{5}\right)
_{f}}{f!}\nonumber\\
&  \times\frac{1}{\hat{k}_{1}\cdot k_{2}+\left(  a+b+c+1\right)  }\frac
{1}{\left(  \hat{k}_{1}+k_{2}+k_{3}\right)  \cdot k_{4}+\left(
f-b-d+1\right)  }.\label{same2}%
\end{align}
We see that Eq.(\ref{same2}) above is the same as Eq.(\ref{2R44}) since%
\begin{equation}
-\hat{s}_{12}\left(  z_{3,M}\right)  +2\left(  a+b+c-1\right)  =2\hat{k}%
_{1}\cdot k_{2}+2\left(  a+b+c+1\right)
\end{equation}
and%
\begin{equation}
-\hat{s}_{1234}(z_{3,M})+2\left(  c+e+f-1\right)  =2\left(  \hat{k}_{1}%
+k_{2}+k_{3}\right)  \cdot k_{4}+2\left(  f-b-d+1\right)  .
\end{equation}
This completes the proof that the proposals of the subamplitudes in
Eq.(\ref{LL2}), Eq.(\ref{RR2}) and thus the result of the residue calculated
in Eq.(\ref{R44}) are indeed the correct ones.\bigskip

For the residue $R_{5,3}^{M}$, we apply Eq.(\ref{id1}) for the case of $n=3$
and rewrite Eq.(\ref{R53}) as%
\begin{align}
R_{5,3}^{M} &  =\int_{0}^{1}dz_{3}\int_{0}^{z_{3}}dz_{2}z_{2}^{\hat{k}%
_{1}\cdot k_{2}}z_{3}^{\hat{k}_{1}\cdot k_{3}}\left(  z_{3}-z_{2}\right)
^{k_{2}\cdot k_{3}}\left(  1-z_{2}\right)  ^{k_{2}\cdot k_{4}}\left(
1-z_{3}\right)  ^{k_{3}\cdot k_{4}}\nonumber\\
&  \underset{%
\begin{array}
[c]{c}%
M=c+e+f\\
c,e,f=0
\end{array}
}{\sum}\frac{\left(  -k_{2}\cdot k_{5}\right)  _{c}}{c!}\frac{\left(
-k_{3}\cdot k_{5}\right)  _{e}}{e!}\frac{\left(  -k_{4}\cdot k_{5}\right)
_{f}}{f!}\left(  z_{2}\right)  ^{c}\left(  z_{3}\right)  ^{e}\left(  1\right)
^{f},
\end{align}
which after binomial expansions and integrations can be further written as%
\begin{align}
R_{5,3}^{M} &  =\underset{%
\begin{array}
[c]{c}%
M=c+e+f\\
a,b,c,d,e,f=0
\end{array}
}{\sum}\frac{\left(  -k_{2}\cdot k_{3}\right)  _{a}}{a!}\frac{\left(
-k_{2}\cdot k_{4}\right)  _{b}}{b!}\frac{\left(  -k_{2}\cdot k_{5}\right)
_{c}}{c!}\frac{\left(  -k_{3}\cdot k_{4}\right)  _{c}}{d!}\frac{\left(
-k_{3}\cdot k_{5}\right)  _{e}}{e!}\frac{\left(  -k_{4}\cdot k_{5}\right)
_{f}}{f!}\nonumber\\
&  \times\frac{1}{\hat{k}_{1}\cdot k_{2}+a+b+c+1}\frac{1}{\hat{k}_{1}\cdot
k_{2}+\hat{k}_{1}\cdot k_{3}+k_{2}\cdot k_{3}+b+c+d+e+2}\label{same3}%
\end{align}
We see that Eq.(\ref{same3}) above is the same as Eq.(\ref{2R53}) since%
\begin{equation}
-\hat{s}_{12}\left(  z_{4,M}\right)  +2\left(  a+b+c-1\right)  =2\hat{k}%
_{1}\cdot k_{2}+2\left(  a+b+c+1\right)  \label{cc2}%
\end{equation}
and%
\begin{equation}
-\hat{s}_{123}\left(  z_{4,M}\right)  +2\left(  b+c+d+e-1\right)  =2\hat
{k}_{1}\cdot k_{2}+2\hat{k}_{1}\cdot k_{3}+2k_{2}\cdot k_{3}%
+2(b+c+d+e+2).\label{cc3}%
\end{equation}

This completes the proof that the proposals of the subamplitudes in
Eq.(\ref{LL3}), Eq.(\ref{RR3}) and thus the result of the residue calculated
in Eq.(\ref{R53}) are indeed the correct ones.\bigskip

\subsection{The seven-point KN amplitude}

For the case of $7$-point KN amplitude. There are four types of residues
$R_{3,6}^{M}$, $R_{4,5}^{M}$, $R_{5,4}^{M}$ and $R_{6,3}^{M}$. We will work
out residues $R_{3,6}^{M}$and $R_{4,5}^{M}$. Similar calculation can be done
for $R_{5,4}^{M}$ and $R_{6,3}^{M}$. For the residue $R_{3,6}^{M}$, we propose
the left and right subamplitudes as following%
\begin{equation}
A\left(  \hat{1},2,-P\right)  =R_{3,6}^{P,L}=\prod_{r=1}\frac{\left(
\underset{\sigma=1}{\overset{N_{r}}{%
%TCIMACRO{\dprod }%
%BeginExpansion
{\displaystyle\prod}
%EndExpansion
}}\epsilon_{r}^{(\sigma)}\cdot k_{2}\right)  }{\sqrt{N_{r}!r^{N_{r}}}},
\label{LL7}%
\end{equation}%
\begin{align}
&  A\left(  P,3,4,5,6,\hat{7}\right)  =R_{3,6}^{P,R}\nonumber\\
&  =\int_{0}^{1}dw_{5}\int_{0}^{w_{5}}dw_{4}\int_{0}^{w_{4}}dw_{3}w_{3}%
^{k_{P}\cdot k_{3}}w_{4}^{k_{P}\cdot k_{4}}w_{5}^{\left(  \hat{k}_{1}%
+k_{2}\right)  \cdot k_{5}}\left(  1-w_{3}\right)  ^{k_{3}\cdot k_{6}}\left(
1-w_{4}\right)  ^{k_{4}\cdot k_{6}}\nonumber\\
&  \cdot\left(  1-w_{5}\right)  ^{k_{5}\cdot k_{6}}\left(  w_{4}-w_{3}\right)
^{k_{3}\cdot k_{4}}\left(  w_{5}-w_{3}\right)  ^{k_{3}\cdot k_{5}}\left(
w_{5}-w_{4}\right)  ^{k_{4}\cdot k_{5}}\nonumber\\
&  \times\prod_{r=1}\frac{1}{\sqrt{N_{r}!r^{N_{r}}}}\underset{\sigma
=1}{\overset{N_{r}}{%
%TCIMACRO{\dprod }%
%BeginExpansion
{\displaystyle\prod}
%EndExpansion
}}\left(  -\epsilon_{r}^{(\sigma)}\right)  \cdot\left[  k_{3}\left(  \frac
{1}{w_{3}}\right)  ^{r}+k_{4}\left(  \frac{1}{w_{4}}\right)  ^{r}+k_{5}\left(
\frac{1}{w_{5}}\right)  ^{r}+k_{6}\left(  \frac{1}{1}\right)  ^{r}\right]
\label{RR7}%
\end{align}
where $P$ is a normalized arbitrary string fock state defined in Eq.(\ref{pp})
and $k_{P}=\hat{k}_{1}+k_{2}$. In Eq.(\ref{LL7}), we have chosen $z_{1}=0$,
$z_{2}=1$ and $z_{P}=\infty$ to fix the $SL(2,R)$ gauge. Similarly, in
Eq.(\ref{RR7}), we set $w_{P}=0$, $w_{6}=1$ and $w_{7}=\infty$. After summing
over all $P$, we obtain the residue%

\begin{align}
&  R_{3,6}^{M}=\underset{\left\{  N_{r}\right\}  }{\sum}\int_{0}^{1}dw_{5}%
\int_{0}^{w_{5}}dw_{4}\int_{0}^{w_{4}}dw_{3}w_{3}^{\left(  \hat{k}_{1}%
+k_{2}\right)  \cdot k_{3}}w_{4}^{\left(  \hat{k}_{1}+k_{2}\right)  \cdot
k_{4}}w_{5}^{\left(  \hat{k}_{1}+k_{2}\right)  \cdot k_{5}}\nonumber\\
&  \cdot\left(  1-w_{3}\right)  ^{k_{3}\cdot k_{5}}\left(  1-w_{4}\right)
^{k_{4}\cdot k_{5}}\left(  w_{4}-w_{3}\right)  ^{k_{3}\cdot k_{4}}\nonumber\\
&  \times\prod_{r=1}\frac{\left[  \left(  -k_{2}\cdot k_{3}\right)  \left(
\frac{1}{w_{3}}\right)  ^{r}+\left(  -k_{2}\cdot k_{4}\right)  \left(
\frac{1}{w_{4}}\right)  ^{r}+\left(  -k_{2}\cdot k_{5}\right)  \left(
\frac{1}{w_{5}}\right)  ^{r}+\left(  -k_{2}\cdot k_{6}\right)  \left(
\frac{1}{1}\right)  ^{r}\right]  }{N_{r}!r^{N_{r}}}^{N_{r}}. \label{R36}%
\end{align}

To justify the result in Eq.(\ref{R36}), we follow the standard string
on-shell recursion calculation. First of all, we use the identity
Eq.(\ref{id1}) to transform the last line of Eq.(\ref{R36}) to a summation,
and then do binomial expansions before performing integrations. Finally we end
up with%
\begin{align}
R_{3,6}^{M} &  =\underset{%
\begin{array}
[c]{c}%
M=a+b+c+d\\
a,b,c,d,e,f,g,h,i,j=0
\end{array}
}{\sum}\frac{\left(  -k_{2}\cdot k_{3}\right)  _{a}}{a!}\frac{\left(
-k_{2}\cdot k_{4}\right)  _{b}}{b!}\frac{\left(  -k_{2}\cdot k_{5}\right)
_{c}}{c!}\frac{\left(  -k_{2}\cdot k_{6}\right)  _{d}}{d!}\nonumber\\
&  \cdot\frac{\left(  -k_{3}\cdot k_{4}\right)  _{e}}{e!}\frac{\left(
-k_{3}\cdot k_{5}\right)  _{f}}{f!}\frac{\left(  -k_{3}\cdot k_{6}\right)
_{g}}{g!}\frac{\left(  -k_{4}\cdot k_{5}\right)  _{h}}{h!}\frac{\left(
-k_{4}\cdot k_{6}\right)  _{i}}{i!}\frac{\left(  -k_{5}\cdot k_{6}\right)
_{j}}{j!}\nonumber\\
&  \times\frac{1}{\left(  \hat{k}_{1}+k_{2}\right)  \cdot k_{3}+(e+f+g-a+1)}%
\frac{1}{\left(  \hat{k}_{1}+k_{2}\right)  \cdot\left(  k_{3}+k_{4}\right)
+k_{3}\cdot k_{4}+\left(  f+g+h+i-a-b+2\right)  }\nonumber\\
&  \times\frac{1}{\left(  \hat{k}_{1}+k_{2}\right)  \cdot\left(  k_{3}%
+k_{4}+k_{5}\right)  +k_{3}\cdot k_{4}+k_{3}\cdot k_{5}+k_{4}\cdot
k_{5}+\left(  g+i+j-a-b-c+3\right)  }.\label{500}%
\end{align}

On the other hand, the residue $R_{3,6}^{M}$ can be extracted from the pole of
the $7$-point KN amplitude%
\begin{equation}
z_{2,M}=\frac{\left(  k_{1}+k_{2}\right)  ^{2}+2\left(  M-1\right)  }%
{-2q\cdot\left(  k_{1}+k_{2}\right)  },M=a+b+c+d=0,1,2,...\text{,}%
\end{equation}
and calculated to be%
\begin{align}
R_{3,6}^{M}  &  =\underset{%
\begin{array}
[c]{c}%
M=a+b+c+d\\
a,b,c,d,e,f,g,h,i,j=0
\end{array}
}{\sum}\frac{\left(  -k_{2}\cdot k_{3}\right)  _{a}}{a!}\frac{\left(
-k_{2}\cdot k_{4}\right)  _{b}}{b!}\frac{\left(  -k_{2}\cdot k_{5}\right)
_{c}}{c!}\frac{\left(  -k_{2}\cdot k_{6}\right)  _{d}}{d!}\nonumber\\
&  \cdot\frac{\left(  -k_{3}\cdot k_{4}\right)  _{e}}{e!}\frac{\left(
-k_{3}\cdot k_{5}\right)  _{f}}{f!}\frac{\left(  -k_{3}\cdot k_{6}\right)
_{g}}{g!}\frac{\left(  -k_{4}\cdot k_{5}\right)  _{h}}{h!}\frac{\left(
-k_{4}\cdot k_{6}\right)  _{i}}{i!}\frac{\left(  -k_{5}\cdot k_{6}\right)
_{j}}{j!}\nonumber\\
&  \times\frac{2}{-\hat{s}_{123}\left(  z_{2,M}\right)  +2\left(
b+c+d+e+f+g-1\right)  }\frac{2}{-\hat{s}_{1234}\left(  z_{2,M}\right)
+2\left(  c+d+f+g+h+i-1\right)  }\nonumber\\
&  \times\frac{2}{-\hat{s}_{12345}\left(  z_{2,M}\right)  +2\left(
d+g+i+j-1\right)  }. \label{501}%
\end{align}
It is straightforwad to verify%
\begin{align}
-\hat{s}_{123}\left(  z_{2,M}\right)  +2\left(  b+c+d+e+f+g-1\right)   &
=2\left(  \hat{k}_{1}+k_{2}\right)  \cdot k_{3}+2(e+f+g-a+1),\\
-\hat{s}_{1234}\left(  z_{2,M}\right)  +2\left(  c+d+f+g+h+i-1\right)   &
=2\left(  \hat{k}_{1}+k_{2}\right)  \cdot\left(  k_{3}+k_{4}\right)
+2k_{3}\cdot k_{4}\nonumber\\
&  +2\left(  f+g+h+i-a-b+2\right)  ,\\
-\hat{s}_{12345}\left(  z_{2,M}\right)  +2\left(  d+g+i+j-1\right)   &
=2\left(  \hat{k}_{1}+k_{2}\right)  \cdot\left(  k_{3}+k_{4}+k_{5}\right)
+2k_{3}\cdot k_{4}+2k_{3}\cdot k_{5}\nonumber\\
&  +2k_{4}\cdot k_{5}+2\left(  g+i+j-a-b-c+3\right)  .
\end{align}
This shows that Eq.(\ref{500}) is the same as Eq.(\ref{501}), and the proposal
in Eq.(\ref{LL7}) and Eq.(\ref{RR7}) are consistent ones.

\bigskip For the residue $R_{4,5}^{M}$, we propose the left and right
subamplitudes as following%
\begin{align}
A\left(  \hat{1},2,3,-P\right)   &  =R_{4,5}^{P,L}=\prod_{r=1}\int_{0}%
^{1}dz_{2}z_{2}^{\hat{k}_{1}\cdot k_{2}}\left(  1-z_{2}\right)  ^{k_{2}\cdot
k_{3}}\frac{\underset{\sigma=1}{\overset{N_{r}}{%
%TCIMACRO{\dprod }%
%BeginExpansion
{\displaystyle\prod}
%EndExpansion
}}\epsilon_{r}^{(\sigma)}\cdot\left[  k_{2}\left(  z_{2}\right)  ^{r}%
+k_{3}\left(  1\right)  ^{r}\right]  }{\sqrt{N_{r}!r^{N_{r}}}},\label{LL77}\\
A\left(  P,4,5,6,\hat{7}\right)   &  =R_{4,5}^{P,R}=\prod_{r=1}\int_{0}%
^{1}dw_{4}\int_{0}^{w_{4}}dw_{5}w_{4}^{k_{P}\cdot k_{4}}w_{5}^{k_{P}\cdot
k_{5}}\left(  1-w_{4}\right)  ^{k_{4}\cdot k_{6}}\left(  1-w_{5}\right)
^{k_{5}\cdot k_{6}}\left(  w_{5}-w_{4}\right)  ^{k_{4}\cdot k_{5}}\nonumber\\
&  \times\frac{\underset{\sigma=1}{\overset{N_{r}}{%
%TCIMACRO{\dprod }%
%BeginExpansion
{\displaystyle\prod}
%EndExpansion
}}\left(  -\epsilon_{r}^{(\sigma)}\right)  \cdot\left[  k_{4}\left(  \frac
{1}{w_{4}}\right)  ^{r}+k_{5}\left(  \frac{1}{w_{5}}\right)  ^{r}+k_{6}\left(
\frac{1}{1}\right)  ^{r}\right]  }{\sqrt{N_{r}!r^{N_{r}}}}\label{RR77}%
\end{align}
where $k_{P}=\hat{k}_{1}+k_{2}+k_{3}$. After summing over all $P$, we obtain
the residue%
\begin{align}
R_{4,5}^{M} &  =\underset{\left\{  N_{r}\right\}  }{\sum}\int_{0}^{1}%
dw_{4}\int_{0}^{w_{4}}dw_{5}\int_{0}^{1}dz_{2}z_{2}^{\hat{k}_{1}\cdot k_{2}%
}\left(  1-z_{2}\right)  ^{k_{2}\cdot k_{3}}w_{4}^{k_{P}\cdot k_{4}}%
w_{5}^{k_{P}\cdot k_{5}}\left(  1-w_{4}\right)  ^{k_{4}\cdot k_{6}}\left(
1-w_{5}\right)  ^{k_{5}\cdot k_{6}}\left(  w_{5}-w_{4}\right)  ^{k_{4}\cdot
k_{5}}\nonumber\\
&  \times\prod_{r=1}\frac{\left[
\begin{array}
[c]{c}%
\left(  -k_{2}\cdot k_{4}\right)  \left(  \frac{z_{2}}{w_{4}}\right)
^{r}+\left(  -k_{2}\cdot k_{5}\right)  \left(  \frac{z_{2}}{w_{5}}\right)
^{r}+\left(  -k_{2}\cdot k_{6}\right)  \left(  z_{2}\right)  ^{r}\\
+\left(  -k_{3}\cdot k_{4}\right)  \left(  \frac{1}{w_{4}}\right)
^{r}+\left(  -k_{3}\cdot k_{5}\right)  \left(  \frac{1}{w_{5}}\right)
^{r}+\left(  k_{3}\cdot k_{6}\right)  \left(  \frac{1}{1}\right)  ^{r}%
\end{array}
\right]  ^{N_{r}}}{N_{r}!r^{N_{r}}}.\label{R45}%
\end{align}
The next step is to use the identity in Eq.(\ref{id1}) and then do binomial
expansions before performing integrations on Eq.(\ref{R45}), we finally end up
with%
\begin{align}
R_{4,5}^{M} &  =\underset{%
\begin{array}
[c]{c}%
M=b+c+d+e+f+g\\
a,b,c,d,e,f,g,h,i,j=0
\end{array}
}{\sum}\frac{\left(  -k_{2}\cdot k_{3}\right)  _{a}}{a!}\frac{\left(
-k_{2}\cdot k_{4}\right)  _{b}}{b!}\frac{\left(  -k_{2}\cdot k_{5}\right)
_{c}}{c!}\frac{\left(  -k_{2}\cdot k_{6}\right)  _{d}}{d!}\nonumber\\
&  \cdot\frac{\left(  -k_{3}\cdot k_{4}\right)  _{e}}{e!}\frac{\left(
-k_{3}\cdot k_{5}\right)  _{f}}{f!}\frac{\left(  -k_{3}\cdot k_{6}\right)
_{g}}{g!}\frac{\left(  -k_{4}\cdot k_{5}\right)  _{h}}{h!}\frac{\left(
-k_{4}\cdot k_{6}\right)  _{i}}{i!}\frac{\left(  -k_{5}\cdot k_{6}\right)
_{j}}{j!}\nonumber\\
&  \times\frac{1}{\left(  \hat{k}_{1}+k_{2}+k_{3}\right)  \cdot k_{4}%
+(h+i-b-e+1)}\nonumber\\
&  \times\frac{1}{\left(  \hat{k}_{1}+k_{2}+k_{3}\right)  \cdot\left(
k_{4}+k_{5}\right)  +k_{4}\cdot k_{5}+\left(  i+j-b-c-e-f+2\right)
}\nonumber\\
&  \times\frac{1}{\hat{k}_{1}\cdot k_{2}+\left(  a+b+c+d+1\right)
}.\label{2R45}%
\end{align}

On the other hand, the residue $R_{4,5}^{M}$ can be extracted from the pole of
the $7$-point KN amplitude%
\begin{equation}
z_{3,M}=\frac{\left(  k_{1}+k_{2}+k_{3}\right)  ^{2}+2\left(  M-1\right)
}{-2q\cdot\left(  k_{1}+k_{2}+k_{3}\right)  },M=b+c+d+e+f+g=0,1,2,...\text{,}%
\end{equation}
and calculated to be%
\begin{align}
R_{4,5}^{M} &  =\underset{%
\begin{array}
[c]{c}%
M=b+c+d+e+f+g\\
a,b,c,d,e,f,g,h,i,j=0
\end{array}
}{\sum}\frac{\left(  -k_{2}\cdot k_{3}\right)  _{a}}{a!}\frac{\left(
-k_{2}\cdot k_{4}\right)  _{b}}{b!}\frac{\left(  -k_{2}\cdot k_{5}\right)
_{c}}{c!}\frac{\left(  -k_{2}\cdot k_{6}\right)  _{d}}{d!}\nonumber\\
&  \cdot\frac{\left(  -k_{3}\cdot k_{4}\right)  _{e}}{e!}\frac{\left(
-k_{3}\cdot k_{5}\right)  _{f}}{f!}\frac{\left(  -k_{3}\cdot k_{6}\right)
_{g}}{g!}\frac{\left(  -k_{4}\cdot k_{5}\right)  _{h}}{h!}\frac{\left(
-k_{4}\cdot k_{6}\right)  _{i}}{i!}\frac{\left(  -k_{5}\cdot k_{6}\right)
_{j}}{j!}\nonumber\\
&  \times\frac{1}{-\hat{s}_{1234}\left(  z_{3,M}\right)  +2\left(
c+d+f+g+h+i-1\right)  }\nonumber\\
&  \times\frac{1}{-\hat{s}_{12345}\left(  z_{3,M}\right)  +2\left(
d+g+i+j-1\right)  }\times\frac{1}{-\hat{s}_{12}\left(  z_{3,M}\right)
+2\left(  a+b+c+d-1\right)  }.\label{3R45}%
\end{align}
It is straightforward to verify%
\begin{align}
-\hat{s}_{1234}\left(  z_{3,M}\right)  +2\left(  c+d+f+g+h+i-1\right)   &
=2\left(  \hat{k}_{1}+k_{2}+k_{3}\right)  \cdot k_{4}+2(h+i-b-e+1),\label{111}%
\\
-\hat{s}_{12345}\left(  z_{3,M}\right)  +2\left(  d+g+i+j-1\right)   &
=2\left(  \hat{k}_{1}+k_{2}+k_{3}\right)  \cdot\left(  k_{4}+k_{5}\right)
+2k_{4}\cdot k_{5}\nonumber\\
&  +2\left(  i+j-b-c-e-f+2\right)  ,\label{222}\\
-\hat{s}_{12}\left(  z_{3,M}\right)  +2\left(  a+b+c+d-1\right)   &  =2\hat
{k}_{1}\cdot k_{2}+2\left(  a+b+c+d+1\right)  .\label{555}%
\end{align}
This shows that Eq.(\ref{2R45}) is the same as Eq.(\ref{3R45}), and the
proposal in Eq.(\ref{LL77}) and Eq.(\ref{RR77}) are consistent ones.

\subsection{The n-point KN amplitude}

In this section, we will adopt two methods to calculate the residues of
general $n$-point KN amplitudes. We will first use the method of subamplitudes
similar to the previous subsections. We then use a direct calculation method.
The result that residues of all $n$-point SSA can be expressed in terms of
Lauricella functions will be discussed in section V.

\subsubsection{Method of subamplitudes}

To calculate the residues $R_{m+1,n-m+1}^{M}$ of $A_{n}$ in Eq.(\ref{AJ}), we
can generalize the previous calculations and propose the following two
subamplitudes $R_{m+1,n-m+1}^{P,L}\left(  \hat{1},2,...,m,-P\right)  $ and
$R_{m+1,n-m+1}^{P,R}\left(  P,m+1,\cdots,n-1,\hat{n}\right)  $%
\begin{align}
&  R_{m+1,n-m+1}^{P,L}\left(  \underset{z_{1}=0}{\underbrace{\hat{1}}%
},\underset{z_{2}}{\underbrace{2}},\cdots,\underset{z_{m-1}}{\underbrace{m-1}%
},\underset{z_{m}=1}{\underbrace{m}},\underset{\infty}{\underbrace{-P}}\right)
\nonumber\\
&  =\int_{0}^{1}dz_{m-1}\cdots\int_{0}^{z_{3}}dz_{2}\underset{1\leq j<i\leq
m}{\prod}\left(  z_{i}-z_{j}\right)  ^{k_{i}\cdot k_{j}}\prod_{r=1}%
\frac{\left[  \underset{\sigma=1}{\overset{N_{r}}{%
%TCIMACRO{\dprod }%
%BeginExpansion
{\displaystyle\prod}
%EndExpansion
}}\epsilon_{r}^{(\sigma)}\cdot\underset{2\leq i\leq m}{\sum}k_{i}\left(
z_{i}\right)  ^{r}\right]  }{\sqrt{N_{r}!r^{N_{r}}}},\label{SLL}\\
&  R_{m+1,n-m+1}^{P,R}\left(  \underset{w_{m}=0}{\underbrace{P}}%
,\underset{_{w_{m+1}}}{\underbrace{m+1}},\cdots,\underset{w_{n-2}%
}{\underbrace{n-2}},\underset{w_{n-1}=1}{\underbrace{n-1}},\underset{w_{n}%
=\infty}{\underbrace{\hat{n}}}\right) \nonumber\\
&  =\int_{0}^{1}dw_{n-2}\cdots\int_{0}^{w_{m+2}}dw_{m+1}\underset{P=m\leq
j<i\leq n-1}{\prod}\left(  w_{i}-w_{j}\right)  ^{k_{i}\cdot k_{j}}\prod
_{r=1}\frac{\left[  \underset{\sigma=1}{\overset{N_{r}}{%
%TCIMACRO{\dprod }%
%BeginExpansion
{\displaystyle\prod}
%EndExpansion
}}\left(  -\epsilon_{r}^{(\sigma)}\right)  \cdot\underset{m+1\leq i\leq
n-1}{\sum}k_{i}\left(  \frac{1}{w_{i}}\right)  ^{r}\right]  }{\sqrt
{N_{r}!r^{N_{r}}}} \label{SRR}%
\end{align}
where $k_{P}=\hat{k}_{1}+\underset{i=2}{\overset{m}{\sum}}k_{i}$. It is
understood that $k_{1}$ in Eq.(\ref{SLL}) should be replaced by $\hat{k}_{1}%
$.
%TCIMACRO{\FRAME{ftbpFU}{4.6337in}{2.3212in}{0pt}{\Qcb{The left and right
%subamplitudes}}{\Qlb{subamp}}{residue.eps}%
%width 4.6337in;  height 2.3212in;  depth 0pt;  original-width 4.5818in;
%original-height 2.2805in;  cropleft "0";  croptop "1";  cropright "1";
%cropbottom "0";  filename 'residue.eps';file-properties "XNPEU";}} }%
%BeginExpansion
\begin{figure}[ptb]%
\centering
\includegraphics[
height=2.3212in,
width=4.6337in
]%
{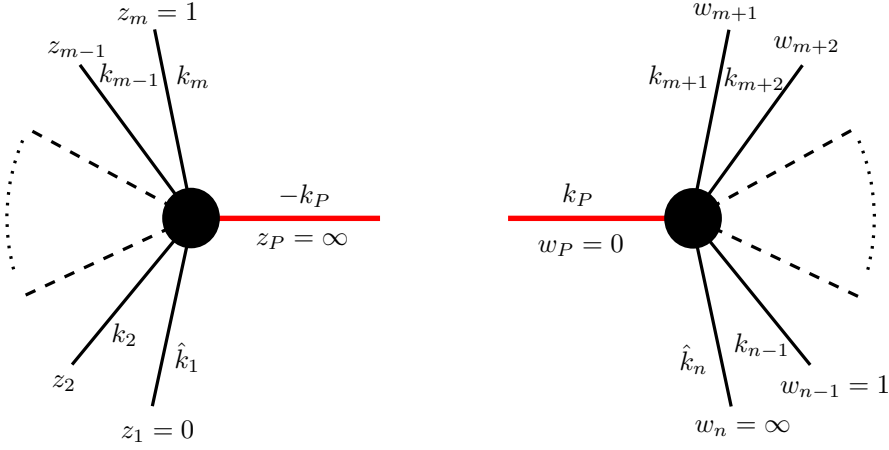}%
\caption{The left and right subamplitudes}%
\label{subamp}%
\end{figure}
%EndExpansion
We then sum over all $P$ to obtain the residue%
\begin{align}
&  R_{m+1,n-m+1}^{M}\left(  \hat{1},2,\cdots,m,m+1,\cdots,n-1,\hat{n}\right)
\nonumber\\
&  =\int_{0}^{1}dz_{m-1}\cdots\int_{0}^{z_{3}}dz_{2}\int_{0}^{1}dw_{n-2}%
\cdots\int_{0}^{w_{m+2}}dw_{m+1}\underset{1\leq j<i\leq m}{\prod}\left(
z_{i}-z_{j}\right)  ^{k_{i}\cdot k_{j}}\nonumber\\
&  \times\underset{P=m\leq j<i\leq n-1}{\prod}\left(  w_{i}-w_{j}\right)
^{k_{i}\cdot k_{j}}\underset{\left\{  N_{r}\right\}  }{\sum}\prod_{r=1}%
\frac{\left[  \underset{2\leq j\leq m,m+1\leq i\leq n-1}{\sum}\left(
-k_{j}\cdot k_{i}\right)  \left(  \frac{z_{j}}{w_{i}}\right)  ^{r}\right]
}{N_{r}!r^{N_{r}}}^{N_{r}}. \label{mm}%
\end{align}
We will show in section V that all residues $R_{m+1,n-m+1}^{M}$ in
Eq.(\ref{mm}) can be expressed in terms of sum over LSSA.

\subsubsection{Direct calculation}

In the previous method of subamplitudes calculation, we did not do a
consistency check similar to Eq.(\ref{111}), Eq.(\ref{222}) and Eq.(\ref{555}%
). In this section, we adopt a direct calculation starting from the general
$n$-point KN amplitude. The $n$-point KN amplitude is%
\begin{equation}
A_{n}^{KN}=\int_{0}^{1}dz_{n-2}\int_{0}^{z_{n-2}}dz_{n-3}\cdots\int_{0}%
^{z_{3}}dz_{2}\text{ }\prod_{i>j=1}^{n-2}\left(  z_{i}-z_{j}\right)
^{k_{i}\cdot k_{j}}\text{, }z_{i}\geq z_{j}%
\end{equation}
which has $\left(  n-3\right)  $ series of poles for each integral on
$z_{2},\cdots,z_{n-2}$.

To investigate the poles for the integral on $z_{m}$, which splits the
$n$-point Koba-Nielsen amplitude to a $\left(  m+1\right)  $-point and a
$\left(  n-m+1\right)  $-point amplitudes connected by a string propagator, we
express $A_{n}^{KN}$ in the following form%
\begin{align}
A_{n}^{KN}  &  =\int_{0}^{1}dz_{n-2}\cdots\int_{0}^{z_{m+2}}dz_{m+1}\int%
_{0}^{z_{m+1}}dz_{m}\int_{0}^{z_{m}}dz_{m-1}\cdots\int_{0}^{z_{3}}%
dz_{2}\nonumber\\
&  \cdot\prod_{m\geq i>j}\left(  z_{i}-z_{j}\right)  ^{k_{i}\cdot k_{j}}%
\prod_{i>j>m}\left(  z_{i}-z_{j}\right)  ^{k_{i}\cdot k_{j}}\prod_{i>m\geq
j}\left(  z_{i}-z_{j}\right)  ^{k_{i}\cdot k_{j}} \label{KN}%
\end{align}
where we have taken $\left(  z_{0},z_{n-1},z_{n}\right)  =\left(
0,1,\infty\right)  $ to fix the $SL\left(  2,R\right)  $ invariance.

By defining the new coordinates,%
\begin{align}
\text{for }i  &  >m:y_{i}=z_{i}\text{ with }\left(  y_{m}\equiv y_{P}%
,y_{n-1},y_{n}\right)  =\left(  0,1,\infty\right)  ,\\
\text{for }i  &  =m:z_{P}=\frac{z_{m}}{z_{m+1}},\\
\text{for }i  &  <m:x_{i}=\frac{z_{i}}{z_{m}}\text{ with }\left(  x_{1}%
,x_{m},x_{m+1}\right)  =\left(  0,1,\infty\right)  ,
\end{align}
Eq.(\ref{KN}) becomes%
\begin{align}
A_{n}^{KN}  &  =\int_{0}^{1}dy_{n-2}\cdots\int_{0}^{y_{m+2}}dy_{m+1}\text{
}\prod_{i=m+1}^{n-2}y_{i}^{k_{i}\cdot k_{P}}\prod_{i>j=m+1}^{n-2}\left(
y_{i}-y_{j}\right)  ^{k_{i}\cdot k_{j}}\nonumber\\
&  \times\int_{0}^{1}dx_{m-1}\cdots\int_{0}^{x_{3}}dx_{2}\text{ }\prod
_{i>j=1}^{m-1}\left(  x_{i}-x_{j}\right)  ^{k_{i}\cdot k_{j}}\nonumber\\
&  \times\int_{0}^{1}dz_{P}\text{ }\left(  z_{P}\right)  ^{\sum\limits_{m\geq
i>j}k_{i}\cdot k_{j}+m-2}\left(  y_{m+1}\right)  ^{\sum\limits_{m\geq
i>j}k_{i}\cdot k_{j}+m-1}\prod_{2\leq j\leq m,m+1\leq i\leq n-1}\left(
1-\frac{y_{m+1}z_{P}x_{j}}{y_{i}}\right)  ^{k_{i}\cdot k_{j}}. \label{KNxy}%
\end{align}
We then expand the last product in Eq.(\ref{KNxy}) by using Binomial theorem%
\begin{equation}
\left(  1-\frac{y_{m+1}z_{P}x_{j}}{y_{i}}\right)  ^{k_{i}\cdot k_{j}}%
=\sum_{a_{ij}}\frac{\left(  -k_{i}\cdot k_{j}\right)  _{a_{ij}}}{a_{ij}%
!}\left(  \frac{y_{m+1}z_{P}x_{j}}{y_{i}}\right)  ^{a_{ij}}%
\end{equation}
where $a_{ij}$'s are integers, to get%
\begin{align}
A_{n}^{KN}  &  =\int_{0}^{1}dy_{n-2}\cdots\int_{0}^{y_{m+2}}dy_{m+1}\text{
}\underset{P=m\leq j<i\leq n-1}{\prod}\left(  y_{i}-y_{j}\right)  ^{k_{i}\cdot
k_{j}}\nonumber\\
&  \times\int_{0}^{1}dx_{m-1}\cdots\int_{0}^{x_{3}}dx_{2}\text{ }%
\underset{1\leq j<i\leq m}{\prod}\left(  x_{i}-x_{j}\right)  ^{k_{i}\cdot
k_{j}}\sum_{a_{ij}}\left(  y_{m+1}\right)  ^{\sum\limits_{m\geq i>j}k_{i}\cdot
k_{j}+\sum\limits_{2\leq j\leq m,m+1\leq i\leq n-1}a_{ij}+m-1}\nonumber\\
&  \times\prod_{2\leq j\leq m,m+1\leq i\leq n-1}\frac{\left(  -k_{i}\cdot
k_{j}\right)  _{a_{ij}}}{a_{ij}!}\left(  \frac{x_{j}}{y_{i}}\right)  ^{a_{ij}%
}\int_{0}^{1}dz_{P}\text{ }\left(  z_{P}\right)  ^{\sum\limits_{m\geq
i>j}k_{i}\cdot k_{j}\sum\limits_{2\leq j\leq m,m+1\leq i\leq n-1}a_{ij}+m-2}.
\end{align}
In the above calculation, we have defined the momentum of the propagator as
$k_{P}=\sum\limits_{j=1}^{m}k_{j}$.

By performing the integral on $z_{P}$, we finally express $A_{n}^{KN}$ in the
following form%
\begin{align}
A_{n}^{KN}  &  =\int_{0}^{1}dy_{n-2}\cdots\int_{0}^{y_{m+2}}dy_{m+1}\text{
}\underset{P=m\leq j<i\leq n-1}{\prod}\left(  y_{i}-y_{j}\right)  ^{k_{i}\cdot
k_{j}}\nonumber\\
&  \times\int_{0}^{1}dx_{m-1}\cdots\int_{0}^{x_{3}}dx_{2}\text{ }%
\underset{1\leq j<i\leq m}{\prod}\left(  x_{i}-x_{j}\right)  ^{k_{i}\cdot
k_{j}}\sum_{a_{ij}}\left(  y_{m+1}\right)  ^{\sum\limits_{m\geq i>j}k_{i}\cdot
k_{j}+\sum\limits_{2\leq j\leq m,m+1\leq i\leq n-1}a_{ij}+m-1}\nonumber\\
&  \times\prod_{2\leq j\leq m,m+1\leq i\leq n-1}\frac{\left(  -k_{i}\cdot
k_{j}\right)  _{a_{ij}}}{a_{ij}!}\left(  \frac{x_{j}}{y_{i}}\right)  ^{a_{ij}%
}\frac{2}{-s_{1\cdots m}+2\left(  \sum\limits_{2\leq j\leq m,m+1\leq i\leq
n-1}a_{ij}-1\right)  }%
\end{align}
where we have defined%
\begin{equation}
s_{1\cdots m}=-\left(  k_{1}+\cdots+k_{m}\right)  ^{2}.
\end{equation}
Next, we make the momentum shift as in BCFW,%
\begin{equation}
k_{1}\left(  z\right)  =k_{1}+zq,k_{n}\left(  z\right)  =k_{n}-zq\text{ with
}q^{2}=k_{1}\cdot q=k_{n}\cdot q=0.
\end{equation}
The KN amplitudes can be written as in Eq.(\ref{AJ})
\begin{align}
A_{n}^{KN}  &  =\sum_{m=2}^{n-2}\underset{M=0}{\overset{\infty}{\sum}}\left[
\frac{A_{n}^{KN}(z)}{z}(z-z_{m,M})\right]  _{z=z_{m,M}}\nonumber\\
&  =\sum_{m=2}^{n-2}\underset{M=0}{\overset{\infty}{\sum}}\frac{2}{-s_{1\cdots
m}+2\left(  M-1\right)  }R_{m+1,n-m+1}^{M} \label{AJ1}%
\end{align}
where we have defined the residue $R_{m+1,n-m+1}^{M}$. The poles\ $z_{m,M}%
$\ can be calculated from the equation%
\begin{equation}
-s_{1\cdots m}\left(  z_{m,M}\right)  +2\left(  \sum_{2\leq j\leq m,m+1\leq
i\leq n-1}a_{ij}-1\right)  =0.
\end{equation}
For a fixed mass $M=\sum\limits_{2\leq j\leq m,m+1\leq i\leq n-1}a_{ij}$, the
poles are%
\begin{equation}
z_{m,M}=\frac{-s_{1\cdots m}+2\left(  M-1\right)  }{-2\sum\limits_{i=1}%
^{m}q\cdot k_{i}}.
\end{equation}
One of the term in Eq.(\ref{AJ1}) for the split $m$ can be calculated as%
\begin{align}
&  \underset{M=0}{\overset{\infty}{\sum}}\frac{2}{-s_{1\cdots m}+2\left(
M-1\right)  }R_{m+1,n-m+1}^{M}\nonumber\\
&  =\underset{M=0}{\overset{\infty}{\sum}}\int_{0}^{1}dy_{n-2}\cdots\int%
_{0}^{y_{m+2}}dy_{m+1}\text{ }\underset{P=m\leq j<i\leq n-1}{\prod}\left(
y_{i}-y_{j}\right)  ^{k_{i}\cdot k_{j}}\int_{0}^{1}dx_{m-1}\cdots\int%
_{0}^{x_{3}}dx_{2}\text{ }\underset{1\leq j<i\leq m}{\prod}\left(  x_{i}%
-x_{j}\right)  ^{k_{i}\cdot k_{j}}\nonumber\\
&  \times\frac{2}{-s_{12\cdots n}+2\left(  M-1\right)  }\prod_{2\leq j\leq
m,m+1\leq i\leq n-1}\sum_{\Sigma a_{ij}=M}\frac{\left(  -k_{i}\cdot
k_{j}\right)  _{a_{ij}}}{a_{ij}!}\left(  \frac{x_{j}}{y_{i}}\right)  ^{a_{ij}%
}.
\end{align}
In the above calculation, we have used the identity
\begin{equation}
\left(  \sum\limits_{m\geq i>j}k_{i}\cdot k_{j}\right)  \left(  z_{m,M}%
\right)  +M+m-1=0. \label{dd}%
\end{equation}
A special case of Eq.(\ref{dd}) corresponds to the consistency check of
Eq.(\ref{111}), Eq.(\ref{222}) and Eq.(\ref{555}) in the calculation of the
residue $R_{4,5}^{M}$ of $7$-point KN amplitude. Using the identity
\begin{equation}
\prod_{i=1}^{n}\sum_{\Sigma b_{i}=M}\frac{\left(  T_{i}\right)  _{b_{i}}%
}{b_{i}!}X_{i}^{b_{i}}=\sum_{\left\{  N_{r}\right\}  }\prod_{r=1}\frac
{1}{N_{r}!r^{N_{r}}}\left(  \sum_{i=1}^{n}T_{i}X_{i}^{r}\right)  ^{N_{r}},
\end{equation}
which was proved in the appendix, and identifying the parameter $a_{ij}$ as
$b_{i}$, the residue becomes%
\begin{align}
&  R_{m+1,n-m+1}^{M}\nonumber\\
&  =\int_{0}^{1}dx_{m-1}\cdots\int_{0}^{x_{3}}dx_{2}\int_{0}^{1}dy_{n-2}%
\cdots\int_{0}^{y_{m+2}}dy_{m+1}\text{ }\underset{1\leq j<i\leq m}{\prod
}\left(  x_{i}-x_{j}\right)  ^{k_{i}\cdot k_{j}}\nonumber\\
&  \times\text{ }\underset{P=m\leq j<i\leq n-1}{\prod}\left(  y_{i}%
-y_{j}\right)  ^{k_{i}\cdot k_{j}}\sum_{\left\{  N_{r}\right\}  }\prod
_{r=1}\frac{\left[  \sum\limits_{2\leq j\leq m,m+1\leq i\leq n-1}\left(
-k_{i}\cdot k_{j}\right)  \left(  \frac{x_{j}}{y_{i}}\right)  ^{r}\right]
^{N_{r}}}{N_{r}!r^{N_{r}}}. \label{Ymmm}%
\end{align}
The residue in Eq.(\ref{Ymmm}) can be written as%
\begin{equation}
R_{m+1,n-m+1}^{M}=\sum_{P}A_{m+1}\left(  1,\cdots,m,-P\right)  A_{n-m+1}%
\left(  P,m+1,\cdots,n\right)  \label{ppp}%
\end{equation}
where%
\begin{align}
A_{m+1}\left(  1,\cdots,m,-P\right)   &  =\int_{0}^{1}dx_{m-1}\cdots\int%
_{0}^{x_{3}}dx_{2}\nonumber\\
&  \times\text{ }\underset{1\leq j<i\leq m}{\prod}\left(  x_{i}-x_{j}\right)
^{k_{i}\cdot k_{j}}\prod_{r=1}\frac{\left[  \underset{\sigma=1}{\overset{N_{r}%
}{%
%TCIMACRO{\dprod }%
%BeginExpansion
{\displaystyle\prod}
%EndExpansion
}}\epsilon_{r}^{(\sigma)}\cdot\sum\limits_{2\leq j\leq m}k_{j}\left(
x_{j}\right)  ^{r}\right]  }{\sqrt{N_{r}!r^{N_{r}}}},\label{YLL}\\
A_{n-m+1}\left(  P,m+1,\cdots,n\right)   &  =\int_{0}^{1}dy_{n-2}\cdots
\int_{0}^{y_{m+2}}dy_{m+1}\nonumber\\
&  \times\text{ }\underset{P=m\leq j<i\leq n-1}{\prod}\left(  y_{i}%
-y_{j}\right)  ^{k_{i}\cdot k_{j}}\prod_{r=1}\frac{\left[  \underset{\sigma
=1}{\overset{N_{r}}{%
%TCIMACRO{\dprod }%
%BeginExpansion
{\displaystyle\prod}
%EndExpansion
}}\left(  -\epsilon_{r}^{(\sigma)}\right)  \cdot\sum\limits_{m+1\leq i\leq
n-1}k_{i}\left(  \frac{1}{y_{i}}\right)  ^{r}\right]  }{\sqrt{N_{r}!r^{N_{r}}%
}} \label{YRR}%
\end{align}
are the $m+1$ and $\left(  n-m+1\right)  $-point subamplitudes of $m$ and
$\left(  n-m\right)  $\ tachyons and a string fock state $P$. Note that the
sum over $P$ in Eq.(\ref{ppp}) reduces to $\sum_{\left\{  N_{r}\right\}  }$ in
Eq.(\ref{Ymmm}). The results in Eq.(\ref{Ymmm}), Eq.(\ref{YLL}) and
Eq.(\ref{YRR}) are consistent with Eq.(\ref{mm}), Eq.(\ref{SLL}) and
Eq.(\ref{SRR}) respectively proposed and calculated in the last subsection.%

%TCIMACRO{\TeXButton{equation number}{\setcounter{equation}{0}
%\renewcommand{\theequation}{\arabic{section}.\arabic{equation}}}}%
%BeginExpansion
\setcounter{equation}{0}
\renewcommand{\theequation}{\arabic{section}.\arabic{equation}}%
%EndExpansion

\section{Iteration relations among residues}

To get some feeling of the residue calculation, we give the explicit form of
$R_{4,3}^{4}$ in Eq.(\ref{44}) as one example of residue calculation%
\begin{align}
R_{4,3}^{4}  &  =\frac{1}{4}\int_{0}^{1}dz_{2}z_{2}^{\hat{k}_{1}\cdot k_{2}%
}\left(  1-z_{2}\right)  ^{k_{2}\cdot k_{3}}\left[  -k_{3}\cdot k_{4}%
-(-k_{2}\cdot k_{4})z_{2}^{4}\right] \nonumber\\
&  +\frac{1}{1!1!}\left(  \frac{1}{3}\right)  \left(  \frac{1}{1}\right)
\int_{0}^{1}dz_{2}z_{2}^{\hat{k}_{1}\cdot k_{2}}\left(  1-z_{2}\right)
^{k_{2}\cdot k_{3}}\left[  -k_{3}\cdot k_{4}-(-k_{2}\cdot k_{4})z_{2}%
^{3}\right]  \left[  -k_{3}\cdot k_{4}-(-k_{2}\cdot k_{4})z_{2}\right]
\nonumber\\
&  +\frac{1}{2!}\left(  \frac{1}{2}\right)  ^{2}\int_{0}^{1}dz_{2}z_{2}%
^{\hat{k}_{1}\cdot k_{2}}\left(  1-z_{2}\right)  ^{k_{2}\cdot k_{3}}\left[
-k_{3}\cdot k_{4}-(-k_{2}\cdot k_{4})z_{2}^{2}\right]  ^{2}\nonumber\\
&  +\frac{1}{2!1!}\left(  \frac{1}{1}\right)  ^{2}\left(  \frac{1}{2}\right)
\int_{0}^{1}dz_{2}z_{2}^{\hat{k}_{1}\cdot k_{2}}\left(  1-z_{2}\right)
^{k_{2}\cdot k_{3}}\left[  -k_{3}\cdot k_{4}-(-k_{2}\cdot k_{4})z_{2}\right]
^{2}\left[  -k_{3}\cdot k_{4}-(-k_{2}\cdot k_{4})z_{2}^{2}\right] \nonumber\\
&  +\frac{1}{4!}\left(  \frac{1}{1}\right)  ^{4}\int_{0}^{1}dz_{2}z_{2}%
^{\hat{k}_{1}\cdot k_{2}}\left(  1-z_{2}\right)  ^{k_{2}\cdot k_{3}}\left[
-k_{3}\cdot k_{4}-(-k_{2}\cdot k_{4})z_{2}\right]  ^{4}. \label{R5}%
\end{align}
There are $5$ terms in $R_{4,3}^{4}$ which are all in the form of Eq.(\ref{p})
and thus can be expressed in terms of the Lauricella functions. It is
interesting to note that Eq.(\ref{R5}) is a generalization of the last line of
Eq.(\ref{sum2}) of Veneziano amplitude to $5$-point KN amplitude. In
particular, the coefficients of each of the $5$ terms in both expressions are
the same. Similarly, $R_{4,3}^{1}$, $R_{4,3}^{2}$ and $R_{4,3}^{3}$ can all be
expressed in terms of the Lauricella functions. Moreover, because of the
solvability or the $SL(K+3,C)$ symmetry of the LSSA, we expect relations among
various residues $R_{4,3}^{J}$. Indeed, if we define%
\begin{align}
F_{M}\left(  z_{2},z_{1}\right)   &  =\underset{b+c=M,b,c=0}{\sum}%
\frac{\left(  -k_{2}\cdot k_{4}\right)  _{b}}{b!}\frac{\left(  -k_{3}\cdot
k_{4}\right)  _{c}}{c!}\left(  z_{2}\right)  ^{b}\left(  z_{1}\right)  ^{c}\\
&  =\underset{\left\{  N_{r}\right\}  }{\sum}\left(  -k_{3}\cdot k_{4}\right)
^{N}\prod_{r=1}\frac{1}{N_{r}!r^{N_{r}}}\left[  z_{1}^{r}-\left(  -\frac
{k_{2}\cdot k_{4}}{k_{3}\cdot k_{4}}\right)  z_{2}^{r}\right]  ^{N_{r}},
\end{align}
then by Eq.(\ref{44})%
\begin{equation}
R_{4,3}^{M}=\int_{0}^{1}dz_{2}z_{2}^{\hat{k}_{1}\cdot k_{2}}\left(
1-z_{2}\right)  ^{k_{2}\cdot k_{3}}F_{M}\left(  z_{2},1\right)  .
\end{equation}
Moreover, one can show the following iteration relation%
\begin{equation}
F_{M}\left(  z_{2},z_{1}\right)  =\frac{1}{M}\underset{N=0}{\overset{M-1}{\sum
}}F_{N}\left(  z_{2},z_{1}\right)  \left[  \left(  -k_{2}\cdot k_{4}\right)
\left(  z_{2}\right)  ^{M-N}+\left(  -k_{3}\cdot k_{4}\right)  \left(
z_{1}\right)  ^{M-N}\right]  ,
\end{equation}
which expresses $F_{M}$ in terms of $F_{M-1}$, $F_{M-2}$, $....$, $F_{1}$,
$F_{0}\equiv1$. Similar iteration relation holds for the residue $R_{3,4}^{M}%
$. So one obtains an iteration relation among all residues of $5$-point KN
amplitude. Presumably, these kind of relations among various residues
$R_{4,3}^{J}$ $($or $R_{3,4}^{M})$ resulting from the $SL(K+3,C)$ symmetry
soften the SSA in the hard scattering limit.

We believe that this iteration relation and its implication on the softness of
some kind of hard $5$-point SSA are generalization of the well-known soft
$4$-point Vaneziano amplitude in the hard scattering limit. To further study
this issue, one first needs to define and identify the specific hard
scattering limit of $5$-point SSA as there are $5$ kinematic variables for the
$5$-point function instead of $2$ for the well-known $4$-point function.
Indeed this iteration relation persists for all higher point SSA. We give one
more example here. The explicit form of $R_{5,3}^{3}$ of the $6$-point KN
amplitude can be calculated from Eq.(\ref{R53})%
\begin{align}
R_{5,3}^{3}  &  =\frac{1}{1!3}\int_{0}^{1}dz_{3}\int_{0}^{z_{3}}dz_{2}%
z_{2}^{\hat{k}_{1}\cdot k_{2}}z_{3}^{\hat{k}_{1}\cdot k_{3}}\left(
z_{3}-z_{2}\right)  ^{k_{2}\cdot k_{3}}\left(  1-z_{2}\right)  ^{k_{2}\cdot
k_{4}}\left(  1-z_{3}\right)  ^{k_{3}\cdot k_{4}}\nonumber\\
&  \times\left[  \left(  -k_{2}\cdot k_{5}\right)  \left(  z_{2}\right)
^{3}+\left(  -k_{3}\cdot k_{5}\right)  \left(  z_{3}\right)  ^{3}+\left(
-k_{4}\cdot k_{5}\right)  \left(  1\right)  ^{3}\right] \nonumber\\
&  +\frac{1}{1!1}\frac{1}{1!2}\int_{0}^{1}dz_{3}\int_{0}^{z_{3}}dz_{2}%
z_{2}^{\hat{k}_{1}\cdot k_{2}}z_{3}^{\hat{k}_{1}\cdot k_{3}}\left(
z_{3}-z_{2}\right)  ^{k_{2}\cdot k_{3}}\left(  1-z_{2}\right)  ^{k_{2}\cdot
k_{4}}\left(  1-z_{3}\right)  ^{k_{3}\cdot k_{4}}\nonumber\\
&  \times\left[  \left(  -k_{2}\cdot k_{5}\right)  \left(  z_{2}\right)
+\left(  -k_{3}\cdot k_{5}\right)  \left(  z_{3}\right)  +\left(  -k_{4}\cdot
k_{5}\right)  \left(  1\right)  \right]  ^{2}\nonumber\\
&  \times\left[  \left(  -k_{2}\cdot k_{5}\right)  \left(  z_{2}\right)
^{2}+\left(  -k_{3}\cdot k_{5}\right)  \left(  z_{3}\right)  ^{2}+\left(
-k_{4}\cdot k_{5}\right)  \left(  1\right)  ^{2}\right] \nonumber\\
&  +\frac{1}{3!1}\int_{0}^{1}dz_{3}\int_{0}^{z_{3}}dz_{2}z_{2}^{\hat{k}%
_{1}\cdot k_{2}}z_{3}^{\hat{k}_{1}\cdot k_{3}}\left(  z_{3}-z_{2}\right)
^{k_{2}\cdot k_{3}}\left(  1-z_{2}\right)  ^{k_{2}\cdot k_{4}}\left(
1-z_{3}\right)  ^{k_{3}\cdot k_{4}}\nonumber\\
&  \times\left[  \left(  -k_{2}\cdot k_{5}\right)  \left(  z_{2}\right)
+\left(  -k_{3}\cdot k_{5}\right)  \left(  z_{3}\right)  +\left(  -k_{4}\cdot
k_{5}\right)  \left(  1\right)  \right]  ^{3}. \label{R3}%
\end{align}
There are $3$ terms in $R_{5,3}^{3}$ which can all be expressed in terms of
the Lauricella functions as will be shown in section V. It is interesting to
note that Eq.(\ref{R3}) is a generalization of the third line of
Eq.(\ref{sum2}) of Veneziano amplitude to $6$-point KN amplitude. In
particular, the coefficients of each of the $3$ terms in both expressions are
the same. Similarly, $R_{5,3}^{1}$ and $R_{5,3}^{2}$ can all be expressed in
terms of the Lauricella functions. If we define%
\begin{equation}
F_{M}\left(  z_{2,}z_{3},z_{1}\right)  =\underset{%
\begin{array}
[c]{c}%
M=c+e+f\\
c,e,f=0
\end{array}
}{\sum}\frac{\left(  -k_{2}\cdot k_{5}\right)  _{c}}{c!}\frac{\left(
-k_{3}\cdot k_{5}\right)  _{e}}{e!}\frac{\left(  -k_{4}\cdot k_{5}\right)
_{f}}{f!}\left(  z_{2}\right)  ^{c}\left(  z_{3}\right)  ^{e}\left(
z_{1}\right)  ^{f},
\end{equation}
then
\begin{equation}
R_{5,3}^{M}=\int_{0}^{1}dz_{3}\int_{0}^{z_{3}}dz_{2}z_{2}^{\hat{k}_{1}\cdot
k_{2}}z_{3}^{\hat{k}_{1}\cdot k_{3}}\left(  z_{3}-z_{2}\right)  ^{k_{2}\cdot
k_{3}}\left(  1-z_{2}\right)  ^{k_{2}\cdot k_{4}}\left(  1-z_{3}\right)
^{k_{3}\cdot k_{4}}F_{M}\left(  z_{2,}z_{3},1\right)  .
\end{equation}
One can show the following iteration relation%
\begin{equation}
F_{M}\left(  z_{2},z_{3},z_{1}\right)  =\frac{1}{M}%
\underset{N=0}{\overset{M-1}{\sum}}F_{N}\left(  z_{2},z_{3},z_{1}\right)
\left[  \left(  -k_{2}\cdot k_{5}\right)  \left(  z_{2}\right)  ^{M-N}+\left(
-k_{3}\cdot k_{5}\right)  \left(  z_{3}\right)  ^{M-N}+\left(  -k_{4}\cdot
k_{5}\right)  \left(  z_{1}\right)  ^{M-N}\right]  ,
\end{equation}
which again expresses $F_{M}$ in terms of $F_{M-1}$, $F_{M-2}$, $....$,
$F_{1}$, $F_{0}\equiv1$.

One can easily generalize this iteration property to any residue
$R_{m+1,n-m+1}^{M}$ of $n$-point KN amplitudes. Indeed, by using Eq.(\ref{mm})
and Eq.(\ref{id1}), we can express the residue $R_{m+1,n-m+1}^{M}$ in terms of
$F_{M}$%

\begin{align}
&  R_{m+1,n-m+1}^{M}\left(  \hat{1},2,\cdots,m,m+1,\cdots,n-1,\hat{n}\right)
\nonumber\\
&  =\int_{0}^{1}dz_{m-1}\cdots\int_{0}^{z_{3}}dz_{2}\int_{0}^{1}dw_{n-2}%
\cdots\int_{0}^{w_{m+2}}dw_{m+1}\underset{1\leq j<i\leq m}{\prod}\left(
z_{i}-z_{j}\right)  ^{k_{i}\cdot k_{j}}\nonumber\\
&  \times\underset{P=m\leq j<i\leq n-1}{\prod}\left(  w_{i}-w_{j}\right)
^{k_{i}\cdot k_{j}}\underset{\left\{  N_{r}\right\}  }{\sum}\prod_{r=1}%
\frac{\left[  \underset{2\leq j\leq m,m+1\leq i\leq n-1}{\sum}\left(
-k_{j}\cdot k_{i}\right)  \left(  \frac{z_{j}}{w_{i}}\right)  ^{r}\right]
}{N_{r}!r^{N_{r}}}^{N_{r}}\nonumber\\
&  =\int_{0}^{1}dz_{m-1}\cdots\int_{0}^{z_{3}}dz_{2}\int_{0}^{1}dw_{n-2}%
\cdots\int_{0}^{w_{m+2}}dw_{m+1}\underset{1\leq j<i\leq m}{\prod}\left(
z_{i}-z_{j}\right)  ^{k_{i}\cdot k_{j}}\nonumber\\
&  \times\underset{P=m\leq j<i\leq n-1}{\prod}\left(  w_{i}-w_{j}\right)
^{k_{i}\cdot k_{j}}F_{M}\text{ }(\frac{z_{j}}{w_{i}})
\end{align}
where $j=2,\cdots,m;i=m+1,\cdots,n-1$ and $z_{m}=w_{n-1}=1$. Similarly, one
can show the following iteration relation \newline%
\begin{equation}
F_{M}=\frac{1}{M}\underset{N=0}{\overset{M-1}{\sum}}F_{N}\left[
\underset{2\leq j\leq m,m+1\leq i\leq n-1}{\sum}\left(  -k_{j}\cdot
k_{i}\right)  \left(  \frac{z_{j}}{w_{i}}\right)  ^{M-N}\right]  ,
\end{equation}
which again expresses $F_{M}$ in terms of $F_{M-1}$, $F_{M-2}$, $....$,
$F_{1}$, $F_{0}\equiv1$. More investigation needs to be done on this
interesting issue.%

%TCIMACRO{\TeXButton{equation number}{\setcounter{equation}{0}
%\renewcommand{\theequation}{\arabic{section}.\arabic{equation}}}}%
%BeginExpansion
\setcounter{equation}{0}
\renewcommand{\theequation}{\arabic{section}.\arabic{equation}}%
%EndExpansion

\section{Expressing n-point SSA in terms of the LSSA}

We note from Eq.(\ref{RS}) that the $5$-point KN amplitude can be expressed in
terms of residues $R_{3,4}^{M}$ and $R_{4,3}^{M}$ after doing $1$-step
recursion. These residues can then be written as $4$-point SSA with one tensor
leg (excited string state) or the LSSA. In this section, we will first use the
shifting method to demonstrate that all $5$-point SSA with arbitrary five
tensor legs can be expressed in terms of the LSSA. In general, we can use
mathematical induction, together with the on-shell recursion and the shifting
principle, to show that all $n$-point SSA including the KN amplitudes can be
expressed in terms of the LSSA. We begins with the discussion of the $4$-point
case and introduce the shifting method.

\subsection{Four-point SSA with tensor legs and the Shifting principle}

We first consider the two tachyons and two vectors amplitude \cite{KLT}%
\begin{align}
&  A\left(  \zeta_{1},k_{1};k_{2};\zeta_{3},k_{3};k_{4}\right) \nonumber\\
&  =\int dz_{2}\left\vert z_{13}z_{14}z_{34}\right\vert \left\langle
e^{ik_{1}X_{1}+i\zeta_{1}\partial_{1}X_{1}}e^{ik_{2}X_{2}}e^{ik_{3}%
X_{3}+i\zeta_{3}\partial_{3}X_{3}}e^{ik_{4}X_{4}}\right\rangle |_{m.l.}%
\nonumber\\
&  =\int dz_{2}\left\vert z_{13}z_{14}z_{34}\right\vert \left\vert
z_{12}\right\vert ^{k_{1}\cdot k_{2}}\left\vert z_{13}\right\vert ^{k_{1}\cdot
k_{3}}\left\vert z_{14}\right\vert ^{k_{1}\cdot k_{4}}\left\vert
z_{23}\right\vert ^{k_{2}\cdot k_{3}}\left\vert z_{24}\right\vert ^{k_{2}\cdot
k_{4}}\left\vert z_{34}\right\vert ^{k_{3}\cdot k_{4}}\nonumber\\
&  \times\exp\left[  \zeta_{1}\cdot\left(  \frac{k_{2}}{z_{21}}+\frac{k_{3}%
}{z_{31}}+\frac{k_{4}}{z_{41}}\right)  +\zeta_{3}\cdot\left(  \frac{k_{1}%
}{z_{13}}+\frac{k_{2}}{z_{23}}+\frac{k_{4}}{z_{43}}\right)  +\frac{\zeta
_{1}\cdot\zeta_{3}}{z_{13}^{2}}\right]  .
\end{align}
After the standard $SL(2,R)$ gauge fixing, we obtain%
\begin{align}
&  A\left(  \zeta_{1},k_{1};k_{2};\zeta_{3},k_{3};k_{4}\right) \nonumber\\
&  =\int_{0}^{1}dz_{2}z_{2}^{k_{1}\cdot k_{2}}\left(  1-z_{2}\right)
^{k_{2}\cdot k_{3}}\left[  \left(  \frac{\zeta_{1}\cdot k_{2}}{z_{2}}%
+\frac{\zeta_{1}\cdot k_{3}}{1}\right)  \left(  \frac{-\zeta_{3}\cdot k_{1}%
}{1}-\frac{\zeta_{3}\cdot k_{2}}{1-z_{2}}\right)  +\frac{\zeta_{1}\cdot
\zeta_{3}}{1}\right]  .
\end{align}
Plug in the kinematic $k_{1}\cdot k_{2}=\frac{-s}{2}-1$, $k_{2}\cdot
k_{3}=\frac{-t}{2}-1$ and after some algebra, we get
\begin{align}
&  A\left(  \zeta_{1},k_{1};k_{2};\zeta_{3},k_{3};k_{4}\right) \nonumber\\
=  &  \int_{0}^{1}dz_{2}z_{2}^{\frac{-s}{2}-1}\left(  1-z_{2}\right)
^{\frac{-t}{2}-1}\left[  -\left(  \frac{\zeta_{1}\cdot k_{2}}{z_{2}}%
+\frac{\zeta_{1}\cdot k_{3}}{1}\right)  \left(  \frac{\zeta_{3}\cdot k_{1}}%
{1}+\frac{\zeta_{3}\cdot k_{2}}{1-z_{2}}\right)  +\frac{\zeta_{1}\cdot
\zeta_{3}}{1}\right] \nonumber\\
&  =\left(  \zeta_{1}\cdot k_{2}\right)  \left(  \zeta_{3}\cdot k_{4}\right)
\int_{0}^{1}dz_{2}z_{2}^{\frac{-s}{2}-2}\left(  1-z_{2}\right)  ^{\frac{-t}%
{2}-2}\left(  1-\left(  -\frac{\zeta_{1}\cdot k_{3}}{\zeta_{1}\cdot k_{2}%
}z_{2}\right)  \right)  \left(  1-\left(  -\frac{\zeta_{3}\cdot k_{1}}%
{\zeta_{3}\cdot k_{4}}z_{2}\right)  \right) \label{poly}\\
&  +\zeta_{1}\cdot\zeta_{3}\int_{0}^{1}dz_{2}z_{2}^{\frac{-s}{2}-1}\left(
1-z_{2}\right)  ^{\frac{-t}{2}-1},
\end{align}
which can be written in terms of the Lauricella functions with $K=0,2$
presented in Eq.(\ref{KK}) of section II. It is important to note that in
order to get the $F_{D}^{\left(  2\right)  }$ term in the above amplitude
calculation, we avoid doing expansion for the product of binomial in the
integrand in Eq.(\ref{poly}). This is a great simplification for amplitude
calculation with higher $K$.

The calculation of the above $4$-point SSA with tensor legs are similar to
that of the $4$-tachyon Veneziano amplitude except \textit{shifting} some
appropriate kinematic variables. This is not surprising as the similar
shiftings already show up in the more familiar $4$-point SSA with three
tachyons and one vector SSA which can be calculated to be ($\left(
z_{1},z_{3},z_{4}\right)  =\left(  0,1,\infty\right)  $)%
\begin{align}
&  A_{4}\left(  k_{1};\zeta_{2},k_{2};k_{3};k_{4}\right) \nonumber\\
&  =\int_{0}^{1}dz_{2}z_{2}^{k_{1}\cdot k_{2}}\left(  1-z_{2}\right)
^{k_{2}\cdot k_{3}}\times(-\zeta_{2})\cdot\left[  \frac{k_{1}}{z_{2}}%
-\frac{k_{3}}{1-z_{2}}\right] \nonumber\\
&  =(-\zeta_{2}\cdot k_{1})A_{4}^{(1)}-(-\zeta_{2}\cdot k_{3})A_{4}^{(2)}%
\end{align}
where ($T=-k_{2}\cdot k_{3}$, $s=-2k_{1}\cdot k_{2}-2$)%
\begin{align}
&  A_{4}^{(1)}=\underset{1}{\underbrace{\frac{2}{-s-2}}}+\underset{\alpha
_{-1}}{\underbrace{T\frac{2}{-s+0}}}+(\underset{\alpha_{-2}}{\underbrace{\frac
{T}{2}}}+\underset{\alpha_{-1}^{2}}{\underbrace{\frac{T^{2}}{2!}\frac{1}%
{1^{2}})}}\frac{2}{-s+2}\nonumber\\
&  +\underset{\alpha_{-3}}{\underbrace{(\frac{T}{3}}}+\underset{\alpha
_{-1}\alpha_{-2}}{\underbrace{\frac{T^{2}}{1!1!}\frac{1}{1^{1}}\frac{1}{2^{1}%
}}}+\underset{\alpha_{-1}^{3}}{\underbrace{\frac{T^{3}}{3!})}}\frac{2}%
{-s+4}+.....
\end{align}
and%
\begin{align}
&  A_{4}^{(2)}=\underset{\alpha_{-1}}{\underbrace{\frac{2}{-s+2}}%
}+\underset{\alpha_{-2},\alpha_{-1}^{2}}{\underbrace{(T+1)\frac{2}{-s+4}}%
}+\underset{\alpha_{-3},\alpha_{-1}\alpha_{-2},\alpha_{-1}^{3}%
}{\underbrace{\left[  \frac{(T+1)}{2}+\frac{(T+1)^{2}}{2!}\frac{1}{1^{2}%
}\right]  \frac{2}{-s+6}}}\nonumber\\
&  +\underset{\alpha_{-4},\alpha_{-1}\alpha_{-3},\alpha_{-2}^{2},\alpha
_{-1}^{2}\alpha_{-2},\alpha_{-1}^{4}}{\underbrace{\left[  \frac{(T+1)}%
{3}+\frac{(T+1)^{2}}{1!1!}\frac{1}{1^{1}}\frac{1}{2^{1}}+\frac{(T+1)^{3}}%
{3!}\right]  \frac{2}{-s+8}}}+.....\text{.}%
\end{align}
These results correspond to shift the kinematic variables $T$ and $s$ of
results of the original $4$-point Veneziano amplitude presented in
Eq.(\ref{sum}) to $T+1$ and $s+4$, and they agree with results obtained by
operator method calculation in \cite{stringbcfw}.

The SSA of three vectors and one tachyon can be similarly calculated to be%
\begin{align}
&  A\left(  \zeta_{1},k_{1};\zeta_{2},k_{2};\zeta_{3},k_{3};k_{4}\right)
\nonumber\\
&  =\left(  \zeta_{1}\cdot k_{2}\right)  \left(  \zeta_{2}\cdot k_{2}\right)
\left(  \zeta_{3}\cdot k_{4}\right)  \frac{\Gamma\left(  \frac{-s}%
{2}-2\right)  \Gamma\left(  \frac{-t}{2}-2\right)  }{\Gamma\left(  -\frac
{s}{2}-\frac{t}{2}-4\right)  }\nonumber\\
&  \times F_{D}^{\left(  3\right)  }\left(  \frac{-s}{2}-2;-1,-1,-1;-\frac
{s}{2}-\frac{t}{2}-4;\frac{\zeta_{1}\cdot k_{3}}{\zeta_{1}\cdot k_{2}}%
,-\frac{\zeta_{2}\cdot k_{4}}{\zeta_{2}\cdot k_{1}},-\frac{\zeta_{3}\cdot
k_{1}}{\zeta_{3}\cdot k_{4}}\right) \nonumber\\
&  +\left(  \zeta_{1}\cdot\zeta_{2}\right)  \left(  -\zeta_{3}\cdot
k_{4}\right)  \frac{\Gamma\left(  \frac{-s}{2}-2\right)  \Gamma\left(
\frac{-t}{2}-1\right)  }{\Gamma\left(  -\frac{s}{2}-\frac{t}{2}-3\right)
}F_{D}^{\left(  1\right)  }\left(  \frac{-s}{2}-2;-1;-\frac{s}{2}-\frac{t}%
{2}-3;-\frac{\zeta_{3}\cdot k_{1}}{\zeta_{3}\cdot k_{4}}\right) \nonumber\\
&  +\left(  \zeta_{2}\cdot\zeta_{3}\right)  \left(  -\zeta_{1}\cdot
k_{2}\right)  \frac{\Gamma\left(  \frac{-s}{2}-1\right)  \Gamma\left(
\frac{-t}{2}-2\right)  }{\Gamma\left(  -\frac{s}{2}-\frac{t}{2}-3\right)
}F_{D}^{\left(  1\right)  }\left(  \frac{-s}{2}-1;-1;-\frac{s}{2}-\frac{t}%
{2}-3;\frac{\zeta_{1}\cdot k_{3}}{\zeta_{1}\cdot k_{2}}\right) \nonumber\\
&  +\left(  \zeta_{1}\cdot\zeta_{3}\right)  \left(  -\zeta_{2}\cdot
k_{1}\right)  \frac{\Gamma\left(  \frac{-s}{2}-1\right)  \Gamma\left(
\frac{-t}{2}-1\right)  }{\Gamma\left(  -\frac{s}{2}-\frac{t}{2}-2\right)
}F_{D}^{\left(  1\right)  }\left(  \frac{-s}{2}-1;-1;-\frac{s}{2}-\frac{t}%
{2}-2;-\frac{\zeta_{2}\cdot k_{4}}{\zeta_{2}\cdot k_{1}}\right)  \label{4444}%
\end{align}
which can be written in terms of the Lauricella functions with $K=1,3$. The
expression with $4$ terms in Eq.(\ref{4444}) is to be compared with expression
with $12$ terms calculated in \cite{KLT}.

As another example, the SSA of one rank-$2$ tensor, one vector and one tachyon
can be calculated to be ($\zeta_{\mu\nu}=\zeta_{\mu}^{(1)}\zeta_{\nu}^{(2)}$)%
\begin{align}
&  A\left(  \zeta_{\mu\nu},k_{1};k_{2};\zeta,k_{3};k_{4}\right)  \nonumber\\
&  =-\zeta_{\mu\nu}k_{2}^{\mu}k_{2}^{\nu}\left(  \zeta\cdot k_{4}\right)
\frac{\Gamma\left(  \frac{-s}{2}-2\right)  \Gamma\left(  \frac{-t}%
{2}-2\right)  }{\Gamma\left(  -\frac{s}{2}-\frac{t}{2}-4\right)  }\nonumber\\
&  \times F_{D}^{\left(  3\right)  }\left(  \frac{-s}{2}-2;-1,-1,-1;-\frac
{s}{2}-\frac{t}{2}-4;-\frac{\zeta^{(1)}\cdot k_{3}}{\zeta^{(1)}\cdot k_{2}%
},-\frac{\zeta^{(2)}\cdot k_{3}}{\zeta^{(2)}\cdot k_{2}},-\frac{\zeta\cdot
k_{1}}{\zeta\cdot k_{4}}\right)  \nonumber\\
&  +\zeta_{\mu\nu}\zeta^{\mu}k_{2}^{\nu}\left(  \zeta\cdot k_{4}\right)
\frac{\Gamma\left(  \frac{-s}{2}-1\right)  \Gamma\left(  \frac{-t}%
{2}-1\right)  }{\Gamma\left(  -\frac{s}{2}-\frac{t}{2}-2\right)  }%
F_{D}^{\left(  1\right)  }\left(  \frac{-s}{2}-1;-1;-\frac{s}{2}-\frac{t}%
{2}-2;-\frac{\zeta^{(1)}\cdot k_{3}}{\zeta^{(1)}\cdot k_{2}}\right)
\nonumber\\
&  +\zeta_{\mu\nu}\zeta^{\mu}k_{2}^{\nu}\left(  \zeta\cdot k_{4}\right)
\frac{\Gamma\left(  \frac{-s}{2}-1\right)  \Gamma\left(  \frac{-t}%
{2}-1\right)  }{\Gamma\left(  -\frac{s}{2}-\frac{t}{2}-2\right)  }%
F_{D}^{\left(  1\right)  }\left(  \frac{-s}{2}-1;-1;-\frac{s}{2}-\frac{t}%
{2}-2;-\frac{\zeta^{(2)}\cdot k_{3}}{\zeta^{(2)}\cdot k_{2}}\right)  ,
\end{align}
which again can be written in terms of the Lauricella functions with $K=1,3$.
In conclusion, we see that the calculation of SSA expressed in terms of the
Lauricella functions is much simpler than the traditional calculation.

Finally for the 26D open bosonic string, a general state at mass level $N$%
\begin{equation}
M^{2}=2\left(  N-1\right)  ,\text{ and }N=\sum_{r>0}rn_{r},
\end{equation}
is of the form%
\begin{equation}
\left\vert P\right\rangle =\prod_{r>0}\prod_{\sigma=1}^{n_{r}}\frac
{\varepsilon_{r}^{\left(  \sigma\right)  }\cdot\alpha_{-r}}{\sqrt
{n_{r}!r^{n_{r}}}}|0,k\rangle
\end{equation}
where $\varepsilon_{r}^{\left(  \sigma\right)  }$ are polarizations with
$\sigma=1,\cdots n_{r}$ for each operator $\alpha_{-r}$. The corresponding
string vertex is%
\begin{equation}
V\left(  k,z\right)  =\prod_{r>0}\prod_{\sigma=1}^{n_{r}}\varepsilon
_{r}^{\left(  \sigma\right)  }\cdot\partial_{z}^{r}X\left(  z\right)
e^{ik\cdot X\left(  z\right)  }.
\end{equation}
For 4-point amplitude $i=1,2,3,4$, let%
\begin{equation}
X_{i}=X\left(  z_{i}\right)  \text{ and }k_{i}=\left(  E_{i},\vec{p}%
_{i}\right)  \text{ with }\sum k_{i}=\sum\vec{p}_{i}=0,
\end{equation}
and we define the Mandelstam variables as $s=-\left(  k_{1}+k_{2}\right)
^{2}$, $t=-\left(  k_{2}+k_{3}\right)  ^{2}$. The $4$-point SSA with four
general string states can be calculated as%
\begin{align}
A &  =\int dz_{2}\left\vert z_{13}z_{14}z_{34}\right\vert \left\langle
\prod_{i=1}^{4}V_{i}\left(  k_{i},z_{i}\right)  \right\rangle \nonumber\\
&  =\int dz_{2}\left\vert z_{13}z_{14}z_{34}\right\vert \left\langle
\prod_{i=1}^{4}\exp\left(  ik_{i}\cdot X_{i}+i\sum_{r_{i}>0}\sum_{\sigma
_{i}=1}^{n_{r_{i}}}\varepsilon_{r_{i}}^{\left(  \sigma_{i}\right)  }%
\cdot\partial_{i}^{r_{i}}X_{i}\right)  \right\rangle _{m.l.}\nonumber\\
&  =\int dz_{2}\left\vert z_{13}z_{14}z_{34}\right\vert \left\vert
z_{12}\right\vert ^{k_{1}\cdot k_{2}}\left\vert z_{13}\right\vert ^{k_{1}\cdot
k_{3}}\left\vert z_{14}\right\vert ^{k_{1}\cdot k_{4}}\left\vert
z_{23}\right\vert ^{k_{2}\cdot k_{3}}\left\vert z_{24}\right\vert ^{k_{2}\cdot
k_{4}}\left\vert z_{34}\right\vert ^{k_{3}\cdot k_{4}}\nonumber\\
&  \cdot\exp\left(  \sum_{r_{i}>0}\sum_{\sigma_{i}=1}^{n_{r_{i}}}\sum_{i}%
^{4}\sum_{j\neq i}\frac{-\varepsilon_{r_{i}}^{\left(  \sigma_{i}\right)
}\cdot k_{j}}{z_{ji}^{r_{i}}}+\sum_{r_{i},r_{j}>0}\sum_{\sigma_{i}%
=1}^{n_{r_{i}}}\sum_{\sigma_{j}=1}^{n_{r_{j}}}\sum_{i<j=2}^{4}\frac
{-\varepsilon_{r_{i}}^{\left(  \sigma_{i}\right)  }\varepsilon_{r_{j}%
}^{\left(  \sigma_{j}\right)  }}{z_{ji}^{r_{i}}z_{ij}^{r_{j}}}\right)  _{m.l.}%
\end{align}
where the lower label $m.l.$ means that we only keep multi-linear terms with
each polarization $\varepsilon_{r_{i}}^{\left(  \sigma_{i}\right)  }$. The
amplitude can be expressed as%
\begin{align}
A &  =\int_{0}^{1}dz_{2}z_{2}^{k_{1}\cdot k_{2}}\left(  1-z_{2}\right)
^{k_{2}\cdot k_{3}}\nonumber\\
&  \lim_{z_{4}\rightarrow\infty}\cdot\sum_{\left\{  \varepsilon_{r_{i}%
}^{\left(  \sigma_{i}\right)  }\right\}  }\left[  \prod\limits_{i=1}^{4}%
\prod\limits_{\left\{  r_{i},\sigma_{i}\right\}  }\left(  \sum_{j\neq i}%
\frac{\varepsilon_{r_{i}}^{\left(  \sigma_{i}\right)  }\cdot k_{j}}%
{z_{ji}^{r_{i}}}\right)  \cdot\prod\limits_{i<j=2}^{4}\prod\limits_{\left\{
r_{i},\sigma_{i};r_{j},\sigma_{j}\right\}  }\frac{\varepsilon_{r_{i}}^{\left(
\sigma_{i}\right)  }\varepsilon_{r_{j}}^{\left(  \sigma_{j}\right)  }}%
{z_{ji}^{r_{i}}z_{ij}^{r_{j}}}\right]  _{z_{1}=0,z_{3}=1}\label{line}%
\end{align}
where\ the configurations $\left\{  \varepsilon_{r_{i}}^{\left(  \sigma
_{i}\right)  }\right\}  $\ satisfy%
\begin{equation}
\prod\limits_{i=1}^{4}\prod\limits_{\left\{  r_{i},\sigma_{i}\right\}
}\varepsilon_{r_{i}}^{\left(  \sigma_{i}\right)  }\cdot\prod\limits_{i<j=2}%
^{4}\prod\limits_{\left\{  r_{i},\sigma_{i};r_{j},\sigma_{j}\right\}  }\left(
\varepsilon_{r_{i}}^{\left(  \sigma_{i}\right)  }\varepsilon_{r_{j}}^{\left(
\sigma_{j}\right)  }\right)  =\prod_{i=1}^{4}\prod_{r_{i}>0}\prod_{\sigma
_{i}=1}^{n_{r_{i}}}\varepsilon_{r_{i}}^{\left(  \sigma_{i}\right)  },
\end{equation}
which ensures the multi-linear condition. For each configuration $\left\{
\varepsilon_{r_{i}}^{\left(  \sigma_{i}\right)  }\right\}  $, it is
straightforward to transform Eq.(\ref{line}) to the standard integral form of
the Lauricella function. In the first line of Eq.(\ref{line}), we have fixed
$SL\left(  2,R\right)  $ by choosing $\left(  z_{1},z_{3},z_{4}\right)
=\left(  0,1,\infty\right)  $ and drop a factor which is independent of
momentum $k_{i}$. It is much more tricky to take the $z_{4}\rightarrow\infty$
in the second line of Eq.(\ref{line}). For $V_{4}\left(  k_{i},z_{i}\right)  $
to be a higher spin state, one needs to apply stringy massive Ward identities
or decoupling of massive ZNS to obtain a consistent result \cite{LLY3}.

We conclude this subsection by noting that the calculation of $4$-point SSA
with tensor legs are similar to that of the $4$-tachyon Veneziano amplitude
except \textit{shifting} some appropriate kinematic variables. It is easy to
generalize this shifting method to arbitrary $n$-point SSA and obtain the following

\bigskip

\textit{Shifting principle} : If the $n$-point KN amplitude can be expressed
in terms of the LSSA, then one can use the shifting method to calculate all
$n$-point SSA with tensor legs (excited string states) and express them in
terms of the LSSA.

\bigskip

We will present more examples to demonstrate the shifting method in the
following subsections. We will see that while the string on-shell recursion
relation can be used to reduce higher point KN amplitudes to the lower SSA and
express them in terms of the LSSA, the shifting method can be used to express
SSA with tensor legs in terms of the LSSA.

\subsection{Five-point SSA with tensor legs}

In section III, we have shown that the two residues $R_{3,4}^{M}$ and
$R_{4,3}^{M}$ of $5$-point KN amplitude in Eq.(\ref{RS}) can be expressed in
terms of the Lauricella functions. In this subsection, we will use the
shifting method to calculate more examples of $5$-point SSA with tensor legs
and express them in terms of the LSSA. In general, according to the shifting
principle, all $5$-point SSA can be expressed in terms of LSSA. Unfortunately,
to explicitly calculate the higher point ($n\geq5$) SSA with multi-tensor legs
is quite lengthy. The case of $5$-point SSA with $1$-tensor legs, for example,
will show up after doing $1$-step recursion on the $6$-point KN amplitude. To
begin with, let's consider the four tachyons, one vector SSA ($\left(
z_{1},z_{4},z_{5}\right)  =\left(  0,1,\infty\right)  $)%
\begin{align}
&  A_{5}\left(  \zeta_{1},k_{1};k_{2};k_{3};k_{4};k_{5}\right) \nonumber\\
&  =\int_{0}^{1}dz_{3}\int_{0}^{z_{3}}dz_{2}z_{2}^{k_{2}\cdot k_{1}}%
z_{3}^{k_{3}\cdot k_{1}}\left(  1-z_{3}\right)  ^{k_{4}\cdot k_{3}}\left(
1-z_{2}\right)  ^{k_{4}\cdot k_{2}}\left(  z_{3}-z_{2}\right)  ^{k_{3}\cdot
k_{2}}\nonumber\\
&  \times(-\zeta_{1})\cdot\left[  \frac{k_{2}}{z_{2}}+\frac{k_{3}}{z_{3}%
}+\frac{k_{4}}{1}\right] \nonumber\\
&  =A_{5}^{(1)}+A_{5}^{(2)}+A_{5}^{(3)} \label{three}%
\end{align}
where we have used the result of Eq.(\ref{RR}) by taking $N=1=M$.
Mathematically, all three terms of Eq.(\ref{three}) are similar to the
$5$-point KN amplitude and we can apply the shifting principle to do the
calculation. We will explicitly calculate the three terms separately. We first
note that the original $5$-point KN amplitude in Eq.(\ref{24}) can be
rewritten as%
\begin{align}
A_{5}^{KN}  &  =\underset{a,b,c=0}{\overset{\infty}{%
%TCIMACRO{\dsum }%
%BeginExpansion
{\displaystyle\sum}
%EndExpansion
}}\frac{\left(  -k_{2}\cdot k_{3}\right)  _{a}}{a!}\frac{\left(  -k_{2}\cdot
k_{4}\right)  _{b}}{b!}\frac{\left(  -k_{3}\cdot k_{4}\right)  _{c}}%
{c!}\nonumber\\
&  \times\frac{2}{2k_{1}\cdot k_{2}+2\left(  a+b+1\right)  }\frac{2}%
{2(k_{1}\cdot k_{2}+k_{2}\cdot k_{3}+k_{1}\cdot k_{3})+2\left(  b+c+2\right)
}. \label{KN2}%
\end{align}
Similarly, the three terms in Eq.(\ref{three}) can be calculated to be%
\begin{align}
A_{5}^{(1)}  &  =(-\zeta_{1}\cdot k_{2})\int_{0}^{1}dz_{3}\int_{0}^{z_{3}%
}dz_{2}z_{2}^{k_{2}\cdot k_{1}-1}z_{3}^{k_{3}\cdot k_{1}}\left(
1-z_{3}\right)  ^{k_{4}\cdot k_{3}}\left(  1-z_{2}\right)  ^{k_{4}\cdot k_{2}%
}\left(  z_{3}-z_{2}\right)  ^{k_{3}\cdot k_{2}}\nonumber\\
&  =(-\zeta_{1}\cdot k_{2})\underset{a,b,c=0}{\overset{\infty}{%
%TCIMACRO{\dsum }%
%BeginExpansion
{\displaystyle\sum}
%EndExpansion
}}\frac{\left(  -k_{2}\cdot k_{3}\right)  _{a}}{a!}\frac{\left(  -k_{2}\cdot
k_{4}\right)  _{b}}{b!}\frac{\left(  -k_{3}\cdot k_{4}\right)  _{c}}%
{c!}\nonumber\\
&  \times\frac{2}{2k_{1}\cdot k_{2}+2\left(  a+b\right)  }\frac{2}%
{2(k_{1}\cdot k_{2}+k_{2}\cdot k_{3}+k_{1}\cdot k_{3})+2\left(  b+c+1\right)
},
\end{align}%
\begin{align}
A_{5}^{(2)}  &  =(-\zeta_{1}\cdot k_{3})\int_{0}^{1}dz_{3}\int_{0}^{z_{3}%
}dz_{2}z_{2}^{k_{2}\cdot k_{1}}z_{3}^{k_{3}\cdot k_{1}-1}\left(
1-z_{3}\right)  ^{k_{4}\cdot k_{3}}\left(  1-z_{2}\right)  ^{k_{4}\cdot k_{2}%
}\left(  z_{3}-z_{2}\right)  ^{k_{3}\cdot k_{2}}\nonumber\\
&  =(-\zeta_{1}\cdot k_{3})\underset{a,b,c=0}{\overset{\infty}{%
%TCIMACRO{\dsum }%
%BeginExpansion
{\displaystyle\sum}
%EndExpansion
}}\frac{\left(  -k_{2}\cdot k_{3}\right)  _{a}}{a!}\frac{\left(  -k_{2}\cdot
k_{4}\right)  _{b}}{b!}\frac{\left(  -k_{3}\cdot k_{4}\right)  _{c}}%
{c!}\nonumber\\
&  \times\frac{2}{2k_{1}\cdot k_{2}+2\left(  a+b+1\right)  }\frac{2}%
{2(k_{1}\cdot k_{2}+k_{2}\cdot k_{3}+k_{1}\cdot k_{3})+2\left(  b+c-1\right)
}%
\end{align}
and%
\begin{align}
A_{5}^{(3)}  &  =(-\zeta_{1}\cdot k_{4})\int_{0}^{1}dz_{3}\int_{0}^{z_{3}%
}dz_{2}z_{2}^{k_{2}\cdot k_{1}}z_{3}^{k_{3}\cdot k_{1}}\left(  1-z_{3}\right)
^{k_{4}\cdot k_{3}}\left(  1-z_{2}\right)  ^{k_{4}\cdot k_{2}}\left(
z_{3}-z_{2}\right)  ^{k_{3}\cdot k_{2}}\nonumber\\
&  (-\zeta_{1}\cdot k_{4})\underset{a,b,c=0}{\overset{\infty}{%
%TCIMACRO{\dsum }%
%BeginExpansion
{\displaystyle\sum}
%EndExpansion
}}\frac{\left(  -k_{2}\cdot k_{3}\right)  _{a}}{a!}\frac{\left(  -k_{2}\cdot
k_{4}\right)  _{b}}{b!}\frac{\left(  -k_{3}\cdot k_{4}\right)  _{c}}{c!}\\
&  \times\frac{2}{2k_{1}\cdot k_{2}+2\left(  a+b+1\right)  }\frac{2}%
{2(k_{1}\cdot k_{2}+k_{2}\cdot k_{3}+k_{1}\cdot k_{3})+2\left(  b+c+2\right)
}.
\end{align}

We can now apply string on-shell recursion relation to obtain%
\begin{align}
A_{5}^{(1)}  &  =(-\zeta_{1}\cdot k_{2})\underset{_{M}}{\sum}\frac{2}%
{2k_{1}\cdot k_{2}+2M}R_{3,4}^{(1),M}\nonumber\\
&  +(-\zeta_{1}\cdot k_{2})\underset{_{M}}{\sum}\frac{2}{2(k_{1}\cdot
k_{2}+k_{2}\cdot k_{3}+k_{1}\cdot k_{3})+2\left(  M+1\right)  }R_{4,3}^{(1),M}
\label{mul}%
\end{align}
where the two residues%
\begin{equation}
R_{3,4}^{(1),M}=\underset{\left\{  N_{r}\right\}  }{\sum}\left(  -k_{2}\cdot
k_{3}\right)  ^{N}\int_{0}^{1}dz_{2}z_{2}^{\left(  \hat{k}_{1}+k_{2}\right)
\cdot k_{3}-M}\left(  1-z_{2}\right)  ^{k_{3}\cdot k_{4}}\prod_{r=1}\frac
{1}{N_{r}!r^{N_{r}}}\left[  1-\left(  -\frac{k_{2}\cdot k_{4}}{k_{2}\cdot
k_{3}}\right)  z_{2}^{r}\right]  ^{N_{r}}%
\end{equation}
and%
\begin{equation}
R_{4,3}^{(1),M}=\underset{\left\{  N_{r}\right\}  }{\sum}\left(  -k_{3}\cdot
k_{4}\right)  ^{N}\int_{0}^{1}dz_{2}z_{2}^{\hat{k}_{1}\cdot k_{2}-1}\left(
1-z_{2}\right)  ^{k_{2}\cdot k_{3}}\prod_{r=1}\frac{1}{N_{r}!r^{N_{r}}}\left[
1-\left(  -\frac{k_{2}\cdot k_{4}}{k_{3}\cdot k_{4}}\right)  z_{2}^{r}\right]
^{N_{r}}%
\end{equation}
can all be expressed in terms of the Lauricella functions. Similarly, we can
calculate%
\begin{align}
A_{5}^{(2)}  &  =(-\zeta_{1}\cdot k_{3})\underset{_{M}}{\sum}\frac{2}%
{2k_{1}\cdot k_{2}+2\left(  M+1\right)  }R_{3,4}^{(2),M}\nonumber\\
&  +(-\zeta_{1}\cdot k_{2})\underset{_{M}}{\sum}\frac{2}{2(k_{1}\cdot
k_{2}+k_{2}\cdot k_{3}+k_{1}\cdot k_{3})+2\left(  M+1\right)  }R_{4,3}^{(2),M}%
\end{align}
where%
\begin{align}
R_{3,4}^{(2),M}  &  =\underset{\left\{  N_{r}\right\}  }{\sum}\left(
-k_{2}\cdot k_{3}\right)  ^{N}\int_{0}^{1}dz_{2}z_{2}^{\left(  \hat{k}%
_{1}+k_{2}\right)  \cdot k_{3}-M-1}\left(  1-z_{2}\right)  ^{k_{3}\cdot k_{4}%
}\prod_{r=1}\frac{1}{N_{r}!r^{N_{r}}}\left[  1-\left(  -\frac{k_{2}\cdot
k_{4}}{k_{2}\cdot k_{3}}\right)  z_{2}^{r}\right]  ^{N_{r}},\\
R_{4,3}^{(2),M}  &  =\underset{\left\{  N_{r}\right\}  }{\sum}\left(
-k_{3}\cdot k_{4}\right)  ^{N}\int_{0}^{1}dz_{2}z_{2}^{\hat{k}_{1}\cdot k_{2}%
}\left(  1-z_{2}\right)  ^{k_{2}\cdot k_{3}}\prod_{r=1}\frac{1}{N_{r}%
!r^{N_{r}}}\left[  1-\left(  -\frac{k_{2}\cdot k_{4}}{k_{3}\cdot k_{4}%
}\right)  z_{2}^{r}\right]  ^{N_{r}};
\end{align}
and%
\begin{align}
A_{5}^{(3)}  &  =(-\zeta_{1}\cdot k_{4})\underset{_{M}}{\sum}\frac{2}%
{2k_{1}\cdot k_{2}+2\left(  M+1\right)  }R_{3,4}^{(3),M}\nonumber\\
&  +(-\zeta_{1}\cdot k_{4})\underset{_{M}}{\sum}\frac{2}{2(k_{1}\cdot
k_{2}+k_{2}\cdot k_{3}+k_{1}\cdot k_{3})+2\left(  M+2\right)  }R_{4,3}^{(3),M}%
\end{align}
where%
\begin{align}
R_{3,4}^{(3),M}  &  =\underset{\left\{  N_{r}\right\}  }{\sum}\left(
-k_{2}\cdot k_{3}\right)  ^{N}\int_{0}^{1}dz_{2}z_{2}^{\left(  \hat{k}%
_{1}+k_{2}\right)  \cdot k_{3}-M}\left(  1-z_{2}\right)  ^{k_{3}\cdot k_{4}%
}\prod_{r=1}\frac{1}{N_{r}!r^{N_{r}}}\left[  1-\left(  -\frac{k_{2}\cdot
k_{4}}{k_{2}\cdot k_{3}}\right)  z_{2}^{r}\right]  ^{N_{r}},\\
R_{4,3}^{(3),M}  &  =\underset{\left\{  N_{r}\right\}  }{\sum}\left(
-k_{3}\cdot k_{4}\right)  ^{N}\int_{0}^{1}dz_{2}z_{2}^{\hat{k}_{1}\cdot
k_{2}-1}\left(  1-z_{2}\right)  ^{k_{2}\cdot k_{3}}\prod_{r=1}\frac{1}%
{N_{r}!r^{N_{r}}}\left[  1-\left(  -\frac{k_{2}\cdot k_{4}}{k_{3}\cdot k_{4}%
}\right)  z_{2}^{r}\right]  ^{N_{r}}. \label{last}%
\end{align}

We note that the calculations above are straightforward and the results
correspond to shift the appropriate kinematic variables of results of the
original $5$-point KN amplitude. According to the shifting principle, one can
generalize the above calculation to the four tachyons and one higher tensor
SSA and express it in terms of the LSSA.

It is lengthy to explicitly calculate $5$-point SSA with three tachyons and
two general tensor legs SSA. However, it is straightforward though still
lengthy to explicitly calculate the three tachyons, two vectors SSA%
\begin{align}
&  A_{5}\left(  \zeta_{1},k_{1};k_{2};k_{3};\zeta_{4},k_{4};k_{5}\right)
\nonumber\\
&  =\int_{0}^{1}dz_{3}\int_{0}^{z_{3}}dz_{2}z_{2}^{k_{2}\cdot k_{1}}%
z_{3}^{k_{3}\cdot k_{1}}\left(  1-z_{3}\right)  ^{k_{4}\cdot k_{3}}\left(
1-z_{2}\right)  ^{k_{4}\cdot k_{2}}\left(  z_{3}-z_{2}\right)  ^{k_{3}\cdot
k_{2}}\nonumber\\
&  \times\left[  \zeta_{1}\cdot\left(  \frac{k_{2}}{z_{2}}+\frac{k_{3}}{z_{3}%
}+\frac{k_{4}}{1}\right)  (-\zeta_{4})\cdot\left(  \frac{k_{1}}{1}+\frac
{k_{2}}{1-z_{2}}+\frac{k_{3}}{1-z_{3}}\right)  +\frac{\zeta_{1}\cdot\zeta_{4}%
}{1}\right]  \nonumber\\
&  =A_{5}^{(i,j)}+A_{5}^{(0,0)},\text{ \ }i,j=1,2,3,\label{ij}%
\end{align}
and express it in terms of the sum of LSSA. To do this, one needs to calculate
$3\times3+1=10$ $5$-point SSA. Mathematically, all $10$ amplitude calculations
are similar to the calculation of $5$-point KN amplitude in Eq.(\ref{KN2}).
The first $3$ amplitudes can be identified to be%
\begin{align}
A_{5}^{(1,1)} &  =(\zeta_{4}\cdot k_{1})A_{5}^{(1)},\\
A_{5}^{(2,1)} &  =(\zeta_{4}\cdot k_{1})A_{5}^{(2)},\\
A_{5}^{(3,1)} &  =(\zeta_{4}\cdot k_{1})A_{5}^{(3)}%
\end{align}
where $k_{4}$ of a tachyon with $k_{4}^{2}=2$ in $A_{5}^{(j)}$ is replaced by
$k_{4}$ of a vector with $k_{4}^{2}=0$. The other $7$ amplitudes correspond to
shift some appropriate kinematic variables of Eq.(\ref{KN2}) and can be
calculated to be%
\begin{align}
A_{5}^{(1,2)} &  =(-\zeta_{1}\cdot k_{2})(\zeta_{4}\cdot k_{2}%
)\underset{a,b,c=0}{\overset{\infty}{%
%TCIMACRO{\dsum }%
%BeginExpansion
{\displaystyle\sum}
%EndExpansion
}}\frac{\left(  -k_{2}\cdot k_{3}\right)  _{a}}{a!}\frac{\left(  -k_{2}\cdot
k_{4}+1\right)  _{b}}{b!}\frac{\left(  -k_{3}\cdot k_{4}\right)  _{c}}%
{c!}\nonumber\\
&  \times\frac{2}{2k_{1}\cdot k_{2}+2\left(  a+b\right)  }\frac{2}%
{2(k_{1}\cdot k_{2}+k_{2}\cdot k_{3}+k_{1}\cdot k_{3})+2\left(  b+c+1\right)
},
\end{align}%
\begin{align}
A_{5}^{(2,2)} &  =(-\zeta_{1}\cdot k_{3})(\zeta_{4}\cdot k_{2}%
)\underset{a,b,c=0}{\overset{\infty}{%
%TCIMACRO{\dsum }%
%BeginExpansion
{\displaystyle\sum}
%EndExpansion
}}\frac{\left(  -k_{2}\cdot k_{3}+1\right)  _{a}}{a!}\frac{\left(  -k_{2}\cdot
k_{4}+1\right)  _{b}}{b!}\frac{\left(  -k_{3}\cdot k_{4}\right)  _{c}}%
{c!}\nonumber\\
&  \times\frac{2}{2k_{1}\cdot k_{2}+2\left(  a+b+1\right)  }\frac{2}%
{2(k_{1}\cdot k_{2}+k_{2}\cdot k_{3}+k_{1}\cdot k_{3})+2\left(  b+c+1\right)
},
\end{align}%
\begin{align}
A_{5}^{(3,2)} &  =(-\zeta_{1}\cdot k_{4})(\zeta_{4}\cdot k_{2}%
)\underset{a,b,c=0}{\overset{\infty}{%
%TCIMACRO{\dsum }%
%BeginExpansion
{\displaystyle\sum}
%EndExpansion
}}\frac{\left(  -k_{2}\cdot k_{3}\right)  _{a}}{a!}\frac{\left(  -k_{2}\cdot
k_{4}+1\right)  _{b}}{b!}\frac{\left(  -k_{3}\cdot k_{4}\right)  _{c}}%
{c!}\nonumber\\
&  \times\frac{2}{2k_{1}\cdot k_{2}+2\left(  a+b+1\right)  }\frac{2}%
{2(k_{1}\cdot k_{2}+k_{2}\cdot k_{3}+k_{1}\cdot k_{3})+2\left(  b+c+2\right)
},
\end{align}%
\begin{align}
A_{5}^{(1,3)} &  =(-\zeta_{1}\cdot k_{2})(\zeta_{4}\cdot k_{3}%
)\underset{a,b,c=0}{\overset{\infty}{%
%TCIMACRO{\dsum }%
%BeginExpansion
{\displaystyle\sum}
%EndExpansion
}}\frac{\left(  -k_{2}\cdot k_{3}\right)  _{a}}{a!}\frac{\left(  -k_{2}\cdot
k_{4}\right)  _{b}}{b!}\frac{\left(  -k_{3}\cdot k_{4}+1\right)  _{c}}%
{c!}\nonumber\\
&  \times\frac{2}{2k_{1}\cdot k_{2}+2\left(  a+b\right)  }\frac{2}%
{2(k_{1}\cdot k_{2}+k_{2}\cdot k_{3}+k_{1}\cdot k_{3})+2\left(  b+c+1\right)
},
\end{align}%
\begin{align}
A_{5}^{(2,3)} &  =(-\zeta_{1}\cdot k_{3})(\zeta_{4}\cdot k_{3}%
)\underset{a,b,c=0}{\overset{\infty}{%
%TCIMACRO{\dsum }%
%BeginExpansion
{\displaystyle\sum}
%EndExpansion
}}\frac{\left(  -k_{2}\cdot k_{3}+1\right)  _{a}}{a!}\frac{\left(  -k_{2}\cdot
k_{4}\right)  _{b}}{b!}\frac{\left(  -k_{3}\cdot k_{4}+1\right)  _{c}}%
{c!}\nonumber\\
&  \times\frac{2}{2k_{1}\cdot k_{2}+2\left(  a+b+1\right)  }\frac{2}%
{2(k_{1}\cdot k_{2}+k_{2}\cdot k_{3}+k_{1}\cdot k_{3})+2\left(  b+c+1\right)
},
\end{align}%
\begin{align}
A_{5}^{(3,3)} &  =(-\zeta_{1}\cdot k_{4})(\zeta_{4}\cdot k_{3}%
)\underset{a,b,c=0}{\overset{\infty}{%
%TCIMACRO{\dsum }%
%BeginExpansion
{\displaystyle\sum}
%EndExpansion
}}\frac{\left(  -k_{2}\cdot k_{3}\right)  _{a}}{a!}\frac{\left(  -k_{2}\cdot
k_{4}\right)  _{b}}{b!}\frac{\left(  -k_{3}\cdot k_{4}+1\right)  _{c}}%
{c!}\nonumber\\
&  \times\frac{2}{2k_{1}\cdot k_{2}+2\left(  a+b+1\right)  }\frac{2}%
{2(k_{1}\cdot k_{2}+k_{2}\cdot k_{3}+k_{1}\cdot k_{3})+2\left(  b+c+2\right)
}%
\end{align}
and%
\begin{align}
A_{5}^{(0,0)} &  =(\zeta_{1}\cdot\zeta_{4})\underset{a,b,c=0}{\overset{\infty
}{%
%TCIMACRO{\dsum }%
%BeginExpansion
{\displaystyle\sum}
%EndExpansion
}}\frac{\left(  -k_{2}\cdot k_{3}\right)  _{a}}{a!}\frac{\left(  -k_{2}\cdot
k_{4}\right)  _{b}}{b!}\frac{\left(  -k_{3}\cdot k_{4}\right)  _{c}}%
{c!}\nonumber\\
&  \times\frac{2}{2k_{1}\cdot k_{2}+2\left(  a+b+1\right)  }\frac{2}%
{2(k_{1}\cdot k_{2}+k_{2}\cdot k_{3}+k_{1}\cdot k_{3})+2\left(  b+c+2\right)
}.
\end{align}

We can now apply string on-shell recursion relation to obtain%
\begin{align}
A_{5}^{(1,2)}  &  =\underset{_{M}}{\sum}\frac{2(-\zeta_{1}\cdot k_{2}%
)(\zeta_{4}\cdot k_{2})}{2k_{1}\cdot k_{2}+2M}R_{3,4}^{(1,2),M}\nonumber\\
&  +\underset{_{M}}{\sum}\frac{2(-\zeta_{1}\cdot k_{2})(\zeta_{4}\cdot k_{2}%
)}{2(k_{1}\cdot k_{2}+k_{2}\cdot k_{3}+k_{1}\cdot k_{3})+2\left(  M+1\right)
}R_{4,3}^{(1,2),M}%
\end{align}
where the two residues%
\begin{equation}
R_{3,4}^{(1,2),M}=\underset{\left\{  N_{r}\right\}  }{\sum}\left(  -k_{2}\cdot
k_{3}\right)  ^{N}\int_{0}^{1}dz_{2}z_{2}^{\left(  \hat{k}_{1}+k_{2}\right)
\cdot k_{3}-M}\left(  1-z_{2}\right)  ^{k_{3}\cdot k_{4}}\prod_{r=1}\frac
{1}{N_{r}!r^{N_{r}}}\left[  1-\left(  \frac{-k_{2}\cdot k_{4}+1}{k_{2}\cdot
k_{3}}\right)  z_{2}^{r}\right]  ^{N_{r}}%
\end{equation}
and%
\begin{equation}
R_{4,3}^{(1,2),M}=\underset{\left\{  N_{r}\right\}  }{\sum}\left(  -k_{3}\cdot
k_{4}\right)  ^{N}\int_{0}^{1}dz_{2}z_{2}^{\hat{k}_{1}\cdot k_{2}-1}\left(
1-z_{2}\right)  ^{k_{2}\cdot k_{3}}\prod_{r=1}\frac{1}{N_{r}!r^{N_{r}}}\left[
1-\left(  \frac{-k_{2}\cdot k_{4}+1}{k_{3}\cdot k_{4}}\right)  z_{2}%
^{r}\right]  ^{N_{r}}%
\end{equation}
can all be expressed in terms of the Lauricella functions. Similarly, we can
calculate%
\begin{align}
A_{5}^{(2,2)}  &  =\underset{_{M}}{\sum}\frac{2(-\zeta_{1}\cdot k_{2}%
)(\zeta_{4}\cdot k_{2})}{2k_{1}\cdot k_{2}+2(M+1)}R_{3,4}^{(2,2),M}\nonumber\\
&  +\underset{_{M}}{\sum}\frac{2(-\zeta_{1}\cdot k_{2})(\zeta_{4}\cdot k_{2}%
)}{2(k_{1}\cdot k_{2}+k_{2}\cdot k_{3}+k_{1}\cdot k_{3})+2\left(  M+2\right)
}R_{4,3}^{(2,2),M}%
\end{align}
where%
\begin{align}
R_{3,4}^{(2,2),M}  &  =\underset{\left\{  N_{r}\right\}  }{\sum}\left(
-k_{2}\cdot k_{3}+1\right)  ^{N}\int_{0}^{1}dz_{2}z_{2}^{\left(  \hat{k}%
_{1}+k_{2}\right)  \cdot k_{3}-M-1}\left(  1-z_{2}\right)  ^{k_{3}\cdot k_{4}%
}\prod_{r=1}\frac{1}{N_{r}!r^{N_{r}}}\left[  1-\left(  \frac{-k_{2}\cdot
k_{4}+1}{k_{2}\cdot k_{3}-1}\right)  z_{2}^{r}\right]  ^{N_{r}},\\
R_{4,3}^{(2,2),M}  &  =\underset{\left\{  N_{r}\right\}  }{\sum}\left(
-k_{3}\cdot k_{4}\right)  ^{N}\int_{0}^{1}dz_{2}z_{2}^{\hat{k}_{1}\cdot
k_{2}-1}\left(  1-z_{2}\right)  ^{k_{2}\cdot k_{3}-1}\prod_{r=1}\frac{1}%
{N_{r}!r^{N_{r}}}\left[  1-\left(  \frac{-k_{2}\cdot k_{4}+1}{k_{3}\cdot
k_{4}}\right)  z_{2}^{r}\right]  ^{N_{r}};
\end{align}%
\begin{align}
A_{5}^{(3,2)}  &  =\underset{_{M}}{\sum}\frac{2(-\zeta_{1}\cdot k_{4}%
)(\zeta_{4}\cdot k_{2})}{2k_{1}\cdot k_{2}+2(M+1)}R_{3,4}^{(3,2),M}\nonumber\\
&  +\underset{_{M}}{\sum}\frac{2(-\zeta_{1}\cdot k_{4})(\zeta_{4}\cdot k_{2}%
)}{2(k_{1}\cdot k_{2}+k_{2}\cdot k_{3}+k_{1}\cdot k_{3})+2\left(  M+2\right)
}R_{4,3}^{(3,2),M}%
\end{align}
where%
\begin{align}
R_{3,4}^{(3,2),M}  &  =\underset{\left\{  N_{r}\right\}  }{\sum}\left(
-k_{2}\cdot k_{3}\right)  ^{N}\int_{0}^{1}dz_{2}z_{2}^{\left(  \hat{k}%
_{1}+k_{2}\right)  \cdot k_{3}-M}\left(  1-z_{2}\right)  ^{k_{3}\cdot k_{4}%
}\prod_{r=1}\frac{1}{N_{r}!r^{N_{r}}}\left[  1-\left(  \frac{-k_{2}\cdot
k_{4}+1}{k_{2}\cdot k_{3}}\right)  z_{2}^{r}\right]  ^{N_{r}},\\
R_{4,3}^{(3,2),M}  &  =\underset{\left\{  N_{r}\right\}  }{\sum}\left(
-k_{3}\cdot k_{4}\right)  ^{N}\int_{0}^{1}dz_{2}z_{2}^{\hat{k}_{1}\cdot k_{2}%
}\left(  1-z_{2}\right)  ^{k_{2}\cdot k_{3}}\prod_{r=1}\frac{1}{N_{r}%
!r^{N_{r}}}\left[  1-\left(  \frac{-k_{2}\cdot k_{4}+1}{k_{3}\cdot k_{4}%
}\right)  z_{2}^{r}\right]  ^{N_{r}}.
\end{align}

The next $3$ amplitudes can be calculated to be%
\begin{align}
A_{5}^{(1,3)}  &  =\underset{_{M}}{\sum}\frac{2(-\zeta_{1}\cdot k_{2}%
)(\zeta_{4}\cdot k_{3})}{2k_{1}\cdot k_{2}+2(M+1)}R_{3,4}^{(1,3),M}\nonumber\\
&  +\underset{_{M}}{\sum}\frac{2(-\zeta_{1}\cdot k_{2})(\zeta_{4}\cdot k_{3}%
)}{2(k_{1}\cdot k_{2}+k_{2}\cdot k_{3}+k_{1}\cdot k_{3})+2\left(  M+2\right)
}R_{4,3}^{(1,3),M}%
\end{align}
where%
\begin{align}
R_{3,4}^{(1,3),M}  &  =\underset{\left\{  N_{r}\right\}  }{\sum}\left(
-k_{2}\cdot k_{3}+1\right)  ^{N}\int_{0}^{1}dz_{2}z_{2}^{\left(  \hat{k}%
_{1}+k_{2}\right)  \cdot k_{3}-M}\left(  1-z_{2}\right)  ^{k_{3}\cdot k_{4}%
-1}\prod_{r=1}\frac{1}{N_{r}!r^{N_{r}}}\left[  1-\left(  \frac{-k_{2}\cdot
k_{4}+1}{k_{2}\cdot k_{3}-1}\right)  z_{2}^{r}\right]  ^{N_{r}},\\
R_{4,3}^{(1,3),M}  &  =\underset{\left\{  N_{r}\right\}  }{\sum}\left(
-k_{3}\cdot k_{4}+1\right)  ^{N}\int_{0}^{1}dz_{2}z_{2}^{\hat{k}_{1}\cdot
k_{2}-1}\left(  1-z_{2}\right)  ^{k_{2}\cdot k_{3}}\prod_{r=1}\frac{1}%
{N_{r}!r^{N_{r}}}\left[  1-\left(  \frac{-k_{2}\cdot k_{4}+1}{k_{3}\cdot
k_{4}-1}\right)  z_{2}^{r}\right]  ^{N_{r}};
\end{align}%
\begin{align}
A_{5}^{(2,3)}  &  =\underset{_{M}}{\sum}\frac{2(-\zeta_{1}\cdot k_{3}%
)(\zeta_{4}\cdot k_{3})}{2k_{1}\cdot k_{2}+2(M+1)}R_{3,4}^{(2,3),M}\nonumber\\
&  +\underset{_{M}}{\sum}\frac{2(-\zeta_{1}\cdot k_{3})(\zeta_{4}\cdot k_{3}%
)}{2(k_{1}\cdot k_{2}+k_{2}\cdot k_{3}+k_{1}\cdot k_{3})+2\left(  M+2\right)
}R_{4,3}^{(2,3),M}%
\end{align}
where%
\begin{align}
R_{3,4}^{(2,3),M}  &  =\underset{\left\{  N_{r}\right\}  }{\sum}\left(
-k_{2}\cdot k_{3}+1\right)  ^{N}\int_{0}^{1}dz_{2}z_{2}^{\left(  \hat{k}%
_{1}+k_{2}\right)  \cdot k_{3}-M-1}\left(  1-z_{2}\right)  ^{k_{3}\cdot
k_{4}-1}\prod_{r=1}\frac{1}{N_{r}!r^{N_{r}}}\left[  1-\left(  \frac
{-k_{2}\cdot k_{4}}{k_{2}\cdot k_{3}-1}\right)  z_{2}^{r}\right]  ^{N_{r}},\\
R_{4,3}^{(2,3),M}  &  =\underset{\left\{  N_{r}\right\}  }{\sum}\left(
-k_{3}\cdot k_{4}+1\right)  ^{N}\int_{0}^{1}dz_{2}z_{2}^{\hat{k}_{1}\cdot
k_{2}}\left(  1-z_{2}\right)  ^{k_{2}\cdot k_{3}-1}\prod_{r=1}\frac{1}%
{N_{r}!r^{N_{r}}}\left[  1-\left(  \frac{-k_{2}\cdot k_{4}}{k_{3}\cdot
k_{4}-1}\right)  z_{2}^{r}\right]  ^{N_{r}};
\end{align}
and%
\begin{align}
A_{5}^{(3,3)}  &  =\underset{_{M}}{\sum}\frac{2(-\zeta_{1}\cdot k_{4}%
)(\zeta_{4}\cdot k_{3})}{2k_{1}\cdot k_{2}+2(M+1)}R_{3,4}^{(3,3),M}\nonumber\\
&  +\underset{_{M}}{\sum}\frac{2(-\zeta_{1}\cdot k_{4})(\zeta_{4}\cdot k_{3}%
)}{2(k_{1}\cdot k_{2}+k_{2}\cdot k_{3}+k_{1}\cdot k_{3})+2\left(  M+2\right)
}R_{4,3}^{(3,3),M}%
\end{align}
where%
\begin{align}
R_{3,4}^{(3,3),M}  &  =\underset{\left\{  N_{r}\right\}  }{\sum}\left(
-k_{2}\cdot k_{3}\right)  ^{N}\int_{0}^{1}dz_{2}z_{2}^{\left(  \hat{k}%
_{1}+k_{2}\right)  \cdot k_{3}-M}\left(  1-z_{2}\right)  ^{k_{3}\cdot k_{4}%
-1}\prod_{r=1}\frac{1}{N_{r}!r^{N_{r}}}\left[  1-\left(  \frac{-k_{2}\cdot
k_{4}}{k_{2}\cdot k_{3}-1}\right)  z_{2}^{r}\right]  ^{N_{r}},\\
R_{4,3}^{(3,3),M}  &  =\underset{\left\{  N_{r}\right\}  }{\sum}\left(
-k_{3}\cdot k_{4}+1\right)  ^{N}\int_{0}^{1}dz_{2}z_{2}^{\hat{k}_{1}\cdot
k_{2}}\left(  1-z_{2}\right)  ^{k_{2}\cdot k_{3}}\prod_{r=1}\frac{1}%
{N_{r}!r^{N_{r}}}\left[  1-\left(  \frac{-k_{2}\cdot k_{4}}{k_{3}\cdot
k_{4}-1}\right)  z_{2}^{r}\right]  ^{N_{r}}.
\end{align}

Finally, the last one is%
\begin{align}
A_{5}^{(0,0)} &  =\underset{_{M}}{\sum}\frac{2\zeta_{1}\cdot\zeta_{4}}%
{2k_{1}\cdot k_{2}+2(M+1)}R_{3,4}^{(0,0),M}\nonumber\\
&  +\underset{_{M}}{\sum}\frac{2\zeta_{1}\cdot\zeta_{4}}{2(k_{1}\cdot
k_{2}+k_{2}\cdot k_{3}+k_{1}\cdot k_{3})+2\left(  M+2\right)  }R_{4,3}%
^{(0,0),M}%
\end{align}
where%
\begin{align}
R_{3,4}^{(0,0),M} &  =\underset{\left\{  N_{r}\right\}  }{\sum}\left(
-k_{2}\cdot k_{3}\right)  ^{N}\int_{0}^{1}dz_{2}z_{2}^{\left(  \hat{k}%
_{1}+k_{2}\right)  \cdot k_{3}-M}\left(  1-z_{2}\right)  ^{k_{3}\cdot k_{4}%
}\prod_{r=1}\frac{1}{N_{r}!r^{N_{r}}}\left[  1-\left(  \frac{-k_{2}\cdot
k_{4}}{k_{2}\cdot k_{3}}\right)  z_{2}^{r}\right]  ^{N_{r}},\\
R_{4,3}^{(0,0),M} &  =\underset{\left\{  N_{r}\right\}  }{\sum}\left(
-k_{3}\cdot k_{4}\right)  ^{N}\int_{0}^{1}dz_{2}z_{2}^{\hat{k}_{1}\cdot k_{2}%
}\left(  1-z_{2}\right)  ^{k_{2}\cdot k_{3}}\prod_{r=1}\frac{1}{N_{r}%
!r^{N_{r}}}\left[  1-\left(  \frac{-k_{2}\cdot k_{4}}{k_{3}\cdot k_{4}%
}\right)  z_{2}^{r}\right]  ^{N_{r}}.
\end{align}
This completes the long calculation of Eq.(\ref{ij}), and we have successfully
expressed the $5$-point SSA with three tachyons and two vectors in terms of
the $4$-point LSSA. According to the shifting principle, one can generalize
the above calculation to the three tachyons and two tensor SSA and express it
in terms of the LSSA. In general, it is easy to see that the above
calculations can be generalized to $5$-point SSA with arbitrary five tensor
legs and express them in terms of the LSSA.

\subsection{Six-point SSA with tensor legs}

One important application of the above calculations is to prove that the
$6$-point KN amplitude can be expressed in terms of the Lauricella functions.
Indeed, we note that Eq.(\ref{three}) can be obtained from Eq.(\ref{RR}) by
taking $N=1=M$. So what we have just shown above is that $R_{3,5}^{1}$, the
$M=1$ case of the residue $R_{3,5}^{M}$ in Eq.(\ref{RR}) can be expressed in
terms of the Lauricella functions. In fact, by using Eq.(\ref{R35}) and
Eq.(\ref{three}), we have the identification%
\begin{equation}
R_{3,5}^{1}(\hat{k}_{1}+k_{2},k_{3},k_{4},k_{5},k_{6})=A_{5}(\zeta_{1}%
=k_{2},\hat{k}_{1}+k_{2};k_{3};k_{4};k_{5};k_{6})
\end{equation}
and the following expression%
\begin{align}
&  R_{3,5}^{1}(k=\hat{k}_{1}+k_{2},k_{3},k_{4},k_{5},k_{6})\nonumber\\
&  =\underset{_{M}}{\sum}\frac{-2k_{2}\cdot k_{3}}{2k\cdot k_{3}+2M}%
R_{3,4}^{(1),M}+\underset{_{M}}{\sum}\frac{-2k_{2}\cdot k_{3}}{2(k\cdot
k_{3}+k_{3}\cdot k_{4}+k\cdot k_{4})+2(M+1)}R_{4,3}^{(1),M}\nonumber\\
&  +\underset{_{M}}{\sum}\frac{-2k_{2}\cdot k_{4}}{2k\cdot k_{3}%
+2(M+1)}R_{3,4}^{(2),M}+\underset{_{M}}{\sum}\frac{-2k_{2}\cdot k_{4}%
}{2(k\cdot k_{3}+k_{3}\cdot k_{4}+k\cdot k_{4})+2(M+1)}R_{4,3}^{(2),M}%
\nonumber\\
&  +\underset{_{M}}{\sum}\frac{-2k_{2}\cdot k_{5}}{2k\cdot k_{3}%
+2(M+1)}R_{3,4}^{(3),M}+\underset{_{M}}{\sum}\frac{-2k_{2}\cdot k_{5}%
}{2(k\cdot k_{3}+k_{3}\cdot k_{4}+k\cdot k_{4})+2(M+1)}R_{4,3}^{(3),M},
\end{align}
which expresses one of the residue of the $6$-point KN amplitude $R_{3,5}^{1}$
in terms of the LSSA. Similar consideration applies to $M=2,3,4\cdots$ cases.
For example, for $M=2$ case, by Eq.(\ref{R35}), it is straightforward to
calculate%
\begin{align}
&  R_{3,5}^{2}=\int_{0}^{1}dw_{4}\int_{0}^{w_{4}}dw_{3}w_{3}^{\left(  \hat
{k}_{1}+k_{2}\right)  \cdot k_{3}}w_{4}^{\left(  \hat{k}_{1}+k_{2}\right)
\cdot k_{4}}\left(  1-w_{3}\right)  ^{k_{3}\cdot k_{5}}\left(  1-w_{4}\right)
^{k_{4}\cdot k_{5}}\left(  w_{4}-w_{3}\right)  ^{k_{3}\cdot k_{4}}\nonumber\\
&  \times\left\{
\begin{array}
[c]{c}%
\frac{1}{2!1^{2}}\left[  \left(  -k_{2}\cdot k_{3}\right)  \left(  \frac
{1}{w_{3}}\right)  +\left(  -k_{2}\cdot k_{4}\right)  \left(  \frac{1}{w_{4}%
}\right)  +\left(  -k_{2}\cdot k_{5}\right)  \left(  \frac{1}{1}\right)
\right]  ^{2}\\
+\frac{1}{1!2^{1}}\left[  \left(  -k_{2}\cdot k_{3}\right)  \left(  \frac
{1}{w_{3}^{2}}\right)  +\left(  -k_{2}\cdot k_{4}\right)  \left(  \frac
{1}{w_{4}^{2}}\right)  +\left(  -k_{2}\cdot k_{5}\right)  \left(  \frac
{1}{1^{2}}\right)  \right]
\end{array}
\right\}  ,\label{tt}%
\end{align}
and express it in terms of the LSSA. The calculation of each term in
Eq.(\ref{tt}) is similar to that of the $5$-point KN amplitude except shifting
some kinematic variables. Thus $R_{3,5}^{M}$ are LSSA and so are $R_{5,3}^{M}%
$.
%TCIMACRO{\FRAME{ftbpFU}{6.1056in}{1.2592in}{0pt}{\Qcb{Expressing the 6-point
%KN amplitude in terms of the LSSA by the first recursion}}{\Qlb{6point}%
%}{6-point.eps}{\special{ language "Scientific Word";  type "GRAPHIC";
%width 6.1056in;  height 1.2592in;  depth 0pt;  original-width 6.045in;
%original-height 1.2254in;  cropleft "0";  croptop "1";  cropright "1";
%cropbottom "0";  filename '6-point.eps';file-properties "XNPEU";}} }%
%BeginExpansion
\begin{figure}[ptb]%
\centering
\includegraphics[
height=1.2592in,
width=6.1056in
]%
{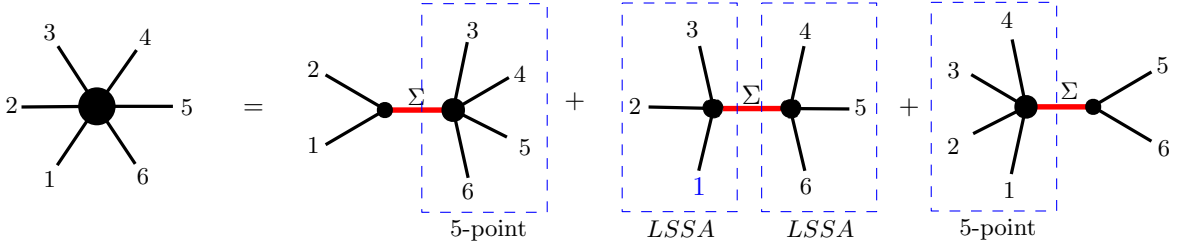}%
\caption{Expressing the 6-point KN amplitude in terms of the LSSA by the first
recursion}%
\label{6point}%
\end{figure}
%EndExpansion
%TCIMACRO{\FRAME{ftbpFU}{4.4797in}{1.286in}{0pt}{\Qcb{Expressing the 6-point KN
%amplitude in terms of the LSSA by the second recursion}}{\Qlb{6point2}%
%}{5-point.eps}{\special{ language "Scientific Word";  type "GRAPHIC";
%width 4.4797in;  height 1.286in;  depth 0pt;  original-width 4.4287in;
%original-height 1.2505in;  cropleft "0";  croptop "1";  cropright "1";
%cropbottom "0";  filename '5-point.eps';file-properties "XNPEU";}} }%
%BeginExpansion
\begin{figure}[ptb]%
\centering
\includegraphics[
height=1.286in,
width=4.4797in
]%
{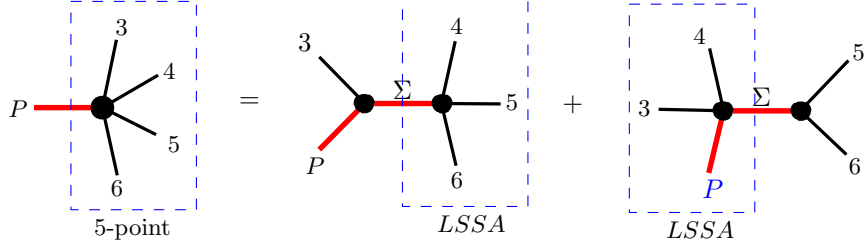}%
\caption{Expressing the 6-point KN amplitude in terms of the LSSA by the
second recursion}%
\label{6point2}%
\end{figure}
%EndExpansion
On the other hand, the result of Eq.(\ref{R44}) tells us that the residue
$R_{4,4}^{M}$ is a sum of products of the Lauricella functions after term by
term integrations for a given $M$. We conclude that the $6$-point KN amplitude
is a LSSA. It is interesting to note that to obtain the LSSA form of the
residue $R_{4,4}^{M}$, one needs only do one-step recursion from the $6$-point
KN amplitude. But to obtain the LSSA form of the residue $R_{3,5}^{M}$
($R_{5,3}^{M}$), one needs to do $2$-step recursion or $(recursion)^{2}$. See
Fig.\ref{6point} and Fig.\ref{6point2}.

In fact, to show that the $6$-point KN amplitude is a LSSA, one needs only do
$1$-step recursion to express it in terms of the lower $5$-point and $4$-point
amplitudes as was shown in Fig.\ref{6point}. Since we have shown that all
$5$-point and $4$-point SSA are LSSA, the $6$-point KN amplitude is a LSSA.
However, to explicitly calculate the LSSA form of the $6$-point KN amplitude,
one needs to do the second recursion as was shown in Fig.\ref{6point2}, and
the calculation will be very lengthy.

The next step is to use the shifting principle to argue that all $6$-point SSA
with arbitrary six tensor legs can be expressed in terms of the LSSA. The
argument is obvious although the detailed calculation will again be very lengthy.

\subsection{Seven-point SSA with tensor legs}

For the case of $7$-point SSA, we first do $1$-step recursion on $7$-point KN
amplitude. The subamplitude $R_{3,6}^{P,R}$ of the residue $R_{3,6}^{M}$ we
obtained is the $6$-point SSA with five tachyons and one tensor leg. So for
simplicity let's first consider the five tachyons, one vector SSA ($\left(
z_{1},z_{5},z_{6}\right)  =\left(  0,1,\infty\right)  $)%
\begin{align}
&  A_{6}\left(  \zeta_{1},k_{1};k_{2};k_{3};k_{4};k_{5};k_{6}\right)
\nonumber\\
&  =\int_{0}^{1}dz_{4}\int_{0}^{z_{4}}dz_{3}\int_{0}^{z_{3}}dz_{2}z_{2}%
^{k_{2}\cdot k_{1}}z_{3}^{k_{3}\cdot k_{1}}z_{4}^{k_{4}\cdot k_{1}}\left(
1-z_{2}\right)  ^{k_{5}\cdot k_{2}}\left(  1-z_{3}\right)  ^{k_{5}\cdot k_{3}%
}\nonumber\\
&  \times\left(  z_{4}-z_{2}\right)  ^{k_{4}\cdot k_{2}}\left(  z_{4}%
-z_{3}\right)  ^{k_{4}\cdot k_{3}}\left(  z_{3}-z_{2}\right)  ^{k_{3}\cdot
k_{2}}\nonumber\\
&  \times(-\zeta_{1})\cdot\left[  \frac{k_{2}}{z_{2}}+\frac{k_{3}}{z_{3}%
}+\frac{k_{4}}{z_{4}}+\frac{k_{5}}{1}\right]  \nonumber\\
&  =A_{6}^{(1)}+A_{6}^{(2)}+A_{6}^{(3)}+A_{6}^{(4)}.\label{four}%
\end{align}
Mathematically, the calculation of all four terms in Eq.(\ref{four}) are
similar to the $6$-point KN amplitude which has been proved to be a LSSA in
the last subsection. So the amplitude in Eq.(\ref{four}) is a LSSA. According
to the shifting principle, one can generalize the above calculation and
consider five tachyons, one arbitrary tensor SSA and express them in terms of
the LSSA. This proves that the residue $R_{3,6}^{M}$ of $7$-point KN amplitude
is a LSSA. Similar consideration applies to the other residue $R_{6,3}^{M}$ of
the $7$-point KN amplitude. The other two residues $R_{4,5}^{M}$, $R_{5,4}%
^{M}$ contain $5$-point and $4$-point SSA which have been shown to be LSSA. We
conclude that the $7$-point KN amplitude is a LSSA. The next step is to use
the shifting principle to argue that all $7$-point SSA with arbitrary seven
tensor legs can be expressed in terms of the LSSA.

It is interesting to note that on the calculation of multi-step recursion
processes, one encounters multi-tensor lower point SSA. For example, in
addition to the $\zeta$ ($6$-point to $5$-point) in Eq.(\ref{mul}), there is
another hidden leg in $R_{3,4}^{(1),M}$ ($R_{4,3}^{(1),M}$) when one performs
the recursion from $5$-point to $4$-point. In sum, one obtains two tensor-legs
in the final $4$-point LSSA when performing $2$-step recursion
$(recursion)^{2}$ on the $6$-point KN amplitude. It is interesting to see that
the multi-tensor SSA are associated with the multi-step recursion processes.

\subsection{N-point SSA with tensor legs}

In general, we can use mathematical induction, together with the on-shell
recursion and the shifting principle, to show that all $n$-point SSA can be
expressed in terms of the LSSA. The procedure goes as following. We assume
that all $k$-point SSA ($k\leq n-1$) are LSSA, and we want to prove that all
$n$-point SSA are LSSA. To prove this, we first apply the on-shell recursion
to express the residue of $n$-point KN amplitude calculated in section III in
terms of the lower point ($k\leq n-1$) SSA which were assumed to be LSSA. See
Fig \ref{recursion} above Eq.(\ref{AJ}). So the $n$-point KN amplitude is a
LSSA. We can then apply the shifting principle to show that all $n$-point SSA
including the $n$-point KN amplitude are LSSA. This completes the proof.

As an example of application of the shifting principle, the $(n-1)$ tachyons,
one vector SSA can be written as ($\left(  z_{1},z_{n-1},z_{n}\right)
=\left(  0,1,\infty\right)  $)%

\begin{align}
&  A_{n}\left(  \zeta_{1},k_{1};k_{2};k_{3};k_{4};\cdot\cdot\cdot;k_{n}\right)
\nonumber\\
&  =\int_{0}^{1}dz_{n-2}\int_{0}^{z_{n-2}}dz_{n-3}\cdots\int_{0}^{z_{3}}%
dz_{2}\text{ }\prod_{i>j=1}^{n-2}\left(  z_{i}-z_{j}\right)  ^{k_{i}\cdot
k_{j}}\nonumber\\
&  \times(-\zeta_{1})\cdot\left[  \frac{k_{2}}{z_{2}}+\frac{k_{3}}{z_{3}%
}+\frac{k_{4}}{z_{4}}+\cdot\cdot\cdot+\frac{k_{n-1}}{1}\right] \nonumber\\
&  =A_{n}^{(1)}+A_{n}^{(2)}+A_{n}^{(3)}+\cdot\cdot\cdot+A_{n}^{(n-2)}%
\end{align}
where the calculation of each $A_{n}^{(j)}$ is similar to the $n$-point KN
amplitude and can be expressed in terms of LSSA.%

%TCIMACRO{\TeXButton{equation number}{\setcounter{equation}{0}
%\renewcommand{\theequation}{\arabic{section}.\arabic{equation}}}}%
%BeginExpansion
\setcounter{equation}{0}
\renewcommand{\theequation}{\arabic{section}.\arabic{equation}}%
%EndExpansion

\section{Conclusion}

In this paper, we calculate explicitly residues of all $n$-point Koba-Nielsen
(KN) amplitudes by using string on-shell recursion relation of string
scattering amplitudes (SSA). We then use mathematical induction, together with
the on-shell recursion and the shifting principle, to show that all $n$-point
SSA including the KN amplitudes can be expressed in terms of the LSSA. In
general, to explicitly calculate the LSSA form of the $n$-point KN amplitude,
one needs to apply up to ($n-4)$-step recursion $(recursion)^{n-4}$ to achieve
a $4$-point LSSA. These results demonstrates the exact $SL(K+3,%
%TCIMACRO{\U{2102} }%
%BeginExpansion
\mathbb{C}
%EndExpansion
)$ symmetry of the tree-level open bosonic string theory.

Moreover, we derive an iteration relation among the residues of a given
$n$-point KN amplitude. This iteration relation is a generalization of the
case of simple Veneziano amplitude to the higher point SSA. We argue that this
iteration relation is related to the $SL(K+3,%
%TCIMACRO{\U{2102} }%
%BeginExpansion
\mathbb{C}
%EndExpansion
)$ symmetry and can presumably, in the $4$-point case, be used to soften the
well-known hard Veneziano amplitude. So we expect the $SL(K+3,%
%TCIMACRO{\U{2102} }%
%BeginExpansion
\mathbb{C}
%EndExpansion
)$ symmetry can be used to soften all higher $n$-point hard SSA. This
mechanism is reminiscent of symmetry principle in QFT.

In the calculation of multi-tensor SSA, we discover that the calculation of
SSA expressed in terms of the Lauricella functions is much simpler than the
traditional calculation \cite{KLT}. It is also interesting to see that to
express the residues of higher point SSA in terms of the Lauricella $4$-point
function, one encounters application of the multi-step recursion processes
which are associated with the multi-tensor SSA calculation.

It will be important to extend the results of this paper to the superstring
theory. However, the known construction of the string fermion vertex involved
only the \textit{leading Regge trajectory} fermion string state of the R
sector \cite{PFSSA,Osch,RRR}. It is a nontrivial task to construct the general
massive fermion string vertex operators \cite{m1,m2,m3,m4}. Many questions
related to the construction of SSA involving the general massive fermion
string states need to be answered before one can further understand the
symmetry of superstring theory.

In addition to the dynamics of $4$-point SSA with higher spin particles, there
are many interesting issues related to higher point SSA. To name some, is
there any evidence of the $SL(K+3,%
%TCIMACRO{\U{2102} }%
%BeginExpansion
\mathbb{C}
%EndExpansion
)$ symmetry other than deriving it from SSA? In a recent publication
\cite{LLYT}, it was shown that the LSSA of three tachyons and one arbitrary
string states can be rederived from the deformed cubic string field theory
(SFT) \cite{Taejin}. But can one use Witten SFT to obtain the $SL(K+3,%
%TCIMACRO{\U{2102} }%
%BeginExpansion
\mathbb{C}
%EndExpansion
)$ symmetry ? How to deal with the $z_{4}\rightarrow\infty$ limit as was shown
in Eq.(\ref{line}) when calculating SSA with higher spin particles putting at
$z_{4}$ ? Are there linear relations among higher point ($n\geq5$) hard SSA
similar to those of the $4$-point case? We hope to address these issues in the
next publications.

\begin{acknowledgments}
We would like to thank H. Kawai and C.I. Tan for discussions which help to
clarify many issues of the LSSA. This work is supported in part by the
Ministry of Science and Technology (MoST) and S.T. Yau center of National Yang
Ming Chiao Tung University (NYCU), Taiwan.
\end{acknowledgments}

\appendix%
%TCIMACRO{\TeXButton{equation number}{\setcounter{equation}{0}
%\renewcommand{\theequation}{A.\arabic{equation}}}}%
%BeginExpansion
\setcounter{equation}{0}
\renewcommand{\theequation}{A.\arabic{equation}}%
%EndExpansion

\section{Generating function}

In this appendix, we prove two relations which we used in the text to do the
residue calculation. We begin with the generating function%
\begin{equation}
e^{x\left[  z+\frac{z^{2}}{2}+\frac{z^{3}}{3}+\cdots\right]  }=e^{\left[
xz+\frac{xz^{2}}{2}+\frac{xz^{3}}{3}+\cdots\right]  }%
\end{equation}
which we used in Eq.(\ref{Vene}) to do the residue calculation of the
Veneziano amplitude. For our purpose here, let's generalize this function and
consider%
\begin{align}
A\left(  x_{1},x_{2},\cdots;z\right)   &  =\exp\left[  B\left(  x_{1}%
,x_{2},\cdots;z\right)  \right]  =e^{\left[  x_{1}z+\frac{x_{2}z^{2}}{2}%
+\frac{x_{3}z^{3}}{3}+\cdots\right]  }\nonumber\\
&  =\underset{m=0}{\overset{\infty}{\sum}}E_{m}\left(  x_{1},x_{2}%
,\cdots,x_{m}\right)  z^{m}.
\end{align}

Two simple applications of the functions $E_{m}\left(  x_{1},x_{2}%
,\cdots,x_{m}\right)  $ follow. If we set $x_{i}=x$ for all $i$, we obtain%

\begin{align}
&  A\left(  x_{1},x_{2},\cdots;z\right)  =\exp\left[  x\left(  z+\frac{z^{2}%
}{2}+\frac{z^{3}}{3}+\frac{z^{4}}{4}+\frac{z^{5}}{5}\cdots\right)  \right]
\nonumber\\
&  =E_{0}+E_{1}\left(  x\right)  z+E_{2}\left(  x\right)  z^{2}+E_{3}\left(
x\right)  z^{3}+E_{4}\left(  x\right)  z^{4}+\cdots
\end{align}
with%

\begin{align}
E_{0}  &  =1=\frac{\left(  x\right)  _{0}}{0!}\nonumber\\
E_{1}  &  =\frac{1}{1}x=x=\frac{x}{1!}=\frac{\left(  x\right)  _{1}}%
{1!}\nonumber\\
E_{2}  &  =\frac{1}{2}x+\frac{1}{2!}\left(  \frac{1}{1}\right)  ^{2}%
x^{2}=\frac{x\left(  x+1\right)  }{2!}=\frac{\left(  x\right)  _{2}}%
{2!}\nonumber\\
E_{3}  &  =\frac{1}{3}x+\frac{1}{1!1!}\left(  \frac{1}{2}\right)  \left(
\frac{1}{1}\right)  x^{2}+\frac{1}{3!}\left(  \frac{1}{1}\right)  ^{3}%
x^{3}=\frac{x\left(  x+1\right)  \left(  x+2\right)  }{3!}=\frac{\left(
x\right)  _{3}}{3!}\nonumber\\
E_{4}  &  =\frac{1}{4}x+\left[  \frac{1}{1!1!}\left(  \frac{1}{3}\right)
\left(  \frac{1}{1}\right)  +\frac{1}{2!}\left(  \frac{1}{2}\right)
^{2}\right]  x^{2}\nonumber\\
+  &  \frac{1}{2!1!}\left(  \frac{1}{1}\right)  ^{2}\left(  \frac{1}%
{2}\right)  x^{3}+\frac{1}{4!}\left(  \frac{1}{1}\right)  ^{4}x^{4}%
=\frac{x\left(  x+1\right)  \left(  x+2\right)  \left(  x+3\right)  }%
{4!}=\frac{\left(  x\right)  _{4}}{4!}\nonumber\\
&  \cdots
\end{align}
The coefficient functions above find application in Eq.(\ref{sum}) which
corresponds to the residues of the Veneziano amplitude. On the other hand, if
we set $\left(  x_{1},x_{2},\cdots;z\right)  =\left(  \alpha_{-1},\alpha
_{-2},\cdots;z\right)  $, we obtain%

\begin{align}
&  A\left(  \alpha_{-1},\alpha_{-2},\cdots;z\right)  =\exp\left[  \alpha
_{-1}z+\frac{\alpha_{-2}z^{2}}{2}+\frac{\alpha_{-3}z^{3}}{3}+\frac{\alpha
_{-4}z^{4}}{4}+\frac{\alpha_{-5}z^{5}}{5}\cdots\right] \nonumber\\
&  =E_{0}+E_{1}\left(  \alpha_{-1}\right)  z+E_{2}\left(  \alpha_{-1}%
,\alpha_{-2}\right)  z^{2}+E_{3}\left(  \alpha_{-1},\alpha_{-2},\alpha
_{-3}\right)  z^{3}\nonumber\\
&  +E_{4}\left(  \alpha_{-1},\alpha_{-2},\alpha_{-3},\alpha_{-4}\right)
z^{4}+...
\end{align}
with%
\begin{align}
E_{0}  &  =1\nonumber\\
E_{1}  &  =\frac{\alpha_{-1}}{1}\nonumber\\
E_{2}  &  =\frac{\alpha_{-2}}{2}+\frac{1}{2!}\left(  \frac{\alpha_{-1}}%
{1}\right)  ^{2}\nonumber\\
E_{3}  &  =\frac{\alpha_{-3}}{3}+\frac{1}{1!1!}\left(  \frac{\alpha_{-2}}%
{2}\right)  \left(  \frac{\alpha_{-1}}{1}\right)  +\frac{1}{3!}\left(
\frac{\alpha_{-1}}{1}\right)  ^{3}\nonumber\\
E_{4}  &  =\frac{\alpha_{-4}}{4}+\frac{1}{1!1!}\left(  \frac{\alpha_{-3}}%
{3}\right)  \left(  \frac{\alpha_{-1}}{1}\right)  +\frac{1}{2!}\left(
\frac{\alpha_{-2}}{2}\right)  ^{2}+\frac{1}{2!1!}\left(  \frac{\alpha_{-1}}%
{1}\right)  ^{2}\left(  \frac{\alpha_{-2}}{2}\right)  +\frac{1}{4!}\left(
\frac{\alpha_{-1}}{1}\right)  ^{4}\nonumber\\
&  \cdots\nonumber\\
E_{M}  &  =\underset{\left\{  N_{t}\right\}  ,M=\underset{t=1}{\overset{\infty
}{%
%TCIMACRO{\dsum }%
%BeginExpansion
{\displaystyle\sum}
%EndExpansion
}}tN_{t}}{\sum}\prod_{r=1}\frac{\left(  \alpha_{-r}\right)  ^{N_{r}}}%
{N_{r}!r^{Nr}} \label{key}%
\end{align}
where in Eq.(\ref{key}) the sum is over partitions of $N$ into $\left\{
N_{r}\right\}  $ with $M=\underset{r=1}{\sum}rN_{r}$ and
$N=\underset{r=1}{\sum}N_{r}$. The coefficient $E_{M}$ above reminds us of the
sum over normalized creation operators of the string spectrum at mass level
$M$.

There are two relations which we used in the text to do the residue
calculation. They are%
\begin{align}
F_{M}\left(  x_{1},x_{2}\right)   &  =\underset{b+c=M,b,c=0}{\sum}%
\frac{\left(  T_{1}\right)  _{b}}{b!}\frac{\left(  T_{2}\right)  _{c}}%
{c!}x_{1}^{b}x_{2}^{c}\nonumber\\
&  =\underset{\left\{  N_{r}\right\}  ,M=\underset{r=1}{\sum}rN_{r}}{\sum
}\prod_{r=1}\left(  \frac{1}{N_{r}!r^{N_{r}}}\left[  T_{1}x_{1}^{r}+T_{2}%
x_{2}^{r}\right]  ^{N_{r}}\right)  , \label{bb}%
\end{align}
and the following recurrence relation%
\begin{equation}
F_{M}\left(  x_{1},x_{2}\right)  =\underset{N=0}{\overset{M-1}{\sum}}%
F_{N}\left(  x_{1},x_{2}\right)  \left[  T_{1}x_{1}^{M-N}+T_{2}x_{2}%
^{M-N}\right]  . \label{ff}%
\end{equation}
To prove Eq.(\ref{bb}), we introduce the generating function%
\begin{align}
G\left(  x_{1},x_{2};z\right)   &  =\left(  1-zx_{1}\right)  ^{-T_{1}}\left(
1-zx_{2}\right)  ^{-T_{2}}\nonumber\\
&  =\underset{b,c=0}{\sum}\frac{\left(  T_{1}\right)  _{b}}{b!}\frac{\left(
T_{2}\right)  _{c}}{c!}x_{1}^{b}x_{2}^{c}z^{b+c},
\end{align}
and%
\begin{align}
F_{M}\left(  x_{1},x_{2}\right)   &  =\frac{1}{M!}\frac{\partial^{M}}{\partial
z^{M}}G\left(  x_{1},x_{2};z\right)  |_{z=0}\nonumber\\
&  =\underset{b+c=M,b,c=0}{\sum}\frac{\left(  T_{1}\right)  _{b}}{b!}%
\frac{\left(  T_{2}\right)  _{c}}{c!}x_{1}^{b}x_{2}^{c}.
\end{align}
On the other hand, we note that%
\begin{align}
G\left(  x_{1},x_{2},z\right)   &  =\left(  1-zx_{1}\right)  ^{-T_{1}}\left(
1-zx_{2}\right)  ^{-T_{2}}=\exp\left[  -T_{1}\ln\left(  1-zx_{1}\right)
-T_{2}\ln\left(  1-zx_{2}\right)  \right] \nonumber\\
&  =\exp\left[  T_{1}\left(  zx_{1}+\frac{\left(  zx_{1}\right)  ^{2}}%
{2}+\frac{\left(  zx_{1}\right)  ^{3}}{3}+\cdots\right)  +T_{2}\left(
zx_{2}+\frac{\left(  zx_{2}\right)  ^{2}}{2}+\frac{\left(  zx_{1}\right)
^{3}}{3}+\cdots\right)  \right] \nonumber\\
&  =\exp\left[  \left(  T_{1}x_{1}+T_{2}x_{2}\right)  z+\left(  T_{1}x_{1}%
^{2}+T_{2}x_{2}^{2}\right)  \frac{z^{2}}{2}+\left(  T_{1}x_{1}^{3}+T_{2}%
x_{2}^{3}\right)  \frac{z^{3}}{3}+\cdots\right] \nonumber\\
&  =A\left(  T_{1}x_{1}+T_{2}x_{2},T_{1}x_{1}^{2}+T_{2}x_{2}^{2},T_{1}%
x_{1}^{3}+T_{2}x_{2}^{3},\cdots;z\right)  ,
\end{align}

so%
\begin{align}
F_{M}\left(  x_{1},x_{2}\right)   &  =E_{M}\left(  T_{1}x_{1}+T_{2}x_{2}%
,T_{1}x_{1}^{2}+T_{2}x_{2}^{2},\cdots,T_{1}x_{1}^{M}+T_{2}x_{2}^{M}\right)
\nonumber\\
&  =\sum_{\left\{  N_{r}\right\}  \text{,with }M=\underset{r}{\sum}rN_{r}%
}\prod_{r=1}\left(  \frac{\left(  T_{1}x_{1}^{r}+T_{2}x_{2}^{r}\right)
^{N_{r}}}{N_{r}!r^{N_{r}}}\right)  \label{cc}%
\end{align}
where we have used Eq.(\ref{key}) to obtain Eq.(\ref{cc}). This completes the
proof of Eq.(\ref{bb}). It is straightforward to generalize Eq.(\ref{bb}) to
\begin{align}
F_{M}\left(  x_{1},\cdots,x_{n}\right)   &  =\underset{%
\begin{array}
[c]{c}%
a_{1}+\cdots+a_{n}=M\\
a_{1},\cdots,a_{n}=0
\end{array}
}{\sum}\frac{\left(  T_{1}\right)  _{a_{1}}}{a_{1}!}\cdots\frac{\left(
T_{n}\right)  _{a_{n}}}{a_{n}!}x_{1}^{a_{1}}\cdots x_{n}^{a_{n}}\nonumber\\
&  =\underset{\left\{  N_{r}\right\}  \text{,with fixed }M=\underset{r}{\sum
}rN_{r}}{\sum}\prod_{r=1}\left(  \frac{1}{N_{r}!r^{N_{r}}}\left[  T_{1}%
x_{1}^{r}+\cdots+T_{n}x_{n}^{r}\right]  ^{N_{r}}\right)  . \label{id1}%
\end{align}

It can be shown that $E_{m}\left(  x_{1},\cdots,x_{m}\right)  $ satisfies a
iteration relation%
\begin{align}
E_{m}\left(  x_{1},\cdots,x_{m}\right)   &  =\frac{1}{m!}\frac{\partial^{m}%
}{\partial z^{m}}A\left(  x_{1},\cdots;z\right)  |_{z=0}=\frac{1}{m!}%
\frac{\partial^{m-1}}{\partial z^{m-1}}\left(  \frac{\partial B}{\partial
z}e^{B}\right)  |_{z=0}\nonumber\\
&  =\frac{1}{m}\underset{n=0}{\overset{m-1}{\sum}}\frac{1}{n!}\left(
\frac{\partial^{n}}{\partial z^{n}}e^{B}\right)  \frac{1}{\left(
m-n-1\right)  !}\frac{\partial^{m-1-n}}{\partial z^{m-1-n}}\left(
\frac{\partial B}{\partial z}\right)  |_{z=0}\nonumber\\
&  =\frac{1}{m}\underset{n=0}{\overset{m-1}{\sum}}\frac{1}{n!}\left(
\frac{\partial^{n}}{\partial z^{n}}e^{B}\right)  |_{z=0}\left[  x_{m-n}%
+O\left(  z\right)  \right]  |_{z=0}=\frac{1}{m}%
\underset{n=0}{\overset{m-1}{\sum}}E_{n}\left(  x_{1},\cdots,x_{n}\right)
x_{m-n}.
\end{align}
If we put $x_{i}=\alpha_{-i}$ for all $i$, we obtain%
\begin{equation}
E_{m}\left(  \alpha_{-1},\alpha_{-2},\cdots;\alpha_{-m}\right)  =\frac{1}%
{m}\underset{n=1}{\overset{m-1}{\sum}}E_{n}\left(  \alpha_{-1},\alpha
_{-2},\cdots;\alpha_{-n}\right)  \alpha_{-\left(  m-n\right)  }.
\end{equation}
Similarly, $F_{M}\left(  x_{1},x_{2}\right)  $ obeys a iteration relation%

\begin{align}
&  F_{M}\left(  x_{1},x_{2}\right)  =\frac{1}{M!}\frac{\partial^{M}}{\partial
z^{M}}G\left(  x_{1},x_{2};z\right)  |_{z=0}\nonumber\\
&  =\frac{1}{M!}\frac{\partial^{M}}{\partial z^{M}}\left(  1-zx_{1}\right)
^{-T_{1}}\left(  1-zx_{2}\right)  ^{-T_{2}}|_{z=0}\nonumber\\
&  =\frac{1}{M!}\frac{\partial^{M-1}}{\partial z^{M-1}}\left[  \left(
1-zx_{1}\right)  ^{-T_{1}}\left(  1-zx_{2}\right)  ^{-T_{2}}\left(
\frac{x_{1}T_{1}}{1-zx_{1}}+\frac{x_{1}T_{2}}{1-zx_{2}}\right)  \right]
|_{z=0},
\end{align}
which can be written as%
\begin{align}
&  =\frac{1}{M!}\underset{N=0}{\overset{M-1}{\sum}}\left[
\begin{array}
[c]{c}%
\frac{\left(  M-1\right)  !}{N!\left(  M-1-N\right)  !}\frac{\partial^{N}%
}{\partial z^{N}}\left(  \left(  1-zx_{1}\right)  ^{-T_{1}}\left(
1-zx_{2}\right)  ^{-T_{2}}\right)  |_{z=0}\\
\times\frac{\partial^{M-1-N}}{\partial z^{M-1-N}}\left(  \frac{x_{1}T_{1}%
}{1-zx_{1}}+\frac{x_{1}T_{2}}{1-zx_{2}}\right)  |_{z=0}%
\end{array}
\right] \nonumber\\
&  =\frac{1}{M}\underset{N=0}{\overset{M-1}{\sum}}\frac{1}{N!}\frac
{\partial^{N}}{\partial z^{N}}\left(  \left(  1-zx_{1}\right)  ^{-T_{1}%
}\left(  1-zx_{2}\right)  ^{-T_{2}}\right)  |_{z=0}\left(  T_{1}x_{1}%
^{M-N}+T_{2}x_{2}^{M-N}\right) \nonumber\\
&  =\frac{1}{M}\underset{N=0}{\overset{M-1}{\sum}}F_{N}\left(  x_{1}%
,x_{2}\right)  \left(  T_{1}x_{1}^{M-N}+T_{2}x_{2}^{M-N}\right)  .
\end{align}
This completes the proof of Eq.(\ref{ff}). It is straightforward to generalize
Eq.(\ref{ff}) to%
\begin{equation}
F_{M}\left(  x_{1},x_{2},\cdots,x_{n}\right)
=\underset{N=0}{\overset{M-1}{\sum}}F_{N}\left(  x_{1},x_{2},\cdots
,x_{n}\right)  \left[  T_{1}x_{1}^{M-N}+T_{2}x_{2}^{M-N}+\cdots+T_{n}%
x_{n}^{M-N}\right]  . \label{FM}%
\end{equation}

%\bibliographystyle{unsrt}
%\bibliography{Review}

\end{document}